\documentclass[]{article}

\usepackage[a4paper,margin=3.5cm]{geometry}

\usepackage{graphicx}
\usepackage{subcaption}
\usepackage[utf8]{inputenc}
\usepackage{amsmath,amsfonts,amssymb}
\usepackage{bbm}
\usepackage{hyperref}
\usepackage{todonotes}
\usepackage{color}
\usepackage{tikz}
\usepackage{gnuplot-lua-tikz}
\usepackage{algorithm}
\usepackage{algorithmic}
\usepackage{booktabs}

\usepackage[square,numbers,sort&compress]{natbib}

\newcommand{\R}{\mathbb{R}}

\newcommand{\Gt}{{\Gamma_\text{top}}}
\newcommand{\Gi}{{\Gamma_\text{int}}}

\newcommand{\Gb}{{\Gamma_\text{bottom}}}
\newcommand{\Oo}{ {\Omega_\text{out}} }
\newcommand{\Oone}{ {\Omega_{\text{cell}, 1}} }

\newcommand{\OiN}{ {\Omega_{\text{cell}, N}} }

\newcommand{\HOneZero}{{H^1_0(\Omega, \R^d)}}

\newcommand{\nuElas}{{\nu_\text{elast}}}
\newcommand{\nuVol}{{\nu_\text{vol}}}
\newcommand{\nuPeri}{{\nu_\text{peri}}}
\newcommand{\nuPenalty}{{\nu_\text{penalty}}}

\newcommand{\JElas}{{J_\text{elast}}}
\newcommand{\JVol}{{J_\text{vol}}}
\newcommand{\JPeri}{{J_\text{peri}}}

\providecommand{\keywords}[1]
{
  \small	
  \textbf{Keywords:} #1
}

\providecommand{\classification}[1]
{
  \small	
  \textbf{Mathematics Subject Classification (2010):} #1
}

\begin{document}

\title{Parallel 3d shape optimization for cellular composites on large distributed-memory clusters}

\author{Jose Pinzon\thanks{High Performance Computing in the Engineering Sciences, Ruhr-Universität Bochum, Universitätsstraße 150, 44801 Bochum} \and Martin Siebenborn\thanks{Department of Mathematics, Universität Hamburg, Bundesstraße 55, 20146 Hamburg} \and Andreas Vogel\footnotemark[1]
}

\date{}

\maketitle

\begin{abstract}
Skin modeling is an ongoing research area that highly benefits from modern parallel algorithms. This article aims at applying shape optimization to compute cell size and arrangement for elastic energy minimization of a cellular composite material model for the upper layer of the human skin. A gradient-penalized shape optimization algorithm is employed and tested on the distributed-memory cluster {\it Hazel Hen}, HLRS, Germany. The performance of the algorithm is studied in two benchmark tests. First, cell structures are optimized with respect to purely geometric aspects. The model is then extended such that the composite is optimized to withstand applied deformations. In both settings, the algorithm is investigated in terms of weak and strong scalability. The results for the geometric test reflect Kelvin's conjecture that the optimal space-filling design of cells with minimal surface is given by tetrakaidecahedrons. The PDE-constrained test case is chosen in order to demonstrate the influence of the deformation gradient penalization on fine inter-cellular channels in the composite and its influence on the multigrid convergence. A scaling study is presented for up to 12,288 cores and 3 billion DoFs.
\end{abstract}

\keywords{Shape optimization, multigrid methods, algorithmic scalability, skin modeling\\[0.2cm]}
\classification{49Q10, 35Q93, 65M55, 65Y05}

\section{Introduction}
Skin modeling and simulation is an active research field which benefits from the use of advanced numerical techniques. For detailed simulations, the finite element method is typically employed to solve complex models that attempt to closely resemble real biological behavior (see, e.g., \cite{Benitez2017,Leyva-Mendivil2015,MCBRIDE2016201}). In such simulations, especially for three-dimensional models, the use of efficient algorithms and large distributed-memory clusters becomes mandatory. Many different skin models have been proposed to address different properties and phenomena to be simulated. Surveys of experimental and modeling approaches for skin mechanics are provided in  \cite{jor2013,Limbert2017MathematicalAC,joodaki2018,Querleux2014} which mostly consider the skin layers as a homogeneous medium described by a constitutive model. The influence of cell geometries on the permeability of skin layers has been considered in several works without addressing mechanical aspects and a survey is presented, e.g., in \cite{Naegel09}. The upper layer of the skin is typically modeled geometrically via brick-and-mortar models \cite{Elias1983,heisig1996non} or employing tetrakaidecahedral cells \cite{Menton76,allen1976significance,Christophers74,Wittum2017a}. The latter choice provides a space-filling arrangement with minimal surface \cite{Thomson1887} which is referred to as Kelvin's conjecture. Only few works are devoted to the mechanical properties of the human skin at cellular scale, such as the work in  \cite{Santoprete14} which couples a mechanical model with a cellular scale model.
In this work, we focus on the mechanical behavior of the skin on a cellular level. We treat the uppermost layer as a composite material with cell inclusions possessing different mechanical properties. We then investigate via techniques from numerical PDE-constrained shape optimization the optimal shape of the cells separated by interfaces while keeping the topology fixed. 
The presented work extends previous results in \cite{VogelSiebenborn2019}, where a 2d cellular, elastic composite material is minimized with respect to the elastic energy with a gradient-penalized optimization algorithm.

The broad field of shape and topology optimization methods has a long history. We refer the reader to~\cite{sokolowski1992introduction,Haslinger2003Introduction,allaire2012shape,delfour2001shapes,bendsoe2011topology} for an overview on the topic.
PDE-constrained shape optimization is nowadays applied to a wide variety of areas
from fluid-dynamics \cite{schmidt2013three, BaLiUl, garcke2016stable}, image restoration and segmentation \cite{hintermuller2004second}, acoustics \cite{udawalpola2008optimization}, interface identification in  transmission  processes \cite{schulz2015structured, harbrecht2013numerical, naegel2015scalable}, electrostatics \cite{langer2015shape} and nano-optics \cite{hiptmair2018large} to composite material identification \cite{siebenborn2017algorithmic, naegel2015scalable}. The main advantage, compared to other optimization techniques, is the reduction of degrees of freedom through a previous knowledge of the domain's topology. Utilizing suitable shape spaces and inner products, it is possible to tailor the optimization methodology around efficient finite element solvers.
The mathematical methodology underlying this article is mainly developed in the work presented in~\cite{langer2015shape,schulz2015pde,schulz2016computational,siebenborn2017algorithmic}, especially the selection of a space of reachable shapes and suitable inner products, which are critical for the optimization setup. This is particularly important for the experiment considered here, where fine geometric structures have to be resolved.

For each step of the shape optimization, numerical approximations of the constrained problem as well as for the shape derivative have to be found. This requires to solve large sparse systems of equations and constitutes the runtime-dominating factor of the overall simulation. For this task, we employ a geometric multigrid method \cite{Hackbusch85} which is a very efficient class of iterative solvers with an optimal linear complexity. In particular, the quasi-constant number of required iterations independent of the mesh size allows for largest discretizations and an optimal weak-scaling. The method can be efficiently parallelized on distributed-memory machines as has been shown, e.g., in~\cite{Bastian96,Bergen06,Sampath10,sundar2012parallel,Gmeiner12,williams2014s} and also for algebraic multigrid approaches in~\cite{Baker11,baker2012scaling,Bastian12}. The parallelization helps to speedup simulations but also to accommodate for memory demands of detailed simulations. In this contribution, we build on our implementation of a hierarchically distributed multigrid approach \cite{Reiter2014gmg,Vogel2014ug4} which has already been efficiently employed to address human skin simulations \cite{Vogel15, Vogel16,hlrsbericht17,NIC18}.

The aim of this work is to optimize the three-dimensional cell shapes in the outermost layer of the human skin with respect to minimal elastic energy in a tensile test. For this purpose, we employ PDE-constrained shape optimization on distributed-memory clusters and efficient numerical methods.
The remainder of this paper is organized as follows: 
In Section~\ref{sec:model}, the 3d model is presented and the mathematical treatment from  \cite{VogelSiebenborn2019} summarized for convenience.
Section~\ref{sec:geom-optim} shows the results for  purely geometrical optimization resulting in tetrakaidecahedral cells.
In Section~\ref{sec:elasticity}, the results for the optimization with respect to mechanical aspects is presented.
Finally, we show the obtainable speedup in scaling studies in Section~\ref{sec:scaling} and conclude in Section~\ref{sec:conclusion}.

\section{Model equations and mathematical background}
\label{sec:model}

\begin{figure}[tbp]
	\begin{center}
\centering	\scalebox{0.4}{\includegraphics{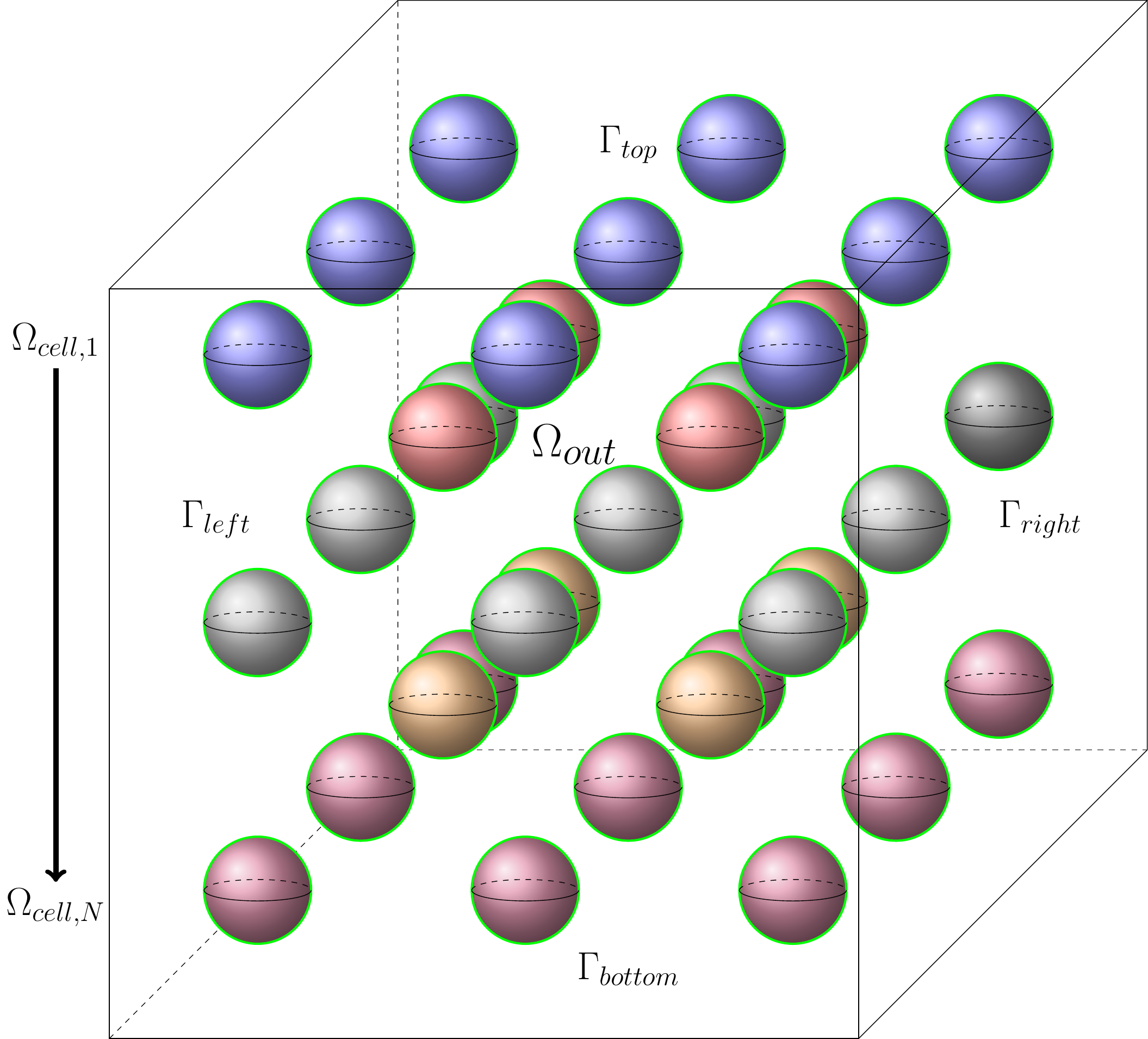}}
\end{center}
	\caption{Schematic view of the domain topology. Shown are cell inclusions with different properties (colors), the remaining void is filled with an outer material (not shown) }
	\label{fig:domain}
\end{figure}

This work extends the model presented in \cite{VogelSiebenborn2019} to a 3d cellular composite material inspired by the outermost layer of the human skin \cite{ADDRSkin2013,Limbert2017MathematicalAC,Naegel09,Querleux2014}. 
Using an initial shape design as depicted in Fig.~\ref{fig:domain}, we apply shape optimization in order to find a cell configuration that minimizes the stored energy by solving
\begin{equation}
\begin{aligned}
\min\limits_{\Gi}\quad & J(u, \Omega) :=  \frac{\nuElas}{2} \int_\Omega \sigma(u):\nabla u\, dx + \nuVol \int_{\Oo}1\, dx +  \nuPeri \int_{\Gi} 1\, ds\\
\text{s.t.}\quad &\int_\Omega \sigma(u) : \nabla w \, dx = \int_\Gt f \cdot w\, ds \quad \forall\, w.\end{aligned}
\label{eq:minimizationproblem}
\end{equation}
Here, $u$ denotes the deformation field which results from a force $f$ applied to the upper surface $\Gt$ and $\sigma(u) = \lambda \, \mathrm{tr}(\nabla u) I + \mu (\nabla u + \nabla u^T)$ is the stress tensor in terms of the Lame parameters $\lambda$ and $\mu$. On the lower boundary $\Gb$ zero Dirichlet conditions are prescribed and on the four side boundaries we assume homogeneous Neumann conditions. The objective function is the sum of the elastic energy and two purely geometric terms which try to minimize the outer volume $\Oo$ and the surface area of the internal interfaces $\Gi$ between materials, respectively. 
The factors $\nuElas,\nuVol,\nuPeri$ in (\ref{eq:minimizationproblem}) weight to which extend the  geometrical condition of a space-filling cell pattern or the energy minimization have to be fulfilled. Please note that the increase in cell volume and minimization of cell surface are two opposing objectives and the weighting factor therefore decide what kind of geometric optimum is found.

\begin{algorithm}[tbp]
	\caption{Gradient-penalized optimization algorithm from \cite{VogelSiebenborn2019}}
	\label{alg:algorithm1}
	\begin{algorithmic}[1]
\STATE Choose initial domain $\Omega^0$; choose $\nuElas, \nuVol, \nuPeri$; choose $\nuPenalty, b, t$
\FOR{$k = 0, 1, 2, \dots $}
\STATE Compute displacement field $u^k$, i.e. solve the elasticity problem, for the current $\Omega^k$
		\STATE Compute shape derivative $dJ(\Omega)[w]$, i.e. assemble
		$dJ(\Omega^k) = \nuElas d\JElas(u^k, \Omega^k) + \nuVol d\JVol(\Omega^k) + \nuPeri d\JPeri(\Omega^k)$\STATE Reset shape derivative in interior, i.e. set $dJ(\Omega^k)[\phi] = 0$ if $\mathrm{supp}(\phi) \cap \Gi = \emptyset$
		\STATE Compute descent direction $v$, i.e. solve $g(v,w) = dJ(\Omega^k)[w]$ for all $w \in \HOneZero$ 
		\STATE Update shape: $\Omega^{k+1} = \lbrace x + tv(x):\, x \in \Omega^k \rbrace$
\ENDFOR
	\end{algorithmic}
\end{algorithm}

In order to obtain a gradient descent method, we first have to consider shape sensitivities and then represent these with respect to an appropriate inner product. An expressions for the shape derivative $dJ(\Omega)[w]$ in direction $w$ of the objective $J$ in \eqref{eq:minimizationproblem} is presented in \cite{siebenborn2017algorithmic,VogelSiebenborn2019} for both the elastic energy and the geometric quantities. We closely follow the approach in \cite{schulz2016efficient} to represent the shape derivative with respect to an inner product defined over the surrounding space of the shape and simultaneously obtain a deformation field for the finite element mesh. Here, the bilinear form of the linear elasticity equation is chosen as an inner product. We thus solve 
\begin{equation}
a(v,w) := \int_{\Omega}\sigma(v):\nabla w\ dx = dJ(\Omega)[w], \quad \forall \; w \in \HOneZero, \label{eq:innerProduct}
\end{equation}
and obtain gradient descent directions $v$, which are applied as deformations to the finite element mesh.

A problem that immediately arises is that optimal solutions are not representable by a mesh, since for a less stiff material in $\Oo$ this domain part will completely vanish, leading to a contact between the inclusions.
From a modeling point of view this should be prevented, since material does not disappear 
and small inter-cellular lipid channels are observed in biological inspections.
We thus adopt the assumption that cells in the skin tissue can move relative to each other but do not overlap and that the lipid material has thus always to be present between cells.
Furthermore, from a computational point of view, the convergence of the multigrid solver is sensitive to the degenerating aspect ratio of elements in the channels of the composite.
In order to overcome these issues, a modification to the inner product of the shape space is proposed in \cite{VogelSiebenborn2019} as
\begin{equation}
g(v,w) := a(v,w) + \nuPenalty \int_\Omega \left( \nabla v: \nabla v - b^2 \right)^+ \, \nabla v : \nabla w \, dx = dJ(\Omega)[w] \quad \forall \; w \in \HOneZero. 
\label{eq:penalty}
\end{equation}
Large deformation gradients $\nabla v$ are thereby avoided through a modification of the descent direction, where $(\cdot)^+$ denotes the positive-part function. Two control parameters are introduced here, a threshold value $b$ and a weight factor $\nuPenalty$. Note that in (\ref{eq:penalty}) the penalization only becomes active, if the Frobenius norm of the gradient exceeds $b$.

In order to find the optimal shape for the cell interfaces in Fig.~\ref{fig:domain}, we apply the gradient-penalized optimization Algorithm \ref{alg:algorithm1} originally presented in \cite{VogelSiebenborn2019}. Here, $d\JElas(u^k, \Omega^k)$ denotes the shape derivative of the compliance energy, $d\JVol(u^k, \Omega^k)$ the derivative of the volume and $d\JPeri(u^k, \Omega^k)$ the derivative of the internal surface area. 

\section{Purely geometrical optimization problem}
\label{sec:geom-optim}

\begin{figure}[tbp]
	\centering
	\begin{subfigure}{0.45\textwidth}
		\centering
		\includegraphics[width=0.8\textwidth]{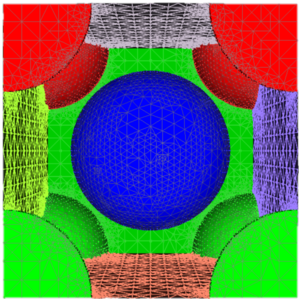}
\end{subfigure}\hfill
	\begin{subfigure}{0.45\textwidth}
		\centering
		\includegraphics[width=0.9\textwidth]{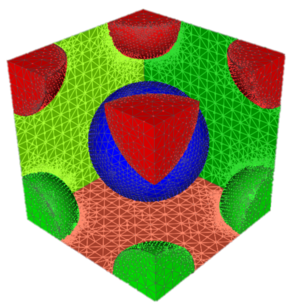}
\end{subfigure}
	\caption{Mesh of the initial shape configuration for the purely geometrical example. The center cell (blue) will be deformed into a tetrakaidecahedron}
	\label{fig:geommesh}
\end{figure}

In a first test, we apply Algorithm~\ref{alg:algorithm1} to a purely geometrical optimization problem by choosing $\nuElas = 0$ in~\eqref{eq:minimizationproblem}.
It is thus only optimized for a space-filling cell design with minimal surfaces.
This enables to compare numerical results with Kelvin's conjecture \cite{Thomson1887} which states that the optimal space-filling arrangement of cells with equal volume but minimal surface area is a 14-sided polyhedron named tetrakaidecahedron (TKD). It consists of 6 square and 8 hexagonal faces.

\begin{figure}[tbp]
	\centering
	\begin{subfigure}[t]{0.9\textwidth}
		\centering
		\makebox[0.04\textwidth][l]{(a)}
		\includegraphics[width=0.3\textwidth]{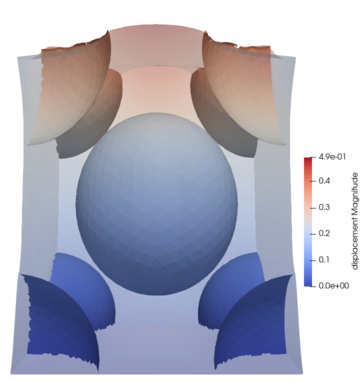}
\hspace{1cm}
		\includegraphics[width=0.35\textwidth]{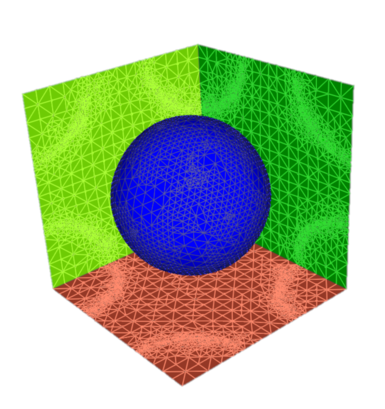}
	\end{subfigure}
	\begin{subfigure}[t]{0.9\textwidth}
		\centering
		\makebox[0.04\textwidth][l]{(b)}
		\includegraphics[width=0.3\textwidth]{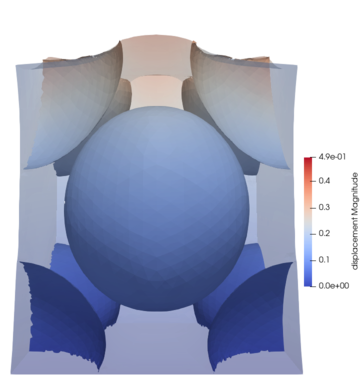}
\hspace{1cm}
		\includegraphics[width=0.35\textwidth]{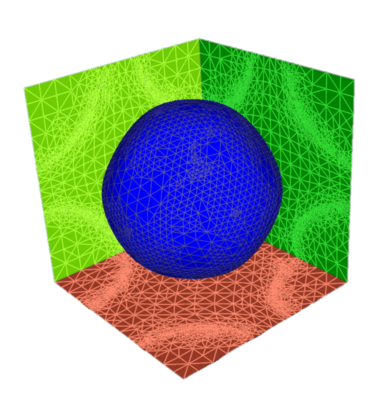}
	\end{subfigure}
	\begin{subfigure}[t]{0.9\textwidth}
		\centering
		\makebox[0.04\textwidth][l]{(c)}
		\includegraphics[width=0.3\textwidth]{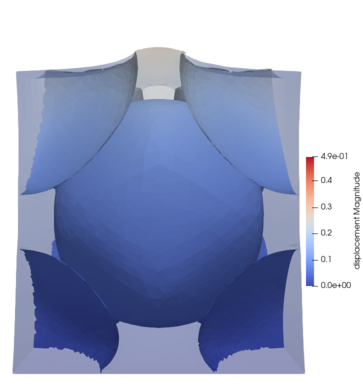}
\hspace{1cm}
		\includegraphics[width=0.35\textwidth]{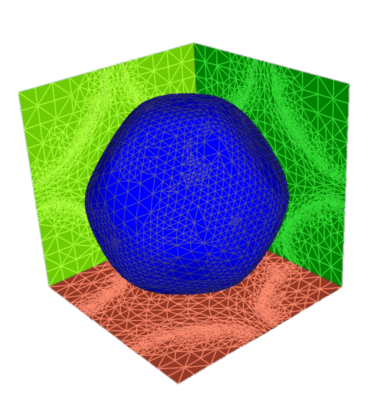}
	\end{subfigure}
	\begin{subfigure}[t]{0.9\textwidth}
		\centering
		\makebox[0.04\textwidth][l]{(d)}
		\includegraphics[width=0.3\textwidth]{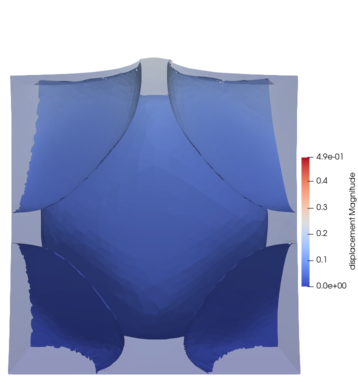}
\hspace{1cm}
		\includegraphics[width=0.35\textwidth]{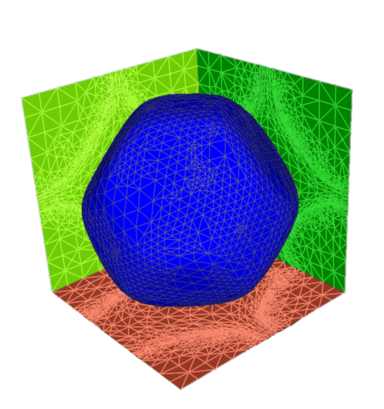}
	\end{subfigure}
	
	\caption{Purely geometrical test case. The deformed configuration of the cell arrangement (left) is shown with the shape of the center cell in reference configuration (right) after (a) 1, (b) 4, (c) 8, and (d) 11 optimization steps}
	\label{fig:geomSequence}
\end{figure}

We use the geometry shown in Fig.~\ref{fig:geommesh} as initial configuration $\Omega^0$. It features a symmetrical arrangement of 9 initially spherical cells. The center cell is surrounded by eight others at the domain's corners.
This is a periodic subset of the domain considered in Section \ref{sec:elasticity} and depicted in Fig.~\ref{fig:9spheresmesh}.
The geometric periodicity and symmetry of this optimization problem is incorporated into the inner products \eqref{eq:innerProduct} and \eqref{eq:penalty}, respectively.
For this purpose the solution space $\HOneZero$ is changed to require only zero deformations orthogonal to the boundaries, e.g., by fixing the first component of the deformation field $v$ to zero on the left and right side of the domain $\Omega$.
Thus, descent directions are allowed to feature tangential displacements, but the outer shape of $\Omega$ remains unchanged.

The parameters in this test are chosen as $\nuElas=0$, $\nuVol=2\cdot10^4$, $\nuPeri=10^2$, $\nuPenalty=10^7$, $b=0.001$, and no refinement levels. 
The values assigned to $\nuVol$ and $\nuPeri$ are chosen to achieve a maximum inner cell volume with minimum outer cell surface. The choice of the gradient penalty factor $\nuPenalty$ and threshold $b$ prevent geometric intersections and overlapping elements in the mesh before the TKD shape is formed.

Figure~\ref{fig:geomSequence} shows the deformation and volume filling for several optimization steps. In the first step (\ref{fig:geomSequence}a), the center cell has a spherical shape. While in the first optimization steps the distance between the inclusions is comparably large, descent directions mostly feature a volume increase. The cells retain a spherical shape due to the perimeter minimization, which can be seen in~Fig.~\ref{fig:geomSequence}b. Further optimization steps lead to an activation of the gradient penalization due to the small distance of the interfaces. We observe the appearance of hexagonal faces in~Fig.~\ref{fig:geomSequence}c. The algorithm is terminated when there is no more progress in the objective function observable due to the penalty in Eq.~\eqref{eq:penalty}. In~Fig.~\ref{fig:geomSequence}d, a TKD cell is already visible and the distance between the cells very small as the outer material has been filled by the inflated inclusions.

\begin{figure}[tbp]
	\centering
	\begin{subfigure}{0.45\textwidth}
		\centering
		\includegraphics[height=\textwidth]{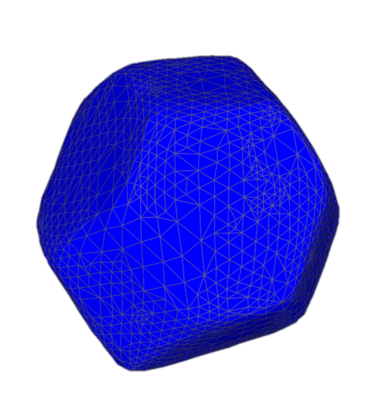}
		\caption{Inner cell with TKD shape}
	\end{subfigure}\hfill
	\begin{subfigure}{0.45\textwidth}
		\centering
		\includegraphics[height=\textwidth]{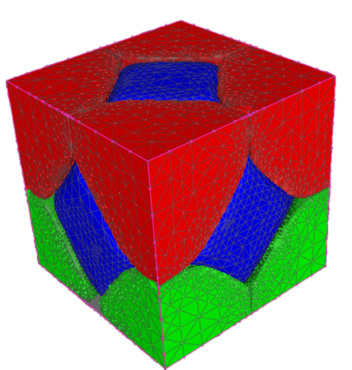}
		\caption{Final geometrical cell arrangement}
	\end{subfigure}
	\caption{Final optimization results for the purely geometrical optimization setting}
	\label{fig:geomresults}
\end{figure}

The final cell arrangement for the purely geometric optimization is shown in Fig.~\ref{fig:geomresults}. As expected, the center cell deforms into a TKD. The square faces result from the center cell volume expansion against the outer material surface. The hexagonal faces appear because of the limiting cells at the corners. As a result of the optimization process, the inflated cells almost occupy the whole inner volume. The outer material is left only in a thin layer, representing the surrounding lipid matrix, separating the cells from each other. Cell overlappings and contact between cells is avoided by the gradient penalization.

\section{Shape optimization for elastic energy minimization}
\label{sec:elasticity}
\begin{figure}[tbp]
	\centering
	\includegraphics[width=0.5\textwidth]{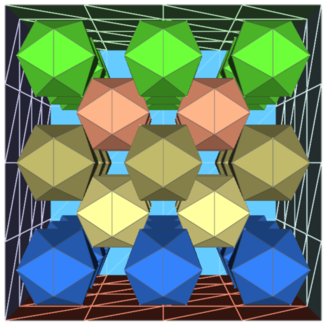}
	
	\caption{Coarse grid discretization of the domain $\Omega$ with 35 cells $\Oone,\dots, \OiN$}
	\label{fig:9spheresmesh}
\end{figure}

For the energy minimizing PDE-constrained shape optimization, the initial mesh as shown in Fig.~\ref{fig:9spheresmesh} is employed. It represents the discretized version of a domain similar to the one shown in Fig.~\ref{fig:domain}. The domain is composed of five cell levels, three with 9 initially spherical cells and two levels with 4 cells. Comparing Fig.~\ref{fig:9spheresmesh} with the domain used for the geometric test in Fig.~\ref{fig:geommesh}, it is evident that the center cell is equivalent to the central cell of the periodic mesh.

For the mechanical setting, the cells in Fig.~\ref{fig:9spheresmesh} are assigned different properties, but cells with the same color have the same material constants. The numbering of cell planes is from top to bottom. The stiffness of different cell types $\Omega_{cell,k}, k=0,...,4$ increase linearly from top to bottom as
\begin{align}
& (\lambda, \mu) = (\lambda_k, \mu_k), \text{ on } \Omega_\text{cell,k}, \text{ with }
\notag 	\\
\lambda_k &= (1 - \frac{k}{4}) \cdot  \lambda_\text{top} + \frac{k}{4}  \cdot \lambda_\text{bottom} , &&
\lambda_\text{top} = 1.0,~~\lambda_\text{bottom} = 10.0,
\\
\mu_k &= (1 - \frac{k}{4}) \cdot \mu_\text{top} +  \frac{k}{4} \cdot \mu_\text{bottom}, &&
\mu_\text{top} = 0.12,~~\mu_\text{bottom} = 0.2.
\end{align}
The material in $\Oo$ is assigned the material properties
\begin{align}
(\lambda, \mu) = (0.5, 0.1), \qquad & \text{ on } \Omega_\text{out},
\end{align}
which favors a cell volume expansion to maximize stiffness.

\begin{figure}[tbp]
	\centering
	\begin{subfigure}[t]{0.9\textwidth}
		\centering
		\makebox[0.04\textwidth][l]{(a)}
		\includegraphics[width=0.3\textwidth]{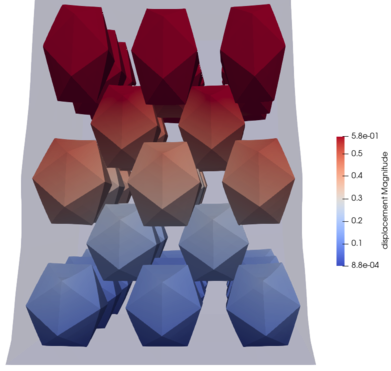}
		\hfill
		\includegraphics[width=0.3\textwidth]{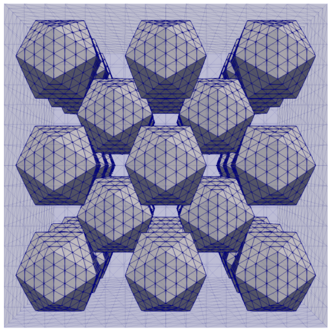}
		\includegraphics[width=0.25\textwidth]{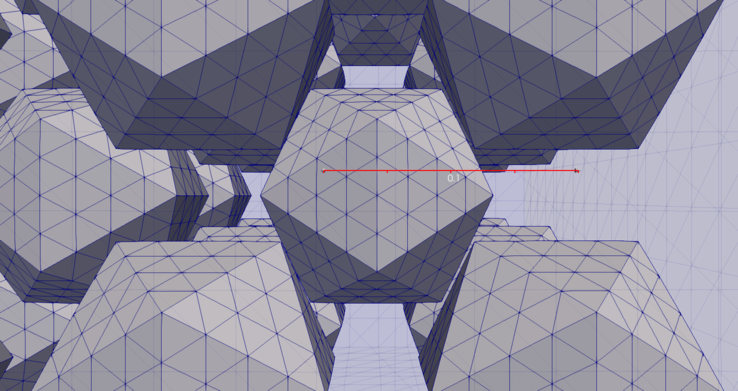}
\end{subfigure}
	\begin{subfigure}[t]{0.9\textwidth}
		\centering
		\makebox[0.04\textwidth][l]{(b)}
		\includegraphics[width=0.3\textwidth]{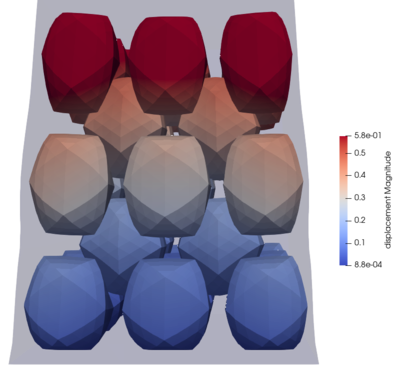}
		\hfill
		\includegraphics[width=0.3\textwidth]{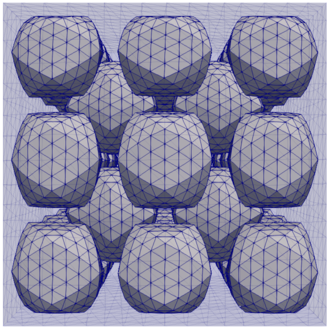}
		\includegraphics[width=0.25\textwidth]{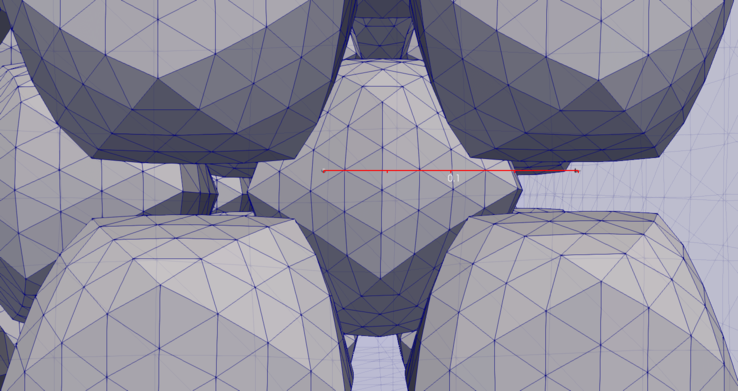}
\end{subfigure}
	\begin{subfigure}[t]{0.9\textwidth}
		\centering
		\makebox[0.04\textwidth][l]{(c)}
		\includegraphics[width=0.3\textwidth]{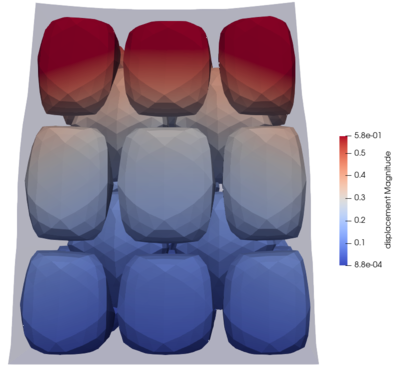}
		\hfill
		\includegraphics[width=0.3\textwidth]{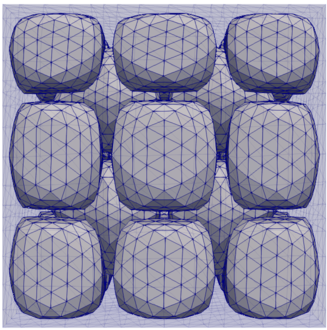}
		\includegraphics[width=0.25\textwidth]{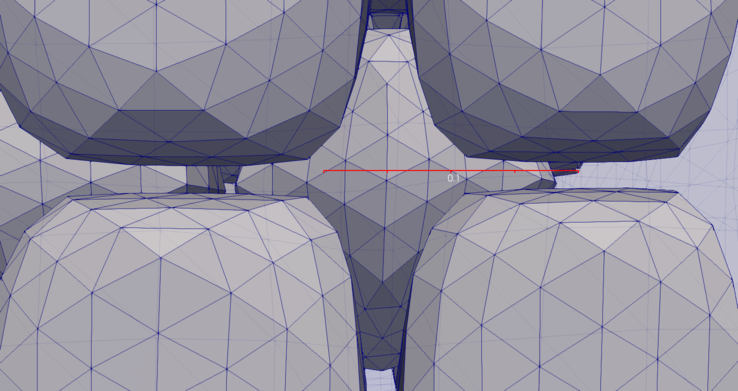}
\end{subfigure}
	\begin{subfigure}[t]{0.9\textwidth}
		\centering
		\makebox[0.04\textwidth][l]{(d)}
		\includegraphics[width=0.3\textwidth]{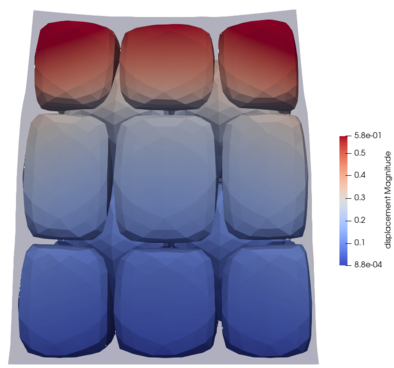}
		\hfill
		\includegraphics[width=0.3\textwidth]{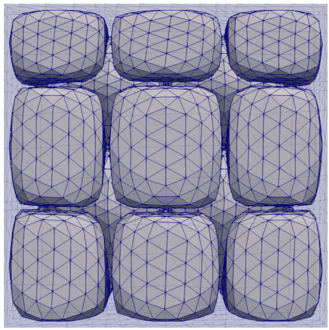}
		\includegraphics[width=0.25\textwidth]{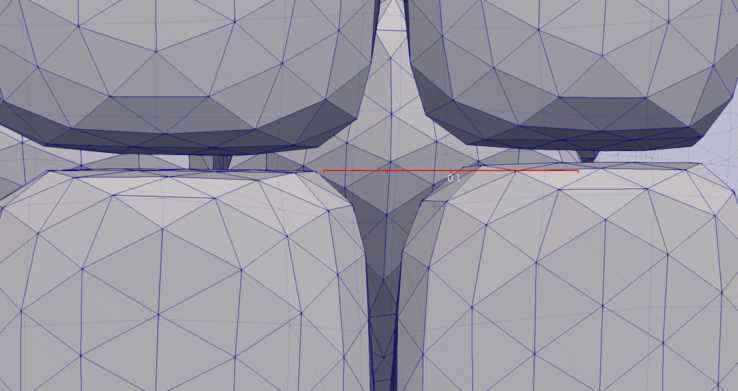}
\end{subfigure}
	\begin{subfigure}[t]{0.9\textwidth}
		\centering
		\makebox[0.04\textwidth][l]{(e)}
		\includegraphics[width=0.3\textwidth]{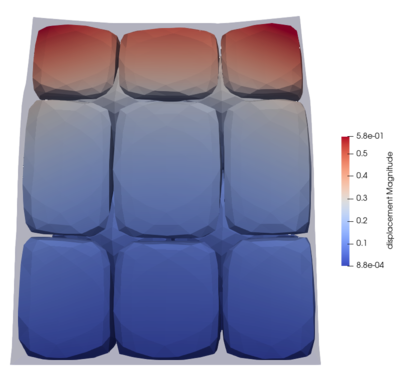}
		\hfill
		\includegraphics[width=0.3\textwidth]{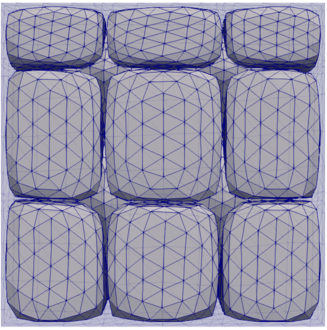}
		\includegraphics[width=0.25\textwidth]{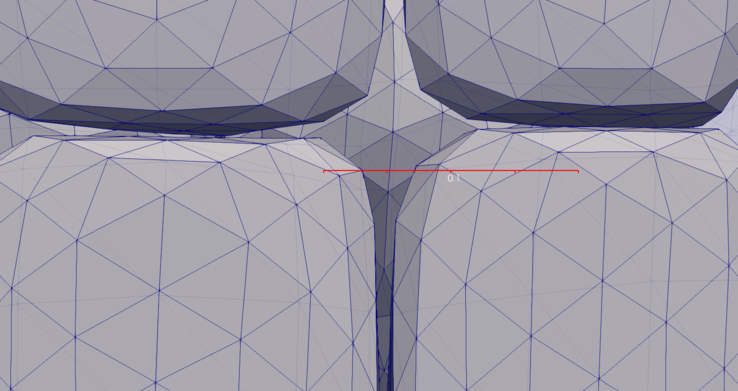}
\end{subfigure}
	\caption{Deformation for the elastic optimization. Deformed configuration (left) and reference configuration (right) shown for optimization step: (a) 1 (b) 11 (c) 21 (d) 31 (e) 41}
	\label{fig:OptimSteps}
\end{figure}

We present results for $\nuElas=10^5$, $\nuVol=10^4$, $\nuPeri=10^3$, $\nuPenalty=10^8$, $b=0.001$, and two levels of refinement. 
The ratio between $\nuVol$ and $\nuPeri$ is such that rapid cell volume and outer surface increase is promoted. Stepsizes in the optimization are controlled and limited via the choice of $b$ and $\nuPenalty$.
The effect of the shape optimization on the resulting displacements due to the applied force $f$ on the upper surface $\Gt$ is visualized in Fig.~\ref{fig:OptimSteps}. The left column of images shows the displacement field $u$ as a deformation to the finite element mesh. The reference configuration for optimization steps is visualized in the right column.
It can be observed that the stepsizes significantly reduce while the distance between interfaces shrinks. A deformation field which further reduces the width of a thin channel exhibits large gradients. Thereby, the gradient penalization in Eq.~\eqref{eq:penalty} is triggered by the positive part becoming active.
Thus, the bilinear form changes to a nonlinear behavior significantly reducing the stepsizes orthogonal to the channels which prevents degeneracy and overlapping cells.

\begin{figure}[tbp]
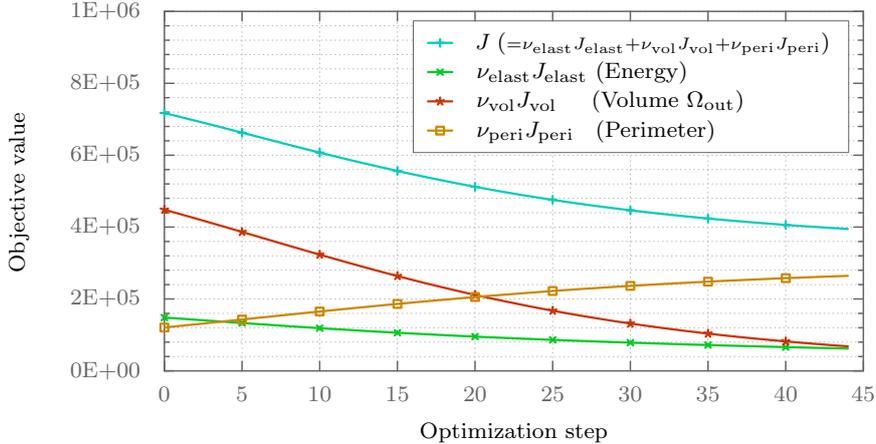

	\centering
	\scalebox{0.7}{\footnotesize

 }
	\caption{Objective function for an elastic optimization using $\nuElas=10^5$, $\nuPenalty=10^8$, $\nuVol=10^4$, $\nuPeri=10^3$, and threshold $b=10^{-3}$}
	\label{fig:elasticObjective}
\end{figure}

The objective function and its individual components are presented in Fig.~\ref{fig:elasticObjective}. Here it can be observed that descent directions are computed for the weighted objective function $J$, but not for its components $\JElas, \JPeri$ and $\JVol$. Thus, while $J$ is monotonically decreasing, the surface of $\Gi$ increases and is balanced against the gain for the outer volume minimization. Note that these particular results depend on the choice of the weighting factors $\nuElas, \nuPeri$ and $\nuVol$.

\begin{figure}[tbp]
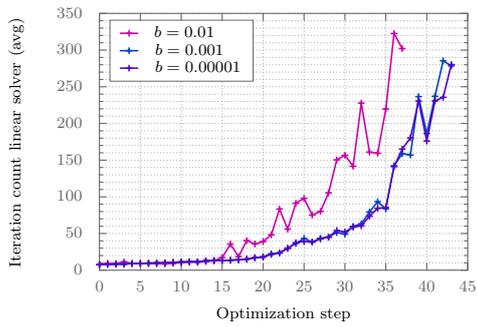
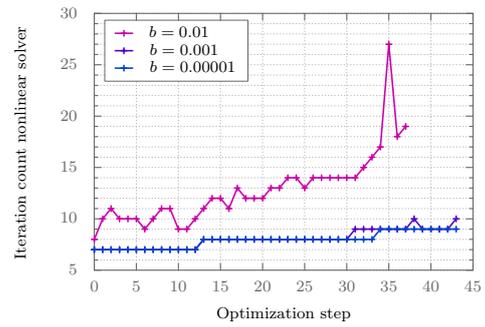
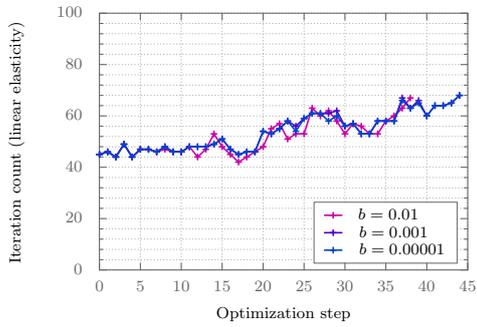
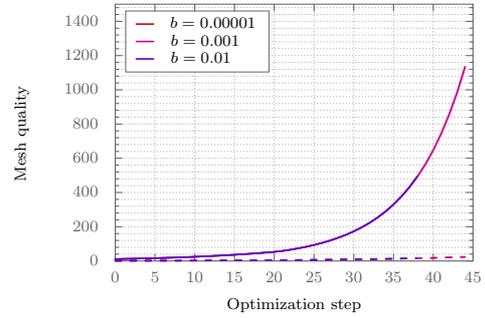

	\centering
	\begin{subfigure}[t]{0.45\textwidth}
		\centering
\scalebox{0.5}{\footnotesize

 }
		\caption{Average number of linear iteration steps to reduce the residuum by a factor of $10^{-3}$ within a Newton step}
	\end{subfigure}
	\hfill
	\begin{subfigure}[t]{0.45\textwidth}
		\centering
\scalebox{0.5}{\footnotesize%
 }
		\caption{Number of steps for the nonlinear iteration (Newton method)}
	\end{subfigure}
	\begin{subfigure}[t]{0.45\textwidth}
		\centering
\scalebox{0.5}{\footnotesize%
 }
		\caption{Number of iterations for the linear solver of the elasticity problem}
	\end{subfigure}
	\hfill
	\begin{subfigure}[t]{0.45\textwidth}
		\centering
\scalebox{0.5}{\footnotesize%
 }
		
		\caption{Mesh quality (aspect ratio); shown are maximal (solid) and average (dashed) values}
	\end{subfigure}
	\caption{Iteration counts for linear and nonlinear solver (initial mesh 2 times refined)}
	\label{fig:IterationCount}
\end{figure}

The representation of shape gradients in Eq.~\eqref{eq:penalty} is solved using a Newton method. The stopping criterion is chosen as an absolute or relative error reduction of $10^{-9}$, whichever occurs first. The linearized problems are solved with a BiCGStab method preconditioned by a geometric multigrid method. The multigrid uses a Block-Jacobi smoother with 3 pre- and postsmoothing steps in a V-cycle, assembled coarse grid matrices, and a LU base solver. The error reduction criteria for the linear elasticity problem is $10^{-10}$, and for the linearizations within the Newton method $10^{-3}$, which is sufficient to retain quadratic convergence. The maximum number of iterations is set to 2000 in both cases.
The performance of the solver is analyzed in Fig.~\ref{fig:IterationCount}. 
We show the iteration counts per optimization step in terms of different values of the threshold $b$ for the Newton solver, the linearized problem within the Newton solver, and the elasticity problem solver. The mesh quality is depicted for every step. A larger deformation per optimization step is allowed for larger values of $b$. It can be seen that the iteration count for the Newton solver, including the linearization, increases significantly over the optimization steps for the largest value of $b$. A sufficiently small deformation gradient threshold prevents mesh degeneration and allows to compute more optimization steps. There is no discernible solver behavior differences once $b$ is small enough. For the linear elastic problem, the iteration counts increase from 40 to approximately 70 for the distorted mesh at the end of the simulation.

\section{Algorithmic scalability}
\label{sec:scaling}
In order to analyze the scalability of Algorithm~\ref{alg:algorithm1}, we utilize the {\it Hazel Hen} supercomputer at the High-Performance Computing Center Stuttgart (HLRS) in Germany. It is equipped with 7712 nodes, each with two Intel Xeon E5-2680 (24 cores per node) and a maximum of 128GB of memory per node. The strong and weak scaling studies measure the time for the first two optimization steps. The timings and speedup, relative to 24 cores, are shown in Fig.~\ref{fig:weak_scaling} and Fig.~\ref{fig:strong_scaling}. The initial core count starts at 24 in order to fully occupy a node, and increments fourfold and eightfold for strong and weak scaling, respectively. For the weak scaling, the mesh is refined once on every core count increment. In order to solve the coarse grid problem, we employ a serial LU solver on a single process. We employ ParMETIS \cite{ParMetis} for the load-balancing and the UG4 \cite{Vogel2014ug4} software package for parallel smoother and multigrid.

In Fig.~\ref{fig:weak_scaling}, we present the weak scaling results which are good for both the linear and nonlinear solvers up to 12,288 cores. There is a slight difference in the number of iterations observed, which are caused by the differences in the slightly different numerical problem due to the refinement. The number of DoFs increase  from 6 million to more than 3 billion. The time for assembly, solver setup and solver execution remains very constant over the whole range of tested cores. This corresponds to a quasi ideal speedup as seen from the plots. 

The results for the strong scaling are presented in Fig.~\ref{fig:strong_scaling}. The assembly shows a quasi-optimal  strong scaling due to the good load-balancing. The linear solve (sol) and initialization (init) results show a reduction in the speedup. This is due to the serial LU base solver which requires a considerable portion of the execution time at higher core counts. This imposes an upper limit to the maximum attainable speedup. The nonlinear solver results are almost ideal for the Newton phase (newton), but the initialization and solution of the linearization experience the same limitation imposed by the serial base solver. This suggests to replace the base solver with a more efficient parallel solver in future works.
Since we used a Jacobi smoother, we expect to have equal iteration counts for the strong scaling. However a slight increase is present, which is caused by round-off errors in the different parallel partitions.

\begin{figure}[tbp]
	\begin{subfigure}[t]{\textwidth}
		\centering
		\scalebox{0.5}{\footnotesize\begin{tikzpicture}[gnuplot]
\tikzset{every node/.append style={scale=1.50}}
\path (0.000,0.000) rectangle (12.500,8.750);
\gpcolor{rgb color={0.702,0.702,0.702}}
\gpsetlinetype{gp lt axes}
\gpsetdashtype{gp dt axes}
\gpsetlinewidth{1.00}
\draw[gp path] (2.115,1.613)--(11.671,1.613);
\gpcolor{rgb color={0.400,0.400,0.400}}
\gpsetlinetype{gp lt border}
\gpsetdashtype{gp dt solid}
\gpsetlinewidth{2.00}
\draw[gp path] (2.115,1.613)--(1.980,1.613);
\node[gp node right,font={\fontsize{8.0pt}{9.6pt}\selectfont}] at (1.980,1.613) {0};
\gpcolor{rgb color={0.702,0.702,0.702}}
\gpsetlinetype{gp lt axes}
\gpsetdashtype{gp dt axes}
\gpsetlinewidth{1.00}
\draw[gp path] (2.115,1.791)--(11.671,1.791);
\gpcolor{rgb color={0.400,0.400,0.400}}
\gpsetlinetype{gp lt border}
\gpsetdashtype{gp dt solid}
\gpsetlinewidth{2.00}
\draw[gp path] (2.115,1.791)--(2.048,1.791);
\gpcolor{rgb color={0.702,0.702,0.702}}
\gpsetlinetype{gp lt axes}
\gpsetdashtype{gp dt axes}
\gpsetlinewidth{1.00}
\draw[gp path] (2.115,1.969)--(11.671,1.969);
\gpcolor{rgb color={0.400,0.400,0.400}}
\gpsetlinetype{gp lt border}
\gpsetdashtype{gp dt solid}
\gpsetlinewidth{2.00}
\draw[gp path] (2.115,1.969)--(2.048,1.969);
\gpcolor{rgb color={0.702,0.702,0.702}}
\gpsetlinetype{gp lt axes}
\gpsetdashtype{gp dt axes}
\gpsetlinewidth{1.00}
\draw[gp path] (2.115,2.147)--(11.671,2.147);
\gpcolor{rgb color={0.400,0.400,0.400}}
\gpsetlinetype{gp lt border}
\gpsetdashtype{gp dt solid}
\gpsetlinewidth{2.00}
\draw[gp path] (2.115,2.147)--(2.048,2.147);
\gpcolor{rgb color={0.702,0.702,0.702}}
\gpsetlinetype{gp lt axes}
\gpsetdashtype{gp dt axes}
\gpsetlinewidth{1.00}
\draw[gp path] (2.115,2.325)--(11.671,2.325);
\gpcolor{rgb color={0.400,0.400,0.400}}
\gpsetlinetype{gp lt border}
\gpsetdashtype{gp dt solid}
\gpsetlinewidth{2.00}
\draw[gp path] (2.115,2.325)--(2.048,2.325);
\gpcolor{rgb color={0.702,0.702,0.702}}
\gpsetlinetype{gp lt axes}
\gpsetdashtype{gp dt axes}
\gpsetlinewidth{1.00}
\draw[gp path] (2.115,2.503)--(11.671,2.503);
\gpcolor{rgb color={0.400,0.400,0.400}}
\gpsetlinetype{gp lt border}
\gpsetdashtype{gp dt solid}
\gpsetlinewidth{2.00}
\draw[gp path] (2.115,2.503)--(1.980,2.503);
\node[gp node right,font={\fontsize{8.0pt}{9.6pt}\selectfont}] at (1.888,2.503) {20};
\gpcolor{rgb color={0.702,0.702,0.702}}
\gpsetlinetype{gp lt axes}
\gpsetdashtype{gp dt axes}
\gpsetlinewidth{1.00}
\draw[gp path] (2.115,2.681)--(11.671,2.681);
\gpcolor{rgb color={0.400,0.400,0.400}}
\gpsetlinetype{gp lt border}
\gpsetdashtype{gp dt solid}
\gpsetlinewidth{2.00}
\draw[gp path] (2.115,2.681)--(2.048,2.681);
\gpcolor{rgb color={0.702,0.702,0.702}}
\gpsetlinetype{gp lt axes}
\gpsetdashtype{gp dt axes}
\gpsetlinewidth{1.00}
\draw[gp path] (2.115,2.859)--(11.671,2.859);
\gpcolor{rgb color={0.400,0.400,0.400}}
\gpsetlinetype{gp lt border}
\gpsetdashtype{gp dt solid}
\gpsetlinewidth{2.00}
\draw[gp path] (2.115,2.859)--(2.048,2.859);
\gpcolor{rgb color={0.702,0.702,0.702}}
\gpsetlinetype{gp lt axes}
\gpsetdashtype{gp dt axes}
\gpsetlinewidth{1.00}
\draw[gp path] (2.115,3.037)--(11.671,3.037);
\gpcolor{rgb color={0.400,0.400,0.400}}
\gpsetlinetype{gp lt border}
\gpsetdashtype{gp dt solid}
\gpsetlinewidth{2.00}
\draw[gp path] (2.115,3.037)--(2.048,3.037);
\gpcolor{rgb color={0.702,0.702,0.702}}
\gpsetlinetype{gp lt axes}
\gpsetdashtype{gp dt axes}
\gpsetlinewidth{1.00}
\draw[gp path] (2.115,3.215)--(11.671,3.215);
\gpcolor{rgb color={0.400,0.400,0.400}}
\gpsetlinetype{gp lt border}
\gpsetdashtype{gp dt solid}
\gpsetlinewidth{2.00}
\draw[gp path] (2.115,3.215)--(2.048,3.215);
\gpcolor{rgb color={0.702,0.702,0.702}}
\gpsetlinetype{gp lt axes}
\gpsetdashtype{gp dt axes}
\gpsetlinewidth{1.00}
\draw[gp path] (2.115,3.393)--(11.671,3.393);
\gpcolor{rgb color={0.400,0.400,0.400}}
\gpsetlinetype{gp lt border}
\gpsetdashtype{gp dt solid}
\gpsetlinewidth{2.00}
\draw[gp path] (2.115,3.393)--(1.980,3.393);
\node[gp node right,font={\fontsize{8.0pt}{9.6pt}\selectfont}] at (1.888,3.393) {40};
\gpcolor{rgb color={0.702,0.702,0.702}}
\gpsetlinetype{gp lt axes}
\gpsetdashtype{gp dt axes}
\gpsetlinewidth{1.00}
\draw[gp path] (2.115,3.571)--(11.671,3.571);
\gpcolor{rgb color={0.400,0.400,0.400}}
\gpsetlinetype{gp lt border}
\gpsetdashtype{gp dt solid}
\gpsetlinewidth{2.00}
\draw[gp path] (2.115,3.571)--(2.048,3.571);
\gpcolor{rgb color={0.702,0.702,0.702}}
\gpsetlinetype{gp lt axes}
\gpsetdashtype{gp dt axes}
\gpsetlinewidth{1.00}
\draw[gp path] (2.115,3.749)--(11.671,3.749);
\gpcolor{rgb color={0.400,0.400,0.400}}
\gpsetlinetype{gp lt border}
\gpsetdashtype{gp dt solid}
\gpsetlinewidth{2.00}
\draw[gp path] (2.115,3.749)--(2.048,3.749);
\gpcolor{rgb color={0.702,0.702,0.702}}
\gpsetlinetype{gp lt axes}
\gpsetdashtype{gp dt axes}
\gpsetlinewidth{1.00}
\draw[gp path] (2.115,3.927)--(11.671,3.927);
\gpcolor{rgb color={0.400,0.400,0.400}}
\gpsetlinetype{gp lt border}
\gpsetdashtype{gp dt solid}
\gpsetlinewidth{2.00}
\draw[gp path] (2.115,3.927)--(2.048,3.927);
\gpcolor{rgb color={0.702,0.702,0.702}}
\gpsetlinetype{gp lt axes}
\gpsetdashtype{gp dt axes}
\gpsetlinewidth{1.00}
\draw[gp path] (2.115,4.105)--(11.671,4.105);
\gpcolor{rgb color={0.400,0.400,0.400}}
\gpsetlinetype{gp lt border}
\gpsetdashtype{gp dt solid}
\gpsetlinewidth{2.00}
\draw[gp path] (2.115,4.105)--(2.048,4.105);
\gpcolor{rgb color={0.702,0.702,0.702}}
\gpsetlinetype{gp lt axes}
\gpsetdashtype{gp dt axes}
\gpsetlinewidth{1.00}
\draw[gp path] (2.115,4.283)--(11.671,4.283);
\gpcolor{rgb color={0.400,0.400,0.400}}
\gpsetlinetype{gp lt border}
\gpsetdashtype{gp dt solid}
\gpsetlinewidth{2.00}
\draw[gp path] (2.115,4.283)--(1.980,4.283);
\node[gp node right,font={\fontsize{8.0pt}{9.6pt}\selectfont}] at (1.888,4.283) {60};
\gpcolor{rgb color={0.702,0.702,0.702}}
\gpsetlinetype{gp lt axes}
\gpsetdashtype{gp dt axes}
\gpsetlinewidth{1.00}
\draw[gp path] (2.115,4.461)--(11.671,4.461);
\gpcolor{rgb color={0.400,0.400,0.400}}
\gpsetlinetype{gp lt border}
\gpsetdashtype{gp dt solid}
\gpsetlinewidth{2.00}
\draw[gp path] (2.115,4.461)--(2.048,4.461);
\gpcolor{rgb color={0.702,0.702,0.702}}
\gpsetlinetype{gp lt axes}
\gpsetdashtype{gp dt axes}
\gpsetlinewidth{1.00}
\draw[gp path] (2.115,4.639)--(11.671,4.639);
\gpcolor{rgb color={0.400,0.400,0.400}}
\gpsetlinetype{gp lt border}
\gpsetdashtype{gp dt solid}
\gpsetlinewidth{2.00}
\draw[gp path] (2.115,4.639)--(2.048,4.639);
\gpcolor{rgb color={0.702,0.702,0.702}}
\gpsetlinetype{gp lt axes}
\gpsetdashtype{gp dt axes}
\gpsetlinewidth{1.00}
\draw[gp path] (2.115,4.817)--(11.671,4.817);
\gpcolor{rgb color={0.400,0.400,0.400}}
\gpsetlinetype{gp lt border}
\gpsetdashtype{gp dt solid}
\gpsetlinewidth{2.00}
\draw[gp path] (2.115,4.817)--(2.048,4.817);
\gpcolor{rgb color={0.702,0.702,0.702}}
\gpsetlinetype{gp lt axes}
\gpsetdashtype{gp dt axes}
\gpsetlinewidth{1.00}
\draw[gp path] (2.115,4.994)--(11.671,4.994);
\gpcolor{rgb color={0.400,0.400,0.400}}
\gpsetlinetype{gp lt border}
\gpsetdashtype{gp dt solid}
\gpsetlinewidth{2.00}
\draw[gp path] (2.115,4.994)--(2.048,4.994);
\gpcolor{rgb color={0.702,0.702,0.702}}
\gpsetlinetype{gp lt axes}
\gpsetdashtype{gp dt axes}
\gpsetlinewidth{1.00}
\draw[gp path] (2.115,5.172)--(11.671,5.172);
\gpcolor{rgb color={0.400,0.400,0.400}}
\gpsetlinetype{gp lt border}
\gpsetdashtype{gp dt solid}
\gpsetlinewidth{2.00}
\draw[gp path] (2.115,5.172)--(1.980,5.172);
\node[gp node right,font={\fontsize{8.0pt}{9.6pt}\selectfont}] at (1.888,5.172) {80};
\gpcolor{rgb color={0.702,0.702,0.702}}
\gpsetlinetype{gp lt axes}
\gpsetdashtype{gp dt axes}
\gpsetlinewidth{1.00}
\draw[gp path] (2.115,5.350)--(11.671,5.350);
\gpcolor{rgb color={0.400,0.400,0.400}}
\gpsetlinetype{gp lt border}
\gpsetdashtype{gp dt solid}
\gpsetlinewidth{2.00}
\draw[gp path] (2.115,5.350)--(2.048,5.350);
\gpcolor{rgb color={0.702,0.702,0.702}}
\gpsetlinetype{gp lt axes}
\gpsetdashtype{gp dt axes}
\gpsetlinewidth{1.00}
\draw[gp path] (2.115,5.528)--(11.671,5.528);
\gpcolor{rgb color={0.400,0.400,0.400}}
\gpsetlinetype{gp lt border}
\gpsetdashtype{gp dt solid}
\gpsetlinewidth{2.00}
\draw[gp path] (2.115,5.528)--(2.048,5.528);
\gpcolor{rgb color={0.702,0.702,0.702}}
\gpsetlinetype{gp lt axes}
\gpsetdashtype{gp dt axes}
\gpsetlinewidth{1.00}
\draw[gp path] (2.115,5.706)--(11.671,5.706);
\gpcolor{rgb color={0.400,0.400,0.400}}
\gpsetlinetype{gp lt border}
\gpsetdashtype{gp dt solid}
\gpsetlinewidth{2.00}
\draw[gp path] (2.115,5.706)--(2.048,5.706);
\gpcolor{rgb color={0.702,0.702,0.702}}
\gpsetlinetype{gp lt axes}
\gpsetdashtype{gp dt axes}
\gpsetlinewidth{1.00}
\draw[gp path] (2.115,5.884)--(11.671,5.884);
\gpcolor{rgb color={0.400,0.400,0.400}}
\gpsetlinetype{gp lt border}
\gpsetdashtype{gp dt solid}
\gpsetlinewidth{2.00}
\draw[gp path] (2.115,5.884)--(2.048,5.884);
\gpcolor{rgb color={0.702,0.702,0.702}}
\gpsetlinetype{gp lt axes}
\gpsetdashtype{gp dt axes}
\gpsetlinewidth{1.00}
\draw[gp path] (2.115,6.062)--(11.671,6.062);
\gpcolor{rgb color={0.400,0.400,0.400}}
\gpsetlinetype{gp lt border}
\gpsetdashtype{gp dt solid}
\gpsetlinewidth{2.00}
\draw[gp path] (2.115,6.062)--(1.980,6.062);
\node[gp node right,font={\fontsize{8.0pt}{9.6pt}\selectfont}] at (1.888,6.062) {100};
\gpcolor{rgb color={0.702,0.702,0.702}}
\gpsetlinetype{gp lt axes}
\gpsetdashtype{gp dt axes}
\gpsetlinewidth{1.00}
\draw[gp path] (2.115,6.240)--(11.671,6.240);
\gpcolor{rgb color={0.400,0.400,0.400}}
\gpsetlinetype{gp lt border}
\gpsetdashtype{gp dt solid}
\gpsetlinewidth{2.00}
\draw[gp path] (2.115,6.240)--(2.048,6.240);
\gpcolor{rgb color={0.702,0.702,0.702}}
\gpsetlinetype{gp lt axes}
\gpsetdashtype{gp dt axes}
\gpsetlinewidth{1.00}
\draw[gp path] (2.115,6.418)--(11.671,6.418);
\gpcolor{rgb color={0.400,0.400,0.400}}
\gpsetlinetype{gp lt border}
\gpsetdashtype{gp dt solid}
\gpsetlinewidth{2.00}
\draw[gp path] (2.115,6.418)--(2.048,6.418);
\gpcolor{rgb color={0.702,0.702,0.702}}
\gpsetlinetype{gp lt axes}
\gpsetdashtype{gp dt axes}
\gpsetlinewidth{1.00}
\draw[gp path] (2.115,6.596)--(2.391,6.596);
\draw[gp path] (5.607,6.596)--(11.671,6.596);
\gpcolor{rgb color={0.400,0.400,0.400}}
\gpsetlinetype{gp lt border}
\gpsetdashtype{gp dt solid}
\gpsetlinewidth{2.00}
\draw[gp path] (2.115,6.596)--(2.048,6.596);
\gpcolor{rgb color={0.702,0.702,0.702}}
\gpsetlinetype{gp lt axes}
\gpsetdashtype{gp dt axes}
\gpsetlinewidth{1.00}
\draw[gp path] (2.115,6.774)--(2.391,6.774);
\draw[gp path] (5.607,6.774)--(11.671,6.774);
\gpcolor{rgb color={0.400,0.400,0.400}}
\gpsetlinetype{gp lt border}
\gpsetdashtype{gp dt solid}
\gpsetlinewidth{2.00}
\draw[gp path] (2.115,6.774)--(2.048,6.774);
\gpcolor{rgb color={0.702,0.702,0.702}}
\gpsetlinetype{gp lt axes}
\gpsetdashtype{gp dt axes}
\gpsetlinewidth{1.00}
\draw[gp path] (2.115,6.952)--(2.391,6.952);
\draw[gp path] (5.607,6.952)--(11.671,6.952);
\gpcolor{rgb color={0.400,0.400,0.400}}
\gpsetlinetype{gp lt border}
\gpsetdashtype{gp dt solid}
\gpsetlinewidth{2.00}
\draw[gp path] (2.115,6.952)--(1.980,6.952);
\node[gp node right,font={\fontsize{8.0pt}{9.6pt}\selectfont}] at (1.888,6.952) {120};
\gpcolor{rgb color={0.702,0.702,0.702}}
\gpsetlinetype{gp lt axes}
\gpsetdashtype{gp dt axes}
\gpsetlinewidth{1.00}
\draw[gp path] (2.115,7.130)--(2.391,7.130);
\draw[gp path] (5.607,7.130)--(11.671,7.130);
\gpcolor{rgb color={0.400,0.400,0.400}}
\gpsetlinetype{gp lt border}
\gpsetdashtype{gp dt solid}
\gpsetlinewidth{2.00}
\draw[gp path] (2.115,7.130)--(2.048,7.130);
\gpcolor{rgb color={0.702,0.702,0.702}}
\gpsetlinetype{gp lt axes}
\gpsetdashtype{gp dt axes}
\gpsetlinewidth{1.00}
\draw[gp path] (2.115,7.308)--(2.391,7.308);
\draw[gp path] (5.607,7.308)--(11.671,7.308);
\gpcolor{rgb color={0.400,0.400,0.400}}
\gpsetlinetype{gp lt border}
\gpsetdashtype{gp dt solid}
\gpsetlinewidth{2.00}
\draw[gp path] (2.115,7.308)--(2.048,7.308);
\gpcolor{rgb color={0.702,0.702,0.702}}
\gpsetlinetype{gp lt axes}
\gpsetdashtype{gp dt axes}
\gpsetlinewidth{1.00}
\draw[gp path] (2.115,7.486)--(2.391,7.486);
\draw[gp path] (5.607,7.486)--(11.671,7.486);
\gpcolor{rgb color={0.400,0.400,0.400}}
\gpsetlinetype{gp lt border}
\gpsetdashtype{gp dt solid}
\gpsetlinewidth{2.00}
\draw[gp path] (2.115,7.486)--(2.048,7.486);
\gpcolor{rgb color={0.702,0.702,0.702}}
\gpsetlinetype{gp lt axes}
\gpsetdashtype{gp dt axes}
\gpsetlinewidth{1.00}
\draw[gp path] (2.115,7.664)--(2.391,7.664);
\draw[gp path] (5.607,7.664)--(11.671,7.664);
\gpcolor{rgb color={0.400,0.400,0.400}}
\gpsetlinetype{gp lt border}
\gpsetdashtype{gp dt solid}
\gpsetlinewidth{2.00}
\draw[gp path] (2.115,7.664)--(2.048,7.664);
\gpcolor{rgb color={0.702,0.702,0.702}}
\gpsetlinetype{gp lt axes}
\gpsetdashtype{gp dt axes}
\gpsetlinewidth{1.00}
\draw[gp path] (2.115,7.842)--(2.391,7.842);
\draw[gp path] (5.607,7.842)--(11.671,7.842);
\gpcolor{rgb color={0.400,0.400,0.400}}
\gpsetlinetype{gp lt border}
\gpsetdashtype{gp dt solid}
\gpsetlinewidth{2.00}
\draw[gp path] (2.115,7.842)--(1.980,7.842);
\node[gp node right,font={\fontsize{8.0pt}{9.6pt}\selectfont}] at (1.888,7.842) {140};
\gpcolor{rgb color={0.702,0.702,0.702}}
\gpsetlinetype{gp lt axes}
\gpsetdashtype{gp dt axes}
\gpsetlinewidth{1.00}
\draw[gp path] (2.115,8.020)--(2.391,8.020);
\draw[gp path] (5.607,8.020)--(11.671,8.020);
\gpcolor{rgb color={0.400,0.400,0.400}}
\gpsetlinetype{gp lt border}
\gpsetdashtype{gp dt solid}
\gpsetlinewidth{2.00}
\draw[gp path] (2.115,8.020)--(2.048,8.020);
\gpcolor{rgb color={0.702,0.702,0.702}}
\gpsetlinetype{gp lt axes}
\gpsetdashtype{gp dt axes}
\gpsetlinewidth{1.00}
\draw[gp path] (2.115,8.198)--(11.671,8.198);
\gpcolor{rgb color={0.400,0.400,0.400}}
\gpsetlinetype{gp lt border}
\gpsetdashtype{gp dt solid}
\gpsetlinewidth{2.00}
\draw[gp path] (2.115,8.198)--(2.048,8.198);
\gpcolor{rgb color={0.702,0.702,0.702}}
\gpsetlinetype{gp lt axes}
\gpsetdashtype{gp dt axes}
\gpsetlinewidth{1.00}
\draw[gp path] (3.228,1.613)--(3.228,6.490);
\draw[gp path] (3.228,8.107)--(3.228,8.287);
\gpcolor{rgb color={0.400,0.400,0.400}}
\gpsetlinetype{gp lt border}
\gpsetdashtype{gp dt solid}
\gpsetlinewidth{2.00}
\draw[gp path] (3.228,1.613)--(3.228,1.478);
\node[gp node center,font={\fontsize{8.0pt}{9.6pt}\selectfont}] at (3.228,1.016) {24};
\gpcolor{rgb color={0.702,0.702,0.702}}
\gpsetlinetype{gp lt axes}
\gpsetdashtype{gp dt axes}
\gpsetlinewidth{1.00}
\draw[gp path] (5.870,1.613)--(5.870,8.287);
\gpcolor{rgb color={0.400,0.400,0.400}}
\gpsetlinetype{gp lt border}
\gpsetdashtype{gp dt solid}
\gpsetlinewidth{2.00}
\draw[gp path] (5.870,1.613)--(5.870,1.478);
\node[gp node center,font={\fontsize{8.0pt}{9.6pt}\selectfont}] at (5.870,1.016) {192};
\gpcolor{rgb color={0.702,0.702,0.702}}
\gpsetlinetype{gp lt axes}
\gpsetdashtype{gp dt axes}
\gpsetlinewidth{1.00}
\draw[gp path] (8.513,1.613)--(8.513,8.287);
\gpcolor{rgb color={0.400,0.400,0.400}}
\gpsetlinetype{gp lt border}
\gpsetdashtype{gp dt solid}
\gpsetlinewidth{2.00}
\draw[gp path] (8.513,1.613)--(8.513,1.478);
\node[gp node center,font={\fontsize{8.0pt}{9.6pt}\selectfont}] at (8.513,1.016) {1536};
\gpcolor{rgb color={0.702,0.702,0.702}}
\gpsetlinetype{gp lt axes}
\gpsetdashtype{gp dt axes}
\gpsetlinewidth{1.00}
\draw[gp path] (11.156,1.613)--(11.156,8.287);
\gpcolor{rgb color={0.400,0.400,0.400}}
\gpsetlinetype{gp lt border}
\gpsetdashtype{gp dt solid}
\gpsetlinewidth{2.00}
\draw[gp path] (11.156,1.613)--(11.156,1.478);
\node[gp node center,font={\fontsize{8.0pt}{9.6pt}\selectfont}] at (11.156,1.016) {12288};
\draw[gp path] (2.115,8.287)--(2.115,1.613)--(11.671,1.613)--(11.671,8.287)--cycle;
\gpcolor{color=gp lt color border}
\node[gp node center,rotate=-270] at (0.921,4.950) {Time [s]};
\node[gp node center] at (6.893,0.400) {Number of processes};
\gpcolor{rgb color={0.400,0.400,0.400}}
\gpsetlinewidth{1.00}
\draw[gp path] (2.391,6.490)--(2.391,8.107)--(5.607,8.107)--(5.607,6.490)--cycle;
\gpcolor{color=gp lt color border}
\node[gp node left] at (3.399,7.760) {$T_{\text{ass}}$};
\gpcolor{rgb color={0.800,0.000,0.063}}
\gpsetlinewidth{3.00}
\draw[gp path] (2.667,7.760)--(3.123,7.760);
\draw[gp path] (3.228,2.377)--(5.870,2.523)--(8.513,2.438)--(11.156,2.296);
\gpsetpointsize{5.20}
\gppoint{gp mark 1}{(3.228,2.377)}
\gppoint{gp mark 1}{(5.870,2.523)}
\gppoint{gp mark 1}{(8.513,2.438)}
\gppoint{gp mark 1}{(11.156,2.296)}
\gppoint{gp mark 1}{(2.895,7.760)}
\gpcolor{color=gp lt color border}
\node[gp node left] at (3.399,7.298) {$T_{\text{init}}$};
\gpcolor{rgb color={0.800,0.000,0.663}}
\draw[gp path] (2.667,7.298)--(3.123,7.298);
\draw[gp path] (3.228,1.750)--(5.870,1.779)--(8.513,1.764)--(11.156,1.761);
\gppoint{gp mark 2}{(3.228,1.750)}
\gppoint{gp mark 2}{(5.870,1.779)}
\gppoint{gp mark 2}{(8.513,1.764)}
\gppoint{gp mark 2}{(11.156,1.761)}
\gppoint{gp mark 2}{(2.895,7.298)}
\gpcolor{color=gp lt color border}
\node[gp node left] at (3.399,6.836) {$T_{\text{solve}}$};
\gpcolor{rgb color={0.329,0.000,0.800}}
\draw[gp path] (2.667,6.836)--(3.123,6.836);
\draw[gp path] (3.228,3.519)--(5.870,4.410)--(8.513,4.198)--(11.156,4.396);
\gppoint{gp mark 3}{(3.228,3.519)}
\gppoint{gp mark 3}{(5.870,4.410)}
\gppoint{gp mark 3}{(8.513,4.198)}
\gppoint{gp mark 3}{(11.156,4.396)}
\gppoint{gp mark 3}{(2.895,6.836)}
\gpdefrectangularnode{gp plot 1}{\pgfpoint{2.115cm}{1.613cm}}{\pgfpoint{11.671cm}{8.287cm}}
\end{tikzpicture}
 }
\scalebox{0.5}{\footnotesize\begin{tikzpicture}[gnuplot]
\tikzset{every node/.append style={scale=1.50}}
\path (0.000,0.000) rectangle (12.500,8.750);
\gpcolor{rgb color={0.702,0.702,0.702}}
\gpsetlinetype{gp lt axes}
\gpsetdashtype{gp dt axes}
\gpsetlinewidth{1.00}
\draw[gp path] (2.115,2.240)--(11.671,2.240);
\gpcolor{rgb color={0.400,0.400,0.400}}
\gpsetlinetype{gp lt border}
\gpsetdashtype{gp dt solid}
\gpsetlinewidth{2.00}
\draw[gp path] (2.115,2.240)--(1.980,2.240);
\node[gp node right,font={\fontsize{8.0pt}{9.6pt}\selectfont}] at (1.980,2.240) {1};
\gpcolor{rgb color={0.702,0.702,0.702}}
\gpsetlinetype{gp lt axes}
\gpsetdashtype{gp dt axes}
\gpsetlinewidth{1.00}
\draw[gp path] (2.115,3.032)--(11.671,3.032);
\gpcolor{rgb color={0.400,0.400,0.400}}
\gpsetlinetype{gp lt border}
\gpsetdashtype{gp dt solid}
\gpsetlinewidth{2.00}
\draw[gp path] (2.115,3.032)--(2.048,3.032);
\gpcolor{rgb color={0.702,0.702,0.702}}
\gpsetlinetype{gp lt axes}
\gpsetdashtype{gp dt axes}
\gpsetlinewidth{1.00}
\draw[gp path] (2.115,3.448)--(11.671,3.448);
\gpcolor{rgb color={0.400,0.400,0.400}}
\gpsetlinetype{gp lt border}
\gpsetdashtype{gp dt solid}
\gpsetlinewidth{2.00}
\draw[gp path] (2.115,3.448)--(2.048,3.448);
\gpcolor{rgb color={0.702,0.702,0.702}}
\gpsetlinetype{gp lt axes}
\gpsetdashtype{gp dt axes}
\gpsetlinewidth{1.00}
\draw[gp path] (2.115,3.731)--(11.671,3.731);
\gpcolor{rgb color={0.400,0.400,0.400}}
\gpsetlinetype{gp lt border}
\gpsetdashtype{gp dt solid}
\gpsetlinewidth{2.00}
\draw[gp path] (2.115,3.731)--(2.048,3.731);
\gpcolor{rgb color={0.702,0.702,0.702}}
\gpsetlinetype{gp lt axes}
\gpsetdashtype{gp dt axes}
\gpsetlinewidth{1.00}
\draw[gp path] (2.115,3.947)--(11.671,3.947);
\gpcolor{rgb color={0.400,0.400,0.400}}
\gpsetlinetype{gp lt border}
\gpsetdashtype{gp dt solid}
\gpsetlinewidth{2.00}
\draw[gp path] (2.115,3.947)--(2.048,3.947);
\gpcolor{rgb color={0.702,0.702,0.702}}
\gpsetlinetype{gp lt axes}
\gpsetdashtype{gp dt axes}
\gpsetlinewidth{1.00}
\draw[gp path] (2.115,4.121)--(11.671,4.121);
\gpcolor{rgb color={0.400,0.400,0.400}}
\gpsetlinetype{gp lt border}
\gpsetdashtype{gp dt solid}
\gpsetlinewidth{2.00}
\draw[gp path] (2.115,4.121)--(1.980,4.121);
\node[gp node right,font={\fontsize{8.0pt}{9.6pt}\selectfont}] at (1.888,4.121) {8};
\gpcolor{rgb color={0.702,0.702,0.702}}
\gpsetlinetype{gp lt axes}
\gpsetdashtype{gp dt axes}
\gpsetlinewidth{1.00}
\draw[gp path] (2.115,4.913)--(11.671,4.913);
\gpcolor{rgb color={0.400,0.400,0.400}}
\gpsetlinetype{gp lt border}
\gpsetdashtype{gp dt solid}
\gpsetlinewidth{2.00}
\draw[gp path] (2.115,4.913)--(2.048,4.913);
\gpcolor{rgb color={0.702,0.702,0.702}}
\gpsetlinetype{gp lt axes}
\gpsetdashtype{gp dt axes}
\gpsetlinewidth{1.00}
\draw[gp path] (2.115,5.329)--(11.671,5.329);
\gpcolor{rgb color={0.400,0.400,0.400}}
\gpsetlinetype{gp lt border}
\gpsetdashtype{gp dt solid}
\gpsetlinewidth{2.00}
\draw[gp path] (2.115,5.329)--(2.048,5.329);
\gpcolor{rgb color={0.702,0.702,0.702}}
\gpsetlinetype{gp lt axes}
\gpsetdashtype{gp dt axes}
\gpsetlinewidth{1.00}
\draw[gp path] (2.115,5.613)--(11.671,5.613);
\gpcolor{rgb color={0.400,0.400,0.400}}
\gpsetlinetype{gp lt border}
\gpsetdashtype{gp dt solid}
\gpsetlinewidth{2.00}
\draw[gp path] (2.115,5.613)--(2.048,5.613);
\gpcolor{rgb color={0.702,0.702,0.702}}
\gpsetlinetype{gp lt axes}
\gpsetdashtype{gp dt axes}
\gpsetlinewidth{1.00}
\draw[gp path] (2.115,5.828)--(11.671,5.828);
\gpcolor{rgb color={0.400,0.400,0.400}}
\gpsetlinetype{gp lt border}
\gpsetdashtype{gp dt solid}
\gpsetlinewidth{2.00}
\draw[gp path] (2.115,5.828)--(2.048,5.828);
\gpcolor{rgb color={0.702,0.702,0.702}}
\gpsetlinetype{gp lt axes}
\gpsetdashtype{gp dt axes}
\gpsetlinewidth{1.00}
\draw[gp path] (2.115,6.002)--(11.671,6.002);
\gpcolor{rgb color={0.400,0.400,0.400}}
\gpsetlinetype{gp lt border}
\gpsetdashtype{gp dt solid}
\gpsetlinewidth{2.00}
\draw[gp path] (2.115,6.002)--(1.980,6.002);
\node[gp node right,font={\fontsize{8.0pt}{9.6pt}\selectfont}] at (1.888,6.002) {64};
\gpcolor{rgb color={0.702,0.702,0.702}}
\gpsetlinetype{gp lt axes}
\gpsetdashtype{gp dt axes}
\gpsetlinewidth{1.00}
\draw[gp path] (2.115,6.794)--(2.391,6.794);
\draw[gp path] (5.331,6.794)--(11.671,6.794);
\gpcolor{rgb color={0.400,0.400,0.400}}
\gpsetlinetype{gp lt border}
\gpsetdashtype{gp dt solid}
\gpsetlinewidth{2.00}
\draw[gp path] (2.115,6.794)--(2.048,6.794);
\gpcolor{rgb color={0.702,0.702,0.702}}
\gpsetlinetype{gp lt axes}
\gpsetdashtype{gp dt axes}
\gpsetlinewidth{1.00}
\draw[gp path] (2.115,7.210)--(2.391,7.210);
\draw[gp path] (5.331,7.210)--(11.671,7.210);
\gpcolor{rgb color={0.400,0.400,0.400}}
\gpsetlinetype{gp lt border}
\gpsetdashtype{gp dt solid}
\gpsetlinewidth{2.00}
\draw[gp path] (2.115,7.210)--(2.048,7.210);
\gpcolor{rgb color={0.702,0.702,0.702}}
\gpsetlinetype{gp lt axes}
\gpsetdashtype{gp dt axes}
\gpsetlinewidth{1.00}
\draw[gp path] (2.115,7.494)--(2.391,7.494);
\draw[gp path] (5.331,7.494)--(11.671,7.494);
\gpcolor{rgb color={0.400,0.400,0.400}}
\gpsetlinetype{gp lt border}
\gpsetdashtype{gp dt solid}
\gpsetlinewidth{2.00}
\draw[gp path] (2.115,7.494)--(2.048,7.494);
\gpcolor{rgb color={0.702,0.702,0.702}}
\gpsetlinetype{gp lt axes}
\gpsetdashtype{gp dt axes}
\gpsetlinewidth{1.00}
\draw[gp path] (2.115,7.709)--(2.391,7.709);
\draw[gp path] (5.331,7.709)--(11.671,7.709);
\gpcolor{rgb color={0.400,0.400,0.400}}
\gpsetlinetype{gp lt border}
\gpsetdashtype{gp dt solid}
\gpsetlinewidth{2.00}
\draw[gp path] (2.115,7.709)--(2.048,7.709);
\gpcolor{rgb color={0.702,0.702,0.702}}
\gpsetlinetype{gp lt axes}
\gpsetdashtype{gp dt axes}
\gpsetlinewidth{1.00}
\draw[gp path] (2.115,7.883)--(2.391,7.883);
\draw[gp path] (5.331,7.883)--(11.671,7.883);
\gpcolor{rgb color={0.400,0.400,0.400}}
\gpsetlinetype{gp lt border}
\gpsetdashtype{gp dt solid}
\gpsetlinewidth{2.00}
\draw[gp path] (2.115,7.883)--(1.980,7.883);
\node[gp node right,font={\fontsize{8.0pt}{9.6pt}\selectfont}] at (1.888,7.883) {512};
\gpcolor{rgb color={0.702,0.702,0.702}}
\gpsetlinetype{gp lt axes}
\gpsetdashtype{gp dt axes}
\gpsetlinewidth{1.00}
\draw[gp path] (3.228,1.613)--(3.228,6.028);
\draw[gp path] (3.228,8.107)--(3.228,8.287);
\gpcolor{rgb color={0.400,0.400,0.400}}
\gpsetlinetype{gp lt border}
\gpsetdashtype{gp dt solid}
\gpsetlinewidth{2.00}
\draw[gp path] (3.228,1.613)--(3.228,1.478);
\node[gp node center,font={\fontsize{8.0pt}{9.6pt}\selectfont}] at (3.228,1.016) {24};
\gpcolor{rgb color={0.702,0.702,0.702}}
\gpsetlinetype{gp lt axes}
\gpsetdashtype{gp dt axes}
\gpsetlinewidth{1.00}
\draw[gp path] (5.870,1.613)--(5.870,8.287);
\gpcolor{rgb color={0.400,0.400,0.400}}
\gpsetlinetype{gp lt border}
\gpsetdashtype{gp dt solid}
\gpsetlinewidth{2.00}
\draw[gp path] (5.870,1.613)--(5.870,1.478);
\node[gp node center,font={\fontsize{8.0pt}{9.6pt}\selectfont}] at (5.870,1.016) {192};
\gpcolor{rgb color={0.702,0.702,0.702}}
\gpsetlinetype{gp lt axes}
\gpsetdashtype{gp dt axes}
\gpsetlinewidth{1.00}
\draw[gp path] (8.513,1.613)--(8.513,8.287);
\gpcolor{rgb color={0.400,0.400,0.400}}
\gpsetlinetype{gp lt border}
\gpsetdashtype{gp dt solid}
\gpsetlinewidth{2.00}
\draw[gp path] (8.513,1.613)--(8.513,1.478);
\node[gp node center,font={\fontsize{8.0pt}{9.6pt}\selectfont}] at (8.513,1.016) {1536};
\gpcolor{rgb color={0.702,0.702,0.702}}
\gpsetlinetype{gp lt axes}
\gpsetdashtype{gp dt axes}
\gpsetlinewidth{1.00}
\draw[gp path] (11.156,1.613)--(11.156,8.287);
\gpcolor{rgb color={0.400,0.400,0.400}}
\gpsetlinetype{gp lt border}
\gpsetdashtype{gp dt solid}
\gpsetlinewidth{2.00}
\draw[gp path] (11.156,1.613)--(11.156,1.478);
\node[gp node center,font={\fontsize{8.0pt}{9.6pt}\selectfont}] at (11.156,1.016) {12288};
\draw[gp path] (2.115,8.287)--(2.115,1.613)--(11.671,1.613)--(11.671,8.287)--cycle;
\gpcolor{color=gp lt color border}
\node[gp node center,rotate=-270] at (1.105,4.950) {Speedup (weak)};
\node[gp node center] at (6.893,0.400) {Number of processes};
\gpcolor{rgb color={0.400,0.400,0.400}}
\gpsetlinewidth{1.00}
\draw[gp path] (2.391,6.028)--(2.391,8.107)--(5.331,8.107)--(5.331,6.028)--cycle;
\gpcolor{color=gp lt color border}
\node[gp node left] at (3.399,7.760) {$S_{\text{ass}}$};
\gpcolor{rgb color={0.800,0.000,0.063}}
\gpsetlinewidth{3.00}
\draw[gp path] (2.667,7.760)--(3.123,7.760);
\draw[gp path] (3.228,2.240)--(5.870,3.963)--(8.513,5.932)--(11.156,7.985);
\gpsetpointsize{5.20}
\gppoint{gp mark 1}{(3.228,2.240)}
\gppoint{gp mark 1}{(5.870,3.963)}
\gppoint{gp mark 1}{(8.513,5.932)}
\gppoint{gp mark 1}{(11.156,7.985)}
\gppoint{gp mark 1}{(2.895,7.760)}
\gpcolor{color=gp lt color border}
\node[gp node left] at (3.399,7.298) {$S_{\text{init}}$};
\gpcolor{rgb color={0.800,0.000,0.663}}
\draw[gp path] (2.667,7.298)--(3.123,7.298);
\draw[gp path] (3.228,2.240)--(5.870,3.947)--(8.513,5.916)--(11.156,7.815);
\gppoint{gp mark 2}{(3.228,2.240)}
\gppoint{gp mark 2}{(5.870,3.947)}
\gppoint{gp mark 2}{(8.513,5.916)}
\gppoint{gp mark 2}{(11.156,7.815)}
\gppoint{gp mark 2}{(2.895,7.298)}
\gpcolor{color=gp lt color border}
\node[gp node left] at (3.399,6.836) {$S_{\text{solve}}$};
\gpcolor{rgb color={0.329,0.000,0.800}}
\draw[gp path] (2.667,6.836)--(3.123,6.836);
\draw[gp path] (3.228,2.240)--(5.870,3.774)--(8.513,5.727)--(11.156,7.541);
\gppoint{gp mark 3}{(3.228,2.240)}
\gppoint{gp mark 3}{(5.870,3.774)}
\gppoint{gp mark 3}{(8.513,5.727)}
\gppoint{gp mark 3}{(11.156,7.541)}
\gppoint{gp mark 3}{(2.895,6.836)}
\gpcolor{color=gp lt color border}
\node[gp node left] at (3.399,6.374) {$S_{\text{ideal}}$};
\gpcolor{rgb color={0.800,0.533,0.000}}
\draw[gp path] (2.667,6.374)--(3.123,6.374);
\draw[gp path] (3.228,2.240)--(5.870,4.121)--(8.513,6.002)--(11.156,7.883);
\gpdefrectangularnode{gp plot 1}{\pgfpoint{2.115cm}{1.613cm}}{\pgfpoint{11.671cm}{8.287cm}}
\end{tikzpicture}
 }
		\caption{Linear Elastic Problem Solver}
		\label{tab:weakElast}
	\end{subfigure}
	\centering

	\begin{subfigure}[t]{\textwidth}
		\centering
		\scalebox{0.5}{\footnotesize\begin{tikzpicture}[gnuplot]
\tikzset{every node/.append style={scale=1.50}}
\path (0.000,0.000) rectangle (12.500,8.750);
\gpcolor{rgb color={0.702,0.702,0.702}}
\gpsetlinetype{gp lt axes}
\gpsetdashtype{gp dt axes}
\gpsetlinewidth{1.00}
\draw[gp path] (2.391,1.613)--(11.671,1.613);
\gpcolor{rgb color={0.400,0.400,0.400}}
\gpsetlinetype{gp lt border}
\gpsetdashtype{gp dt solid}
\gpsetlinewidth{2.00}
\draw[gp path] (2.391,1.613)--(2.256,1.613);
\node[gp node right,font={\fontsize{8.0pt}{9.6pt}\selectfont}] at (2.256,1.613) {0};
\gpcolor{rgb color={0.702,0.702,0.702}}
\gpsetlinetype{gp lt axes}
\gpsetdashtype{gp dt axes}
\gpsetlinewidth{1.00}
\draw[gp path] (2.391,1.947)--(11.671,1.947);
\gpcolor{rgb color={0.400,0.400,0.400}}
\gpsetlinetype{gp lt border}
\gpsetdashtype{gp dt solid}
\gpsetlinewidth{2.00}
\draw[gp path] (2.391,1.947)--(2.324,1.947);
\gpcolor{rgb color={0.702,0.702,0.702}}
\gpsetlinetype{gp lt axes}
\gpsetdashtype{gp dt axes}
\gpsetlinewidth{1.00}
\draw[gp path] (2.391,2.280)--(11.671,2.280);
\gpcolor{rgb color={0.400,0.400,0.400}}
\gpsetlinetype{gp lt border}
\gpsetdashtype{gp dt solid}
\gpsetlinewidth{2.00}
\draw[gp path] (2.391,2.280)--(2.324,2.280);
\gpcolor{rgb color={0.702,0.702,0.702}}
\gpsetlinetype{gp lt axes}
\gpsetdashtype{gp dt axes}
\gpsetlinewidth{1.00}
\draw[gp path] (2.391,2.614)--(11.671,2.614);
\gpcolor{rgb color={0.400,0.400,0.400}}
\gpsetlinetype{gp lt border}
\gpsetdashtype{gp dt solid}
\gpsetlinewidth{2.00}
\draw[gp path] (2.391,2.614)--(2.324,2.614);
\gpcolor{rgb color={0.702,0.702,0.702}}
\gpsetlinetype{gp lt axes}
\gpsetdashtype{gp dt axes}
\gpsetlinewidth{1.00}
\draw[gp path] (2.391,2.948)--(11.671,2.948);
\gpcolor{rgb color={0.400,0.400,0.400}}
\gpsetlinetype{gp lt border}
\gpsetdashtype{gp dt solid}
\gpsetlinewidth{2.00}
\draw[gp path] (2.391,2.948)--(2.324,2.948);
\gpcolor{rgb color={0.702,0.702,0.702}}
\gpsetlinetype{gp lt axes}
\gpsetdashtype{gp dt axes}
\gpsetlinewidth{1.00}
\draw[gp path] (2.391,3.282)--(11.671,3.282);
\gpcolor{rgb color={0.400,0.400,0.400}}
\gpsetlinetype{gp lt border}
\gpsetdashtype{gp dt solid}
\gpsetlinewidth{2.00}
\draw[gp path] (2.391,3.282)--(2.256,3.282);
\node[gp node right,font={\fontsize{8.0pt}{9.6pt}\selectfont}] at (2.164,3.282) {500};
\gpcolor{rgb color={0.702,0.702,0.702}}
\gpsetlinetype{gp lt axes}
\gpsetdashtype{gp dt axes}
\gpsetlinewidth{1.00}
\draw[gp path] (2.391,3.615)--(11.671,3.615);
\gpcolor{rgb color={0.400,0.400,0.400}}
\gpsetlinetype{gp lt border}
\gpsetdashtype{gp dt solid}
\gpsetlinewidth{2.00}
\draw[gp path] (2.391,3.615)--(2.324,3.615);
\gpcolor{rgb color={0.702,0.702,0.702}}
\gpsetlinetype{gp lt axes}
\gpsetdashtype{gp dt axes}
\gpsetlinewidth{1.00}
\draw[gp path] (2.391,3.949)--(11.671,3.949);
\gpcolor{rgb color={0.400,0.400,0.400}}
\gpsetlinetype{gp lt border}
\gpsetdashtype{gp dt solid}
\gpsetlinewidth{2.00}
\draw[gp path] (2.391,3.949)--(2.324,3.949);
\gpcolor{rgb color={0.702,0.702,0.702}}
\gpsetlinetype{gp lt axes}
\gpsetdashtype{gp dt axes}
\gpsetlinewidth{1.00}
\draw[gp path] (2.391,4.283)--(11.671,4.283);
\gpcolor{rgb color={0.400,0.400,0.400}}
\gpsetlinetype{gp lt border}
\gpsetdashtype{gp dt solid}
\gpsetlinewidth{2.00}
\draw[gp path] (2.391,4.283)--(2.324,4.283);
\gpcolor{rgb color={0.702,0.702,0.702}}
\gpsetlinetype{gp lt axes}
\gpsetdashtype{gp dt axes}
\gpsetlinewidth{1.00}
\draw[gp path] (2.391,4.616)--(11.671,4.616);
\gpcolor{rgb color={0.400,0.400,0.400}}
\gpsetlinetype{gp lt border}
\gpsetdashtype{gp dt solid}
\gpsetlinewidth{2.00}
\draw[gp path] (2.391,4.616)--(2.324,4.616);
\gpcolor{rgb color={0.702,0.702,0.702}}
\gpsetlinetype{gp lt axes}
\gpsetdashtype{gp dt axes}
\gpsetlinewidth{1.00}
\draw[gp path] (2.391,4.950)--(11.671,4.950);
\gpcolor{rgb color={0.400,0.400,0.400}}
\gpsetlinetype{gp lt border}
\gpsetdashtype{gp dt solid}
\gpsetlinewidth{2.00}
\draw[gp path] (2.391,4.950)--(2.256,4.950);
\node[gp node right,font={\fontsize{8.0pt}{9.6pt}\selectfont}] at (2.164,4.950) {1000};
\gpcolor{rgb color={0.702,0.702,0.702}}
\gpsetlinetype{gp lt axes}
\gpsetdashtype{gp dt axes}
\gpsetlinewidth{1.00}
\draw[gp path] (2.391,5.284)--(11.671,5.284);
\gpcolor{rgb color={0.400,0.400,0.400}}
\gpsetlinetype{gp lt border}
\gpsetdashtype{gp dt solid}
\gpsetlinewidth{2.00}
\draw[gp path] (2.391,5.284)--(2.324,5.284);
\gpcolor{rgb color={0.702,0.702,0.702}}
\gpsetlinetype{gp lt axes}
\gpsetdashtype{gp dt axes}
\gpsetlinewidth{1.00}
\draw[gp path] (2.391,5.617)--(11.671,5.617);
\gpcolor{rgb color={0.400,0.400,0.400}}
\gpsetlinetype{gp lt border}
\gpsetdashtype{gp dt solid}
\gpsetlinewidth{2.00}
\draw[gp path] (2.391,5.617)--(2.324,5.617);
\gpcolor{rgb color={0.702,0.702,0.702}}
\gpsetlinetype{gp lt axes}
\gpsetdashtype{gp dt axes}
\gpsetlinewidth{1.00}
\draw[gp path] (2.391,5.951)--(11.671,5.951);
\gpcolor{rgb color={0.400,0.400,0.400}}
\gpsetlinetype{gp lt border}
\gpsetdashtype{gp dt solid}
\gpsetlinewidth{2.00}
\draw[gp path] (2.391,5.951)--(2.324,5.951);
\gpcolor{rgb color={0.702,0.702,0.702}}
\gpsetlinetype{gp lt axes}
\gpsetdashtype{gp dt axes}
\gpsetlinewidth{1.00}
\draw[gp path] (2.391,6.285)--(11.671,6.285);
\gpcolor{rgb color={0.400,0.400,0.400}}
\gpsetlinetype{gp lt border}
\gpsetdashtype{gp dt solid}
\gpsetlinewidth{2.00}
\draw[gp path] (2.391,6.285)--(2.324,6.285);
\gpcolor{rgb color={0.702,0.702,0.702}}
\gpsetlinetype{gp lt axes}
\gpsetdashtype{gp dt axes}
\gpsetlinewidth{1.00}
\draw[gp path] (2.391,6.619)--(2.667,6.619);
\draw[gp path] (6.987,6.619)--(11.671,6.619);
\gpcolor{rgb color={0.400,0.400,0.400}}
\gpsetlinetype{gp lt border}
\gpsetdashtype{gp dt solid}
\gpsetlinewidth{2.00}
\draw[gp path] (2.391,6.619)--(2.256,6.619);
\node[gp node right,font={\fontsize{8.0pt}{9.6pt}\selectfont}] at (2.164,6.619) {1500};
\gpcolor{rgb color={0.702,0.702,0.702}}
\gpsetlinetype{gp lt axes}
\gpsetdashtype{gp dt axes}
\gpsetlinewidth{1.00}
\draw[gp path] (2.391,6.952)--(2.667,6.952);
\draw[gp path] (6.987,6.952)--(11.671,6.952);
\gpcolor{rgb color={0.400,0.400,0.400}}
\gpsetlinetype{gp lt border}
\gpsetdashtype{gp dt solid}
\gpsetlinewidth{2.00}
\draw[gp path] (2.391,6.952)--(2.324,6.952);
\gpcolor{rgb color={0.702,0.702,0.702}}
\gpsetlinetype{gp lt axes}
\gpsetdashtype{gp dt axes}
\gpsetlinewidth{1.00}
\draw[gp path] (2.391,7.286)--(2.667,7.286);
\draw[gp path] (6.987,7.286)--(11.671,7.286);
\gpcolor{rgb color={0.400,0.400,0.400}}
\gpsetlinetype{gp lt border}
\gpsetdashtype{gp dt solid}
\gpsetlinewidth{2.00}
\draw[gp path] (2.391,7.286)--(2.324,7.286);
\gpcolor{rgb color={0.702,0.702,0.702}}
\gpsetlinetype{gp lt axes}
\gpsetdashtype{gp dt axes}
\gpsetlinewidth{1.00}
\draw[gp path] (2.391,7.620)--(2.667,7.620);
\draw[gp path] (6.987,7.620)--(11.671,7.620);
\gpcolor{rgb color={0.400,0.400,0.400}}
\gpsetlinetype{gp lt border}
\gpsetdashtype{gp dt solid}
\gpsetlinewidth{2.00}
\draw[gp path] (2.391,7.620)--(2.324,7.620);
\gpcolor{rgb color={0.702,0.702,0.702}}
\gpsetlinetype{gp lt axes}
\gpsetdashtype{gp dt axes}
\gpsetlinewidth{1.00}
\draw[gp path] (2.391,7.953)--(2.667,7.953);
\draw[gp path] (6.987,7.953)--(11.671,7.953);
\gpcolor{rgb color={0.400,0.400,0.400}}
\gpsetlinetype{gp lt border}
\gpsetdashtype{gp dt solid}
\gpsetlinewidth{2.00}
\draw[gp path] (2.391,7.953)--(2.324,7.953);
\gpcolor{rgb color={0.702,0.702,0.702}}
\gpsetlinetype{gp lt axes}
\gpsetdashtype{gp dt axes}
\gpsetlinewidth{1.00}
\draw[gp path] (2.391,8.287)--(11.671,8.287);
\gpcolor{rgb color={0.400,0.400,0.400}}
\gpsetlinetype{gp lt border}
\gpsetdashtype{gp dt solid}
\gpsetlinewidth{2.00}
\draw[gp path] (2.391,8.287)--(2.256,8.287);
\node[gp node right,font={\fontsize{8.0pt}{9.6pt}\selectfont}] at (2.164,8.287) {2000};
\gpcolor{rgb color={0.702,0.702,0.702}}
\gpsetlinetype{gp lt axes}
\gpsetdashtype{gp dt axes}
\gpsetlinewidth{1.00}
\draw[gp path] (3.471,1.613)--(3.471,6.490);
\draw[gp path] (3.471,8.107)--(3.471,8.287);
\gpcolor{rgb color={0.400,0.400,0.400}}
\gpsetlinetype{gp lt border}
\gpsetdashtype{gp dt solid}
\gpsetlinewidth{2.00}
\draw[gp path] (3.471,1.613)--(3.471,1.478);
\node[gp node center,font={\fontsize{8.0pt}{9.6pt}\selectfont}] at (3.471,1.016) {24};
\gpcolor{rgb color={0.702,0.702,0.702}}
\gpsetlinetype{gp lt axes}
\gpsetdashtype{gp dt axes}
\gpsetlinewidth{1.00}
\draw[gp path] (6.038,1.613)--(6.038,6.490);
\draw[gp path] (6.038,8.107)--(6.038,8.287);
\gpcolor{rgb color={0.400,0.400,0.400}}
\gpsetlinetype{gp lt border}
\gpsetdashtype{gp dt solid}
\gpsetlinewidth{2.00}
\draw[gp path] (6.038,1.613)--(6.038,1.478);
\node[gp node center,font={\fontsize{8.0pt}{9.6pt}\selectfont}] at (6.038,1.016) {192};
\gpcolor{rgb color={0.702,0.702,0.702}}
\gpsetlinetype{gp lt axes}
\gpsetdashtype{gp dt axes}
\gpsetlinewidth{1.00}
\draw[gp path] (8.604,1.613)--(8.604,8.287);
\gpcolor{rgb color={0.400,0.400,0.400}}
\gpsetlinetype{gp lt border}
\gpsetdashtype{gp dt solid}
\gpsetlinewidth{2.00}
\draw[gp path] (8.604,1.613)--(8.604,1.478);
\node[gp node center,font={\fontsize{8.0pt}{9.6pt}\selectfont}] at (8.604,1.016) {1536};
\gpcolor{rgb color={0.702,0.702,0.702}}
\gpsetlinetype{gp lt axes}
\gpsetdashtype{gp dt axes}
\gpsetlinewidth{1.00}
\draw[gp path] (11.171,1.613)--(11.171,8.287);
\gpcolor{rgb color={0.400,0.400,0.400}}
\gpsetlinetype{gp lt border}
\gpsetdashtype{gp dt solid}
\gpsetlinewidth{2.00}
\draw[gp path] (11.171,1.613)--(11.171,1.478);
\node[gp node center,font={\fontsize{8.0pt}{9.6pt}\selectfont}] at (11.171,1.016) {12288};
\draw[gp path] (2.391,8.287)--(2.391,1.613)--(11.671,1.613)--(11.671,8.287)--cycle;
\gpcolor{color=gp lt color border}
\node[gp node center,rotate=-270] at (0.921,4.950) {Time [s]};
\node[gp node center] at (7.031,0.400) {Number of processes};
\gpcolor{rgb color={0.400,0.400,0.400}}
\gpsetlinewidth{1.00}
\draw[gp path] (2.667,6.490)--(2.667,8.107)--(6.987,8.107)--(6.987,6.490)--cycle;
\gpcolor{color=gp lt color border}
\node[gp node left] at (3.675,7.760) {$T_{\text{lin,init}}$};
\gpcolor{rgb color={0.800,0.000,0.063}}
\gpsetlinewidth{3.00}
\draw[gp path] (2.943,7.760)--(3.399,7.760);
\draw[gp path] (3.471,1.735)--(6.038,1.758)--(8.604,1.739)--(11.171,1.808);
\gpsetpointsize{5.20}
\gppoint{gp mark 1}{(3.471,1.735)}
\gppoint{gp mark 1}{(6.038,1.758)}
\gppoint{gp mark 1}{(8.604,1.739)}
\gppoint{gp mark 1}{(11.171,1.808)}
\gppoint{gp mark 1}{(3.171,7.760)}
\gpcolor{color=gp lt color border}
\node[gp node left] at (3.675,7.298) {$T_{\text{lin,solve}}$};
\gpcolor{rgb color={0.800,0.000,0.663}}
\draw[gp path] (2.943,7.298)--(3.399,7.298);
\draw[gp path] (3.471,1.877)--(6.038,2.045)--(8.604,2.195)--(11.171,2.370);
\gppoint{gp mark 2}{(3.471,1.877)}
\gppoint{gp mark 2}{(6.038,2.045)}
\gppoint{gp mark 2}{(8.604,2.195)}
\gppoint{gp mark 2}{(11.171,2.370)}
\gppoint{gp mark 2}{(3.171,7.298)}
\gpcolor{color=gp lt color border}
\node[gp node left] at (3.675,6.836) {$T_{\text{newton}}$};
\gpcolor{rgb color={0.329,0.000,0.800}}
\draw[gp path] (2.943,6.836)--(3.399,6.836);
\draw[gp path] (3.471,4.554)--(6.038,5.305)--(8.604,5.363)--(11.171,5.590);
\gppoint{gp mark 3}{(3.471,4.554)}
\gppoint{gp mark 3}{(6.038,5.305)}
\gppoint{gp mark 3}{(8.604,5.363)}
\gppoint{gp mark 3}{(11.171,5.590)}
\gppoint{gp mark 3}{(3.171,6.836)}
\gpdefrectangularnode{gp plot 1}{\pgfpoint{2.391cm}{1.613cm}}{\pgfpoint{11.671cm}{8.287cm}}
\end{tikzpicture}
 }
		\scalebox{0.5}{\footnotesize\begin{tikzpicture}[gnuplot]
\tikzset{every node/.append style={scale=1.50}}
\path (0.000,0.000) rectangle (12.500,8.750);
\gpcolor{rgb color={0.702,0.702,0.702}}
\gpsetlinetype{gp lt axes}
\gpsetdashtype{gp dt axes}
\gpsetlinewidth{1.00}
\draw[gp path] (2.115,3.277)--(11.671,3.277);
\gpcolor{rgb color={0.400,0.400,0.400}}
\gpsetlinetype{gp lt border}
\gpsetdashtype{gp dt solid}
\gpsetlinewidth{2.00}
\draw[gp path] (2.115,3.277)--(1.980,3.277);
\node[gp node right,font={\fontsize{8.0pt}{9.6pt}\selectfont}] at (1.980,3.277) {1};
\gpcolor{rgb color={0.702,0.702,0.702}}
\gpsetlinetype{gp lt axes}
\gpsetdashtype{gp dt axes}
\gpsetlinewidth{1.00}
\draw[gp path] (2.115,3.910)--(11.671,3.910);
\gpcolor{rgb color={0.400,0.400,0.400}}
\gpsetlinetype{gp lt border}
\gpsetdashtype{gp dt solid}
\gpsetlinewidth{2.00}
\draw[gp path] (2.115,3.910)--(2.048,3.910);
\gpcolor{rgb color={0.702,0.702,0.702}}
\gpsetlinetype{gp lt axes}
\gpsetdashtype{gp dt axes}
\gpsetlinewidth{1.00}
\draw[gp path] (2.115,4.242)--(11.671,4.242);
\gpcolor{rgb color={0.400,0.400,0.400}}
\gpsetlinetype{gp lt border}
\gpsetdashtype{gp dt solid}
\gpsetlinewidth{2.00}
\draw[gp path] (2.115,4.242)--(2.048,4.242);
\gpcolor{rgb color={0.702,0.702,0.702}}
\gpsetlinetype{gp lt axes}
\gpsetdashtype{gp dt axes}
\gpsetlinewidth{1.00}
\draw[gp path] (2.115,4.469)--(11.671,4.469);
\gpcolor{rgb color={0.400,0.400,0.400}}
\gpsetlinetype{gp lt border}
\gpsetdashtype{gp dt solid}
\gpsetlinewidth{2.00}
\draw[gp path] (2.115,4.469)--(2.048,4.469);
\gpcolor{rgb color={0.702,0.702,0.702}}
\gpsetlinetype{gp lt axes}
\gpsetdashtype{gp dt axes}
\gpsetlinewidth{1.00}
\draw[gp path] (2.115,4.641)--(11.671,4.641);
\gpcolor{rgb color={0.400,0.400,0.400}}
\gpsetlinetype{gp lt border}
\gpsetdashtype{gp dt solid}
\gpsetlinewidth{2.00}
\draw[gp path] (2.115,4.641)--(2.048,4.641);
\gpcolor{rgb color={0.702,0.702,0.702}}
\gpsetlinetype{gp lt axes}
\gpsetdashtype{gp dt axes}
\gpsetlinewidth{1.00}
\draw[gp path] (2.115,4.780)--(11.671,4.780);
\gpcolor{rgb color={0.400,0.400,0.400}}
\gpsetlinetype{gp lt border}
\gpsetdashtype{gp dt solid}
\gpsetlinewidth{2.00}
\draw[gp path] (2.115,4.780)--(1.980,4.780);
\node[gp node right,font={\fontsize{8.0pt}{9.6pt}\selectfont}] at (1.888,4.780) {8};
\gpcolor{rgb color={0.702,0.702,0.702}}
\gpsetlinetype{gp lt axes}
\gpsetdashtype{gp dt axes}
\gpsetlinewidth{1.00}
\draw[gp path] (2.115,5.413)--(11.671,5.413);
\gpcolor{rgb color={0.400,0.400,0.400}}
\gpsetlinetype{gp lt border}
\gpsetdashtype{gp dt solid}
\gpsetlinewidth{2.00}
\draw[gp path] (2.115,5.413)--(2.048,5.413);
\gpcolor{rgb color={0.702,0.702,0.702}}
\gpsetlinetype{gp lt axes}
\gpsetdashtype{gp dt axes}
\gpsetlinewidth{1.00}
\draw[gp path] (2.115,5.745)--(11.671,5.745);
\gpcolor{rgb color={0.400,0.400,0.400}}
\gpsetlinetype{gp lt border}
\gpsetdashtype{gp dt solid}
\gpsetlinewidth{2.00}
\draw[gp path] (2.115,5.745)--(2.048,5.745);
\gpcolor{rgb color={0.702,0.702,0.702}}
\gpsetlinetype{gp lt axes}
\gpsetdashtype{gp dt axes}
\gpsetlinewidth{1.00}
\draw[gp path] (2.115,5.972)--(11.671,5.972);
\gpcolor{rgb color={0.400,0.400,0.400}}
\gpsetlinetype{gp lt border}
\gpsetdashtype{gp dt solid}
\gpsetlinewidth{2.00}
\draw[gp path] (2.115,5.972)--(2.048,5.972);
\gpcolor{rgb color={0.702,0.702,0.702}}
\gpsetlinetype{gp lt axes}
\gpsetdashtype{gp dt axes}
\gpsetlinewidth{1.00}
\draw[gp path] (2.115,6.144)--(2.391,6.144);
\draw[gp path] (6.435,6.144)--(11.671,6.144);
\gpcolor{rgb color={0.400,0.400,0.400}}
\gpsetlinetype{gp lt border}
\gpsetdashtype{gp dt solid}
\gpsetlinewidth{2.00}
\draw[gp path] (2.115,6.144)--(2.048,6.144);
\gpcolor{rgb color={0.702,0.702,0.702}}
\gpsetlinetype{gp lt axes}
\gpsetdashtype{gp dt axes}
\gpsetlinewidth{1.00}
\draw[gp path] (2.115,6.283)--(2.391,6.283);
\draw[gp path] (6.435,6.283)--(11.671,6.283);
\gpcolor{rgb color={0.400,0.400,0.400}}
\gpsetlinetype{gp lt border}
\gpsetdashtype{gp dt solid}
\gpsetlinewidth{2.00}
\draw[gp path] (2.115,6.283)--(1.980,6.283);
\node[gp node right,font={\fontsize{8.0pt}{9.6pt}\selectfont}] at (1.888,6.283) {64};
\gpcolor{rgb color={0.702,0.702,0.702}}
\gpsetlinetype{gp lt axes}
\gpsetdashtype{gp dt axes}
\gpsetlinewidth{1.00}
\draw[gp path] (2.115,6.916)--(2.391,6.916);
\draw[gp path] (6.435,6.916)--(11.671,6.916);
\gpcolor{rgb color={0.400,0.400,0.400}}
\gpsetlinetype{gp lt border}
\gpsetdashtype{gp dt solid}
\gpsetlinewidth{2.00}
\draw[gp path] (2.115,6.916)--(2.048,6.916);
\gpcolor{rgb color={0.702,0.702,0.702}}
\gpsetlinetype{gp lt axes}
\gpsetdashtype{gp dt axes}
\gpsetlinewidth{1.00}
\draw[gp path] (2.115,7.248)--(2.391,7.248);
\draw[gp path] (6.435,7.248)--(11.671,7.248);
\gpcolor{rgb color={0.400,0.400,0.400}}
\gpsetlinetype{gp lt border}
\gpsetdashtype{gp dt solid}
\gpsetlinewidth{2.00}
\draw[gp path] (2.115,7.248)--(2.048,7.248);
\gpcolor{rgb color={0.702,0.702,0.702}}
\gpsetlinetype{gp lt axes}
\gpsetdashtype{gp dt axes}
\gpsetlinewidth{1.00}
\draw[gp path] (2.115,7.475)--(2.391,7.475);
\draw[gp path] (6.435,7.475)--(11.671,7.475);
\gpcolor{rgb color={0.400,0.400,0.400}}
\gpsetlinetype{gp lt border}
\gpsetdashtype{gp dt solid}
\gpsetlinewidth{2.00}
\draw[gp path] (2.115,7.475)--(2.048,7.475);
\gpcolor{rgb color={0.702,0.702,0.702}}
\gpsetlinetype{gp lt axes}
\gpsetdashtype{gp dt axes}
\gpsetlinewidth{1.00}
\draw[gp path] (2.115,7.647)--(2.391,7.647);
\draw[gp path] (6.435,7.647)--(11.671,7.647);
\gpcolor{rgb color={0.400,0.400,0.400}}
\gpsetlinetype{gp lt border}
\gpsetdashtype{gp dt solid}
\gpsetlinewidth{2.00}
\draw[gp path] (2.115,7.647)--(2.048,7.647);
\gpcolor{rgb color={0.702,0.702,0.702}}
\gpsetlinetype{gp lt axes}
\gpsetdashtype{gp dt axes}
\gpsetlinewidth{1.00}
\draw[gp path] (2.115,7.786)--(2.391,7.786);
\draw[gp path] (6.435,7.786)--(11.671,7.786);
\gpcolor{rgb color={0.400,0.400,0.400}}
\gpsetlinetype{gp lt border}
\gpsetdashtype{gp dt solid}
\gpsetlinewidth{2.00}
\draw[gp path] (2.115,7.786)--(1.980,7.786);
\node[gp node right,font={\fontsize{8.0pt}{9.6pt}\selectfont}] at (1.888,7.786) {512};
\gpcolor{rgb color={0.702,0.702,0.702}}
\gpsetlinetype{gp lt axes}
\gpsetdashtype{gp dt axes}
\gpsetlinewidth{1.00}
\draw[gp path] (3.228,1.613)--(3.228,6.028);
\draw[gp path] (3.228,8.107)--(3.228,8.287);
\gpcolor{rgb color={0.400,0.400,0.400}}
\gpsetlinetype{gp lt border}
\gpsetdashtype{gp dt solid}
\gpsetlinewidth{2.00}
\draw[gp path] (3.228,1.613)--(3.228,1.478);
\node[gp node center,font={\fontsize{8.0pt}{9.6pt}\selectfont}] at (3.228,1.016) {24};
\gpcolor{rgb color={0.702,0.702,0.702}}
\gpsetlinetype{gp lt axes}
\gpsetdashtype{gp dt axes}
\gpsetlinewidth{1.00}
\draw[gp path] (5.870,1.613)--(5.870,6.028);
\draw[gp path] (5.870,8.107)--(5.870,8.287);
\gpcolor{rgb color={0.400,0.400,0.400}}
\gpsetlinetype{gp lt border}
\gpsetdashtype{gp dt solid}
\gpsetlinewidth{2.00}
\draw[gp path] (5.870,1.613)--(5.870,1.478);
\node[gp node center,font={\fontsize{8.0pt}{9.6pt}\selectfont}] at (5.870,1.016) {192};
\gpcolor{rgb color={0.702,0.702,0.702}}
\gpsetlinetype{gp lt axes}
\gpsetdashtype{gp dt axes}
\gpsetlinewidth{1.00}
\draw[gp path] (8.513,1.613)--(8.513,8.287);
\gpcolor{rgb color={0.400,0.400,0.400}}
\gpsetlinetype{gp lt border}
\gpsetdashtype{gp dt solid}
\gpsetlinewidth{2.00}
\draw[gp path] (8.513,1.613)--(8.513,1.478);
\node[gp node center,font={\fontsize{8.0pt}{9.6pt}\selectfont}] at (8.513,1.016) {1536};
\gpcolor{rgb color={0.702,0.702,0.702}}
\gpsetlinetype{gp lt axes}
\gpsetdashtype{gp dt axes}
\gpsetlinewidth{1.00}
\draw[gp path] (11.156,1.613)--(11.156,8.287);
\gpcolor{rgb color={0.400,0.400,0.400}}
\gpsetlinetype{gp lt border}
\gpsetdashtype{gp dt solid}
\gpsetlinewidth{2.00}
\draw[gp path] (11.156,1.613)--(11.156,1.478);
\node[gp node center,font={\fontsize{8.0pt}{9.6pt}\selectfont}] at (11.156,1.016) {12288};
\draw[gp path] (2.115,8.287)--(2.115,1.613)--(11.671,1.613)--(11.671,8.287)--cycle;
\gpcolor{color=gp lt color border}
\node[gp node center,rotate=-270] at (1.105,4.950) {Speedup (weak)};
\node[gp node center] at (6.893,0.400) {Number of processes};
\gpcolor{rgb color={0.400,0.400,0.400}}
\gpsetlinewidth{1.00}
\draw[gp path] (2.391,6.028)--(2.391,8.107)--(6.435,8.107)--(6.435,6.028)--cycle;
\gpcolor{color=gp lt color border}
\node[gp node left] at (3.399,7.760) {$S_{\text{lin,init}}$};
\gpcolor{rgb color={0.800,0.000,0.063}}
\gpsetlinewidth{3.00}
\draw[gp path] (2.667,7.760)--(3.123,7.760);
\draw[gp path] (3.228,3.277)--(5.870,4.655)--(8.513,6.261)--(11.156,7.448);
\gpsetpointsize{5.20}
\gppoint{gp mark 1}{(3.228,3.277)}
\gppoint{gp mark 1}{(5.870,4.655)}
\gppoint{gp mark 1}{(8.513,6.261)}
\gppoint{gp mark 1}{(11.156,7.448)}
\gppoint{gp mark 1}{(2.895,7.760)}
\gpcolor{color=gp lt color border}
\node[gp node left] at (3.399,7.298) {$S_{\text{lin,solve}}$};
\gpcolor{rgb color={0.800,0.000,0.663}}
\draw[gp path] (2.667,7.298)--(3.123,7.298);
\draw[gp path] (3.228,3.277)--(5.870,4.425)--(8.513,5.713)--(11.156,7.026);
\gppoint{gp mark 2}{(3.228,3.277)}
\gppoint{gp mark 2}{(5.870,4.425)}
\gppoint{gp mark 2}{(8.513,5.713)}
\gppoint{gp mark 2}{(11.156,7.026)}
\gppoint{gp mark 2}{(2.895,7.298)}
\gpcolor{color=gp lt color border}
\node[gp node left] at (3.399,6.836) {$S_{\text{newton}}$};
\gpcolor{rgb color={0.329,0.000,0.800}}
\draw[gp path] (2.667,6.836)--(3.123,6.836);
\draw[gp path] (3.228,3.277)--(5.870,4.616)--(8.513,6.108)--(11.156,7.568);
\gppoint{gp mark 3}{(3.228,3.277)}
\gppoint{gp mark 3}{(5.870,4.616)}
\gppoint{gp mark 3}{(8.513,6.108)}
\gppoint{gp mark 3}{(11.156,7.568)}
\gppoint{gp mark 3}{(2.895,6.836)}
\gpcolor{color=gp lt color border}
\node[gp node left] at (3.399,6.374) {$S_{\text{ideal}}$};
\gpcolor{rgb color={0.800,0.533,0.000}}
\draw[gp path] (2.667,6.374)--(3.123,6.374);
\draw[gp path] (3.228,3.277)--(5.870,4.780)--(8.513,6.283)--(11.156,7.786);
\gpdefrectangularnode{gp plot 1}{\pgfpoint{2.115cm}{1.613cm}}{\pgfpoint{11.671cm}{8.287cm}}
\end{tikzpicture}
 }
\caption{Shape Derivative Solver}
		\label{tab:weakShapeDer}
	\end{subfigure}
	\centering
\begin{subfigure}[t]{\textwidth}
		\centering
		\begin{tabular}{rrrccc}
			\\
			\toprule
			Procs  & Refs  & DoFs  & Linear solver  & Newton solver   & Linear solver     \\
			&       &        &  (elasticity)      & (shape derivative)    & (shape derivative)\\ 
			\midrule
			24 & 4 & 6,085,035 & 159 & 16 & 293  \\
			192 & 5 & 48,530,259 & 186 & 16 & 382  \\
			1,536 & 6 & 387,647,139 & 190 & 17 & 574  \\
			12,288 & 7 & 3,098,807,619 & 203 & 19 & 740  \\
			\bottomrule
		\end{tabular}
		\caption{Accumulated iteration counts for the weak scaling study}
		\label{tab:weakScaling}
	\end{subfigure}
	\vskip\baselineskip
	
	\caption{\textbf{Weak Scaling:} For the first two optimization steps, the accumulated wallclock time is shown for: (a) the elastic solver, (b) the shape derivative solver. In (c), the accumulated iteration counts are presented for solving the elasticity PDE-constraint with geometric multigrid, the required Newton steps to compute the shape derivative, and the accumulated geometric multigrid steps to solve the linearization within the Newton algorithm }
	\label{fig:weak_scaling}
\end{figure}

\begin{figure}[tbp]
	\begin{subfigure}[t]{\textwidth}
		\centering
		\scalebox{0.5}{\footnotesize\begin{tikzpicture}[gnuplot]
\tikzset{every node/.append style={scale=1.50}}
\path (0.000,0.000) rectangle (12.500,8.750);
\gpcolor{rgb color={0.702,0.702,0.702}}
\gpsetlinetype{gp lt axes}
\gpsetdashtype{gp dt axes}
\gpsetlinewidth{1.00}
\draw[gp path] (2.391,1.613)--(11.671,1.613);
\gpcolor{rgb color={0.400,0.400,0.400}}
\gpsetlinetype{gp lt border}
\gpsetdashtype{gp dt solid}
\gpsetlinewidth{2.00}
\draw[gp path] (2.391,1.613)--(2.256,1.613);
\node[gp node right,font={\fontsize{8.0pt}{9.6pt}\selectfont}] at (2.256,1.613) {0.01};
\gpcolor{rgb color={0.702,0.702,0.702}}
\gpsetlinetype{gp lt axes}
\gpsetdashtype{gp dt axes}
\gpsetlinewidth{1.00}
\draw[gp path] (2.391,2.378)--(11.671,2.378);
\gpcolor{rgb color={0.400,0.400,0.400}}
\gpsetlinetype{gp lt border}
\gpsetdashtype{gp dt solid}
\gpsetlinewidth{2.00}
\draw[gp path] (2.391,2.378)--(2.324,2.378);
\gpcolor{rgb color={0.702,0.702,0.702}}
\gpsetlinetype{gp lt axes}
\gpsetdashtype{gp dt axes}
\gpsetlinewidth{1.00}
\draw[gp path] (2.391,2.746)--(11.671,2.746);
\gpcolor{rgb color={0.400,0.400,0.400}}
\gpsetlinetype{gp lt border}
\gpsetdashtype{gp dt solid}
\gpsetlinewidth{2.00}
\draw[gp path] (2.391,2.746)--(2.324,2.746);
\gpcolor{rgb color={0.702,0.702,0.702}}
\gpsetlinetype{gp lt axes}
\gpsetdashtype{gp dt axes}
\gpsetlinewidth{1.00}
\draw[gp path] (2.391,2.992)--(11.671,2.992);
\gpcolor{rgb color={0.400,0.400,0.400}}
\gpsetlinetype{gp lt border}
\gpsetdashtype{gp dt solid}
\gpsetlinewidth{2.00}
\draw[gp path] (2.391,2.992)--(2.324,2.992);
\gpcolor{rgb color={0.702,0.702,0.702}}
\gpsetlinetype{gp lt axes}
\gpsetdashtype{gp dt axes}
\gpsetlinewidth{1.00}
\draw[gp path] (2.391,3.176)--(11.671,3.176);
\gpcolor{rgb color={0.400,0.400,0.400}}
\gpsetlinetype{gp lt border}
\gpsetdashtype{gp dt solid}
\gpsetlinewidth{2.00}
\draw[gp path] (2.391,3.176)--(2.324,3.176);
\gpcolor{rgb color={0.702,0.702,0.702}}
\gpsetlinetype{gp lt axes}
\gpsetdashtype{gp dt axes}
\gpsetlinewidth{1.00}
\draw[gp path] (2.391,3.323)--(11.671,3.323);
\gpcolor{rgb color={0.400,0.400,0.400}}
\gpsetlinetype{gp lt border}
\gpsetdashtype{gp dt solid}
\gpsetlinewidth{2.00}
\draw[gp path] (2.391,3.323)--(2.256,3.323);
\node[gp node right,font={\fontsize{8.0pt}{9.6pt}\selectfont}] at (2.164,3.323) {0.1};
\gpcolor{rgb color={0.702,0.702,0.702}}
\gpsetlinetype{gp lt axes}
\gpsetdashtype{gp dt axes}
\gpsetlinewidth{1.00}
\draw[gp path] (2.391,4.088)--(11.671,4.088);
\gpcolor{rgb color={0.400,0.400,0.400}}
\gpsetlinetype{gp lt border}
\gpsetdashtype{gp dt solid}
\gpsetlinewidth{2.00}
\draw[gp path] (2.391,4.088)--(2.324,4.088);
\gpcolor{rgb color={0.702,0.702,0.702}}
\gpsetlinetype{gp lt axes}
\gpsetdashtype{gp dt axes}
\gpsetlinewidth{1.00}
\draw[gp path] (2.391,4.456)--(11.671,4.456);
\gpcolor{rgb color={0.400,0.400,0.400}}
\gpsetlinetype{gp lt border}
\gpsetdashtype{gp dt solid}
\gpsetlinewidth{2.00}
\draw[gp path] (2.391,4.456)--(2.324,4.456);
\gpcolor{rgb color={0.702,0.702,0.702}}
\gpsetlinetype{gp lt axes}
\gpsetdashtype{gp dt axes}
\gpsetlinewidth{1.00}
\draw[gp path] (2.391,4.701)--(11.671,4.701);
\gpcolor{rgb color={0.400,0.400,0.400}}
\gpsetlinetype{gp lt border}
\gpsetdashtype{gp dt solid}
\gpsetlinewidth{2.00}
\draw[gp path] (2.391,4.701)--(2.324,4.701);
\gpcolor{rgb color={0.702,0.702,0.702}}
\gpsetlinetype{gp lt axes}
\gpsetdashtype{gp dt axes}
\gpsetlinewidth{1.00}
\draw[gp path] (2.391,4.885)--(11.671,4.885);
\gpcolor{rgb color={0.400,0.400,0.400}}
\gpsetlinetype{gp lt border}
\gpsetdashtype{gp dt solid}
\gpsetlinewidth{2.00}
\draw[gp path] (2.391,4.885)--(2.324,4.885);
\gpcolor{rgb color={0.702,0.702,0.702}}
\gpsetlinetype{gp lt axes}
\gpsetdashtype{gp dt axes}
\gpsetlinewidth{1.00}
\draw[gp path] (2.391,5.033)--(11.671,5.033);
\gpcolor{rgb color={0.400,0.400,0.400}}
\gpsetlinetype{gp lt border}
\gpsetdashtype{gp dt solid}
\gpsetlinewidth{2.00}
\draw[gp path] (2.391,5.033)--(2.256,5.033);
\node[gp node right,font={\fontsize{8.0pt}{9.6pt}\selectfont}] at (2.164,5.033) {1};
\gpcolor{rgb color={0.702,0.702,0.702}}
\gpsetlinetype{gp lt axes}
\gpsetdashtype{gp dt axes}
\gpsetlinewidth{1.00}
\draw[gp path] (2.391,5.797)--(11.671,5.797);
\gpcolor{rgb color={0.400,0.400,0.400}}
\gpsetlinetype{gp lt border}
\gpsetdashtype{gp dt solid}
\gpsetlinewidth{2.00}
\draw[gp path] (2.391,5.797)--(2.324,5.797);
\gpcolor{rgb color={0.702,0.702,0.702}}
\gpsetlinetype{gp lt axes}
\gpsetdashtype{gp dt axes}
\gpsetlinewidth{1.00}
\draw[gp path] (2.391,6.166)--(11.671,6.166);
\gpcolor{rgb color={0.400,0.400,0.400}}
\gpsetlinetype{gp lt border}
\gpsetdashtype{gp dt solid}
\gpsetlinewidth{2.00}
\draw[gp path] (2.391,6.166)--(2.324,6.166);
\gpcolor{rgb color={0.702,0.702,0.702}}
\gpsetlinetype{gp lt axes}
\gpsetdashtype{gp dt axes}
\gpsetlinewidth{1.00}
\draw[gp path] (2.391,6.411)--(11.671,6.411);
\gpcolor{rgb color={0.400,0.400,0.400}}
\gpsetlinetype{gp lt border}
\gpsetdashtype{gp dt solid}
\gpsetlinewidth{2.00}
\draw[gp path] (2.391,6.411)--(2.324,6.411);
\gpcolor{rgb color={0.702,0.702,0.702}}
\gpsetlinetype{gp lt axes}
\gpsetdashtype{gp dt axes}
\gpsetlinewidth{1.00}
\draw[gp path] (2.391,6.595)--(8.179,6.595);
\draw[gp path] (11.395,6.595)--(11.671,6.595);
\gpcolor{rgb color={0.400,0.400,0.400}}
\gpsetlinetype{gp lt border}
\gpsetdashtype{gp dt solid}
\gpsetlinewidth{2.00}
\draw[gp path] (2.391,6.595)--(2.324,6.595);
\gpcolor{rgb color={0.702,0.702,0.702}}
\gpsetlinetype{gp lt axes}
\gpsetdashtype{gp dt axes}
\gpsetlinewidth{1.00}
\draw[gp path] (2.391,6.743)--(8.179,6.743);
\draw[gp path] (11.395,6.743)--(11.671,6.743);
\gpcolor{rgb color={0.400,0.400,0.400}}
\gpsetlinetype{gp lt border}
\gpsetdashtype{gp dt solid}
\gpsetlinewidth{2.00}
\draw[gp path] (2.391,6.743)--(2.256,6.743);
\node[gp node right,font={\fontsize{8.0pt}{9.6pt}\selectfont}] at (2.164,6.743) {10};
\gpcolor{rgb color={0.702,0.702,0.702}}
\gpsetlinetype{gp lt axes}
\gpsetdashtype{gp dt axes}
\gpsetlinewidth{1.00}
\draw[gp path] (2.391,7.507)--(8.179,7.507);
\draw[gp path] (11.395,7.507)--(11.671,7.507);
\gpcolor{rgb color={0.400,0.400,0.400}}
\gpsetlinetype{gp lt border}
\gpsetdashtype{gp dt solid}
\gpsetlinewidth{2.00}
\draw[gp path] (2.391,7.507)--(2.324,7.507);
\gpcolor{rgb color={0.702,0.702,0.702}}
\gpsetlinetype{gp lt axes}
\gpsetdashtype{gp dt axes}
\gpsetlinewidth{1.00}
\draw[gp path] (2.391,7.876)--(8.179,7.876);
\draw[gp path] (11.395,7.876)--(11.671,7.876);
\gpcolor{rgb color={0.400,0.400,0.400}}
\gpsetlinetype{gp lt border}
\gpsetdashtype{gp dt solid}
\gpsetlinewidth{2.00}
\draw[gp path] (2.391,7.876)--(2.324,7.876);
\gpcolor{rgb color={0.702,0.702,0.702}}
\gpsetlinetype{gp lt axes}
\gpsetdashtype{gp dt axes}
\gpsetlinewidth{1.00}
\draw[gp path] (2.391,8.121)--(11.671,8.121);
\gpcolor{rgb color={0.400,0.400,0.400}}
\gpsetlinetype{gp lt border}
\gpsetdashtype{gp dt solid}
\gpsetlinewidth{2.00}
\draw[gp path] (2.391,8.121)--(2.324,8.121);
\gpcolor{rgb color={0.702,0.702,0.702}}
\gpsetlinetype{gp lt axes}
\gpsetdashtype{gp dt axes}
\gpsetlinewidth{1.00}
\draw[gp path] (3.581,1.613)--(3.581,8.287);
\gpcolor{rgb color={0.400,0.400,0.400}}
\gpsetlinetype{gp lt border}
\gpsetdashtype{gp dt solid}
\gpsetlinewidth{2.00}
\draw[gp path] (3.581,1.613)--(3.581,1.478);
\node[gp node center,font={\fontsize{8.0pt}{9.6pt}\selectfont}] at (3.581,1.016) {24};
\gpcolor{rgb color={0.702,0.702,0.702}}
\gpsetlinetype{gp lt axes}
\gpsetdashtype{gp dt axes}
\gpsetlinewidth{1.00}
\draw[gp path] (5.466,1.613)--(5.466,8.287);
\gpcolor{rgb color={0.400,0.400,0.400}}
\gpsetlinetype{gp lt border}
\gpsetdashtype{gp dt solid}
\gpsetlinewidth{2.00}
\draw[gp path] (5.466,1.613)--(5.466,1.478);
\node[gp node center,font={\fontsize{8.0pt}{9.6pt}\selectfont}] at (5.466,1.016) {96};
\gpcolor{rgb color={0.702,0.702,0.702}}
\gpsetlinetype{gp lt axes}
\gpsetdashtype{gp dt axes}
\gpsetlinewidth{1.00}
\draw[gp path] (7.350,1.613)--(7.350,8.287);
\gpcolor{rgb color={0.400,0.400,0.400}}
\gpsetlinetype{gp lt border}
\gpsetdashtype{gp dt solid}
\gpsetlinewidth{2.00}
\draw[gp path] (7.350,1.613)--(7.350,1.478);
\node[gp node center,font={\fontsize{8.0pt}{9.6pt}\selectfont}] at (7.350,1.016) {384};
\gpcolor{rgb color={0.702,0.702,0.702}}
\gpsetlinetype{gp lt axes}
\gpsetdashtype{gp dt axes}
\gpsetlinewidth{1.00}
\draw[gp path] (9.235,1.613)--(9.235,6.490);
\draw[gp path] (9.235,8.107)--(9.235,8.287);
\gpcolor{rgb color={0.400,0.400,0.400}}
\gpsetlinetype{gp lt border}
\gpsetdashtype{gp dt solid}
\gpsetlinewidth{2.00}
\draw[gp path] (9.235,1.613)--(9.235,1.478);
\node[gp node center,font={\fontsize{8.0pt}{9.6pt}\selectfont}] at (9.235,1.016) {1536};
\gpcolor{rgb color={0.702,0.702,0.702}}
\gpsetlinetype{gp lt axes}
\gpsetdashtype{gp dt axes}
\gpsetlinewidth{1.00}
\draw[gp path] (11.120,1.613)--(11.120,6.490);
\draw[gp path] (11.120,8.107)--(11.120,8.287);
\gpcolor{rgb color={0.400,0.400,0.400}}
\gpsetlinetype{gp lt border}
\gpsetdashtype{gp dt solid}
\gpsetlinewidth{2.00}
\draw[gp path] (11.120,1.613)--(11.120,1.478);
\node[gp node center,font={\fontsize{8.0pt}{9.6pt}\selectfont}] at (11.120,1.016) {6144};
\draw[gp path] (2.391,8.287)--(2.391,1.613)--(11.671,1.613)--(11.671,8.287)--cycle;
\gpcolor{color=gp lt color border}
\node[gp node center,rotate=-270] at (1.105,4.950) {Time [s]};
\node[gp node center] at (7.031,0.400) {Number of processes};
\gpcolor{rgb color={0.400,0.400,0.400}}
\gpsetlinewidth{1.00}
\draw[gp path] (8.179,6.490)--(8.179,8.107)--(11.395,8.107)--(11.395,6.490)--cycle;
\gpcolor{color=gp lt color border}
\node[gp node left] at (9.187,7.760) {$T_{\text{ass}}$};
\gpcolor{rgb color={0.800,0.000,0.063}}
\gpsetlinewidth{3.00}
\draw[gp path] (8.455,7.760)--(8.911,7.760);
\draw[gp path] (3.581,7.139)--(5.466,6.161)--(7.350,5.235)--(9.235,4.111)--(11.120,3.154);
\gpsetpointsize{5.20}
\gppoint{gp mark 1}{(3.581,7.139)}
\gppoint{gp mark 1}{(5.466,6.161)}
\gppoint{gp mark 1}{(7.350,5.235)}
\gppoint{gp mark 1}{(9.235,4.111)}
\gppoint{gp mark 1}{(11.120,3.154)}
\gppoint{gp mark 1}{(8.683,7.760)}
\gpcolor{color=gp lt color border}
\node[gp node left] at (9.187,7.298) {$T_{\text{init}}$};
\gpcolor{rgb color={0.800,0.000,0.663}}
\draw[gp path] (8.455,7.298)--(8.911,7.298);
\draw[gp path] (3.581,5.865)--(5.466,5.016)--(7.350,4.498)--(9.235,4.210)--(11.120,4.129);
\gppoint{gp mark 2}{(3.581,5.865)}
\gppoint{gp mark 2}{(5.466,5.016)}
\gppoint{gp mark 2}{(7.350,4.498)}
\gppoint{gp mark 2}{(9.235,4.210)}
\gppoint{gp mark 2}{(11.120,4.129)}
\gppoint{gp mark 2}{(8.683,7.298)}
\gpcolor{color=gp lt color border}
\node[gp node left] at (9.187,6.836) {$T_{\text{solve}}$};
\gpcolor{rgb color={0.329,0.000,0.800}}
\draw[gp path] (8.455,6.836)--(8.911,6.836);
\draw[gp path] (3.581,7.824)--(5.466,6.807)--(7.350,6.134)--(9.235,5.473)--(11.120,5.352);
\gppoint{gp mark 3}{(3.581,7.824)}
\gppoint{gp mark 3}{(5.466,6.807)}
\gppoint{gp mark 3}{(7.350,6.134)}
\gppoint{gp mark 3}{(9.235,5.473)}
\gppoint{gp mark 3}{(11.120,5.352)}
\gppoint{gp mark 3}{(8.683,6.836)}
\gpdefrectangularnode{gp plot 1}{\pgfpoint{2.391cm}{1.613cm}}{\pgfpoint{11.671cm}{8.287cm}}
\end{tikzpicture}
 }
\scalebox{0.5}{\footnotesize\begin{tikzpicture}[gnuplot]
\tikzset{every node/.append style={scale=1.50}}
\path (0.000,0.000) rectangle (12.500,8.750);
\gpcolor{rgb color={0.702,0.702,0.702}}
\gpsetlinetype{gp lt axes}
\gpsetdashtype{gp dt axes}
\gpsetlinewidth{1.00}
\draw[gp path] (2.115,2.280)--(11.671,2.280);
\gpcolor{rgb color={0.400,0.400,0.400}}
\gpsetlinetype{gp lt border}
\gpsetdashtype{gp dt solid}
\gpsetlinewidth{2.00}
\draw[gp path] (2.115,2.280)--(1.980,2.280);
\node[gp node right,font={\fontsize{4.0pt}{4.8pt}\selectfont}] at (1.980,2.280) {1};
\gpcolor{rgb color={0.702,0.702,0.702}}
\gpsetlinetype{gp lt axes}
\gpsetdashtype{gp dt axes}
\gpsetlinewidth{1.00}
\draw[gp path] (2.115,2.733)--(11.671,2.733);
\gpcolor{rgb color={0.400,0.400,0.400}}
\gpsetlinetype{gp lt border}
\gpsetdashtype{gp dt solid}
\gpsetlinewidth{2.00}
\draw[gp path] (2.115,2.733)--(2.048,2.733);
\gpcolor{rgb color={0.702,0.702,0.702}}
\gpsetlinetype{gp lt axes}
\gpsetdashtype{gp dt axes}
\gpsetlinewidth{1.00}
\draw[gp path] (2.115,3.040)--(11.671,3.040);
\gpcolor{rgb color={0.400,0.400,0.400}}
\gpsetlinetype{gp lt border}
\gpsetdashtype{gp dt solid}
\gpsetlinewidth{2.00}
\draw[gp path] (2.115,3.040)--(2.048,3.040);
\gpcolor{rgb color={0.702,0.702,0.702}}
\gpsetlinetype{gp lt axes}
\gpsetdashtype{gp dt axes}
\gpsetlinewidth{1.00}
\draw[gp path] (2.115,3.272)--(11.671,3.272);
\gpcolor{rgb color={0.400,0.400,0.400}}
\gpsetlinetype{gp lt border}
\gpsetdashtype{gp dt solid}
\gpsetlinewidth{2.00}
\draw[gp path] (2.115,3.272)--(2.048,3.272);
\gpcolor{rgb color={0.702,0.702,0.702}}
\gpsetlinetype{gp lt axes}
\gpsetdashtype{gp dt axes}
\gpsetlinewidth{1.00}
\draw[gp path] (2.115,3.459)--(11.671,3.459);
\gpcolor{rgb color={0.400,0.400,0.400}}
\gpsetlinetype{gp lt border}
\gpsetdashtype{gp dt solid}
\gpsetlinewidth{2.00}
\draw[gp path] (2.115,3.459)--(2.048,3.459);
\gpcolor{rgb color={0.702,0.702,0.702}}
\gpsetlinetype{gp lt axes}
\gpsetdashtype{gp dt axes}
\gpsetlinewidth{1.00}
\draw[gp path] (2.115,3.615)--(11.671,3.615);
\gpcolor{rgb color={0.400,0.400,0.400}}
\gpsetlinetype{gp lt border}
\gpsetdashtype{gp dt solid}
\gpsetlinewidth{2.00}
\draw[gp path] (2.115,3.615)--(1.980,3.615);
\node[gp node right,font={\fontsize{4.0pt}{4.8pt}\selectfont}] at (1.888,3.615) {4};
\gpcolor{rgb color={0.702,0.702,0.702}}
\gpsetlinetype{gp lt axes}
\gpsetdashtype{gp dt axes}
\gpsetlinewidth{1.00}
\draw[gp path] (2.115,4.068)--(11.671,4.068);
\gpcolor{rgb color={0.400,0.400,0.400}}
\gpsetlinetype{gp lt border}
\gpsetdashtype{gp dt solid}
\gpsetlinewidth{2.00}
\draw[gp path] (2.115,4.068)--(2.048,4.068);
\gpcolor{rgb color={0.702,0.702,0.702}}
\gpsetlinetype{gp lt axes}
\gpsetdashtype{gp dt axes}
\gpsetlinewidth{1.00}
\draw[gp path] (2.115,4.374)--(11.671,4.374);
\gpcolor{rgb color={0.400,0.400,0.400}}
\gpsetlinetype{gp lt border}
\gpsetdashtype{gp dt solid}
\gpsetlinewidth{2.00}
\draw[gp path] (2.115,4.374)--(2.048,4.374);
\gpcolor{rgb color={0.702,0.702,0.702}}
\gpsetlinetype{gp lt axes}
\gpsetdashtype{gp dt axes}
\gpsetlinewidth{1.00}
\draw[gp path] (2.115,4.607)--(11.671,4.607);
\gpcolor{rgb color={0.400,0.400,0.400}}
\gpsetlinetype{gp lt border}
\gpsetdashtype{gp dt solid}
\gpsetlinewidth{2.00}
\draw[gp path] (2.115,4.607)--(2.048,4.607);
\gpcolor{rgb color={0.702,0.702,0.702}}
\gpsetlinetype{gp lt axes}
\gpsetdashtype{gp dt axes}
\gpsetlinewidth{1.00}
\draw[gp path] (2.115,4.794)--(11.671,4.794);
\gpcolor{rgb color={0.400,0.400,0.400}}
\gpsetlinetype{gp lt border}
\gpsetdashtype{gp dt solid}
\gpsetlinewidth{2.00}
\draw[gp path] (2.115,4.794)--(2.048,4.794);
\gpcolor{rgb color={0.702,0.702,0.702}}
\gpsetlinetype{gp lt axes}
\gpsetdashtype{gp dt axes}
\gpsetlinewidth{1.00}
\draw[gp path] (2.115,4.950)--(11.671,4.950);
\gpcolor{rgb color={0.400,0.400,0.400}}
\gpsetlinetype{gp lt border}
\gpsetdashtype{gp dt solid}
\gpsetlinewidth{2.00}
\draw[gp path] (2.115,4.950)--(1.980,4.950);
\node[gp node right,font={\fontsize{4.0pt}{4.8pt}\selectfont}] at (1.888,4.950) {16};
\gpcolor{rgb color={0.702,0.702,0.702}}
\gpsetlinetype{gp lt axes}
\gpsetdashtype{gp dt axes}
\gpsetlinewidth{1.00}
\draw[gp path] (2.115,5.403)--(11.671,5.403);
\gpcolor{rgb color={0.400,0.400,0.400}}
\gpsetlinetype{gp lt border}
\gpsetdashtype{gp dt solid}
\gpsetlinewidth{2.00}
\draw[gp path] (2.115,5.403)--(2.048,5.403);
\gpcolor{rgb color={0.702,0.702,0.702}}
\gpsetlinetype{gp lt axes}
\gpsetdashtype{gp dt axes}
\gpsetlinewidth{1.00}
\draw[gp path] (2.115,5.709)--(11.671,5.709);
\gpcolor{rgb color={0.400,0.400,0.400}}
\gpsetlinetype{gp lt border}
\gpsetdashtype{gp dt solid}
\gpsetlinewidth{2.00}
\draw[gp path] (2.115,5.709)--(2.048,5.709);
\gpcolor{rgb color={0.702,0.702,0.702}}
\gpsetlinetype{gp lt axes}
\gpsetdashtype{gp dt axes}
\gpsetlinewidth{1.00}
\draw[gp path] (2.115,5.941)--(11.671,5.941);
\gpcolor{rgb color={0.400,0.400,0.400}}
\gpsetlinetype{gp lt border}
\gpsetdashtype{gp dt solid}
\gpsetlinewidth{2.00}
\draw[gp path] (2.115,5.941)--(2.048,5.941);
\gpcolor{rgb color={0.702,0.702,0.702}}
\gpsetlinetype{gp lt axes}
\gpsetdashtype{gp dt axes}
\gpsetlinewidth{1.00}
\draw[gp path] (2.115,6.128)--(2.391,6.128);
\draw[gp path] (5.331,6.128)--(11.671,6.128);
\gpcolor{rgb color={0.400,0.400,0.400}}
\gpsetlinetype{gp lt border}
\gpsetdashtype{gp dt solid}
\gpsetlinewidth{2.00}
\draw[gp path] (2.115,6.128)--(2.048,6.128);
\gpcolor{rgb color={0.702,0.702,0.702}}
\gpsetlinetype{gp lt axes}
\gpsetdashtype{gp dt axes}
\gpsetlinewidth{1.00}
\draw[gp path] (2.115,6.285)--(2.391,6.285);
\draw[gp path] (5.331,6.285)--(11.671,6.285);
\gpcolor{rgb color={0.400,0.400,0.400}}
\gpsetlinetype{gp lt border}
\gpsetdashtype{gp dt solid}
\gpsetlinewidth{2.00}
\draw[gp path] (2.115,6.285)--(1.980,6.285);
\node[gp node right,font={\fontsize{4.0pt}{4.8pt}\selectfont}] at (1.888,6.285) {64};
\gpcolor{rgb color={0.702,0.702,0.702}}
\gpsetlinetype{gp lt axes}
\gpsetdashtype{gp dt axes}
\gpsetlinewidth{1.00}
\draw[gp path] (2.115,6.737)--(2.391,6.737);
\draw[gp path] (5.331,6.737)--(11.671,6.737);
\gpcolor{rgb color={0.400,0.400,0.400}}
\gpsetlinetype{gp lt border}
\gpsetdashtype{gp dt solid}
\gpsetlinewidth{2.00}
\draw[gp path] (2.115,6.737)--(2.048,6.737);
\gpcolor{rgb color={0.702,0.702,0.702}}
\gpsetlinetype{gp lt axes}
\gpsetdashtype{gp dt axes}
\gpsetlinewidth{1.00}
\draw[gp path] (2.115,7.044)--(2.391,7.044);
\draw[gp path] (5.331,7.044)--(11.671,7.044);
\gpcolor{rgb color={0.400,0.400,0.400}}
\gpsetlinetype{gp lt border}
\gpsetdashtype{gp dt solid}
\gpsetlinewidth{2.00}
\draw[gp path] (2.115,7.044)--(2.048,7.044);
\gpcolor{rgb color={0.702,0.702,0.702}}
\gpsetlinetype{gp lt axes}
\gpsetdashtype{gp dt axes}
\gpsetlinewidth{1.00}
\draw[gp path] (2.115,7.276)--(2.391,7.276);
\draw[gp path] (5.331,7.276)--(11.671,7.276);
\gpcolor{rgb color={0.400,0.400,0.400}}
\gpsetlinetype{gp lt border}
\gpsetdashtype{gp dt solid}
\gpsetlinewidth{2.00}
\draw[gp path] (2.115,7.276)--(2.048,7.276);
\gpcolor{rgb color={0.702,0.702,0.702}}
\gpsetlinetype{gp lt axes}
\gpsetdashtype{gp dt axes}
\gpsetlinewidth{1.00}
\draw[gp path] (2.115,7.463)--(2.391,7.463);
\draw[gp path] (5.331,7.463)--(11.671,7.463);
\gpcolor{rgb color={0.400,0.400,0.400}}
\gpsetlinetype{gp lt border}
\gpsetdashtype{gp dt solid}
\gpsetlinewidth{2.00}
\draw[gp path] (2.115,7.463)--(2.048,7.463);
\gpcolor{rgb color={0.702,0.702,0.702}}
\gpsetlinetype{gp lt axes}
\gpsetdashtype{gp dt axes}
\gpsetlinewidth{1.00}
\draw[gp path] (2.115,7.620)--(2.391,7.620);
\draw[gp path] (5.331,7.620)--(11.671,7.620);
\gpcolor{rgb color={0.400,0.400,0.400}}
\gpsetlinetype{gp lt border}
\gpsetdashtype{gp dt solid}
\gpsetlinewidth{2.00}
\draw[gp path] (2.115,7.620)--(1.980,7.620);
\node[gp node right,font={\fontsize{4.0pt}{4.8pt}\selectfont}] at (1.888,7.620) {256};
\gpcolor{rgb color={0.702,0.702,0.702}}
\gpsetlinetype{gp lt axes}
\gpsetdashtype{gp dt axes}
\gpsetlinewidth{1.00}
\draw[gp path] (2.115,8.072)--(2.391,8.072);
\draw[gp path] (5.331,8.072)--(11.671,8.072);
\gpcolor{rgb color={0.400,0.400,0.400}}
\gpsetlinetype{gp lt border}
\gpsetdashtype{gp dt solid}
\gpsetlinewidth{2.00}
\draw[gp path] (2.115,8.072)--(2.048,8.072);
\gpcolor{rgb color={0.702,0.702,0.702}}
\gpsetlinetype{gp lt axes}
\gpsetdashtype{gp dt axes}
\gpsetlinewidth{1.00}
\draw[gp path] (3.341,1.613)--(3.341,6.028);
\draw[gp path] (3.341,8.107)--(3.341,8.287);
\gpcolor{rgb color={0.400,0.400,0.400}}
\gpsetlinetype{gp lt border}
\gpsetdashtype{gp dt solid}
\gpsetlinewidth{2.00}
\draw[gp path] (3.341,1.613)--(3.341,1.478);
\node[gp node center,font={\fontsize{4.0pt}{4.8pt}\selectfont}] at (3.341,1.016) {24};
\gpcolor{rgb color={0.702,0.702,0.702}}
\gpsetlinetype{gp lt axes}
\gpsetdashtype{gp dt axes}
\gpsetlinewidth{1.00}
\draw[gp path] (5.281,1.613)--(5.281,6.028);
\draw[gp path] (5.281,8.107)--(5.281,8.287);
\gpcolor{rgb color={0.400,0.400,0.400}}
\gpsetlinetype{gp lt border}
\gpsetdashtype{gp dt solid}
\gpsetlinewidth{2.00}
\draw[gp path] (5.281,1.613)--(5.281,1.478);
\node[gp node center,font={\fontsize{4.0pt}{4.8pt}\selectfont}] at (5.281,1.016) {96};
\gpcolor{rgb color={0.702,0.702,0.702}}
\gpsetlinetype{gp lt axes}
\gpsetdashtype{gp dt axes}
\gpsetlinewidth{1.00}
\draw[gp path] (7.222,1.613)--(7.222,8.287);
\gpcolor{rgb color={0.400,0.400,0.400}}
\gpsetlinetype{gp lt border}
\gpsetdashtype{gp dt solid}
\gpsetlinewidth{2.00}
\draw[gp path] (7.222,1.613)--(7.222,1.478);
\node[gp node center,font={\fontsize{4.0pt}{4.8pt}\selectfont}] at (7.222,1.016) {384};
\gpcolor{rgb color={0.702,0.702,0.702}}
\gpsetlinetype{gp lt axes}
\gpsetdashtype{gp dt axes}
\gpsetlinewidth{1.00}
\draw[gp path] (9.163,1.613)--(9.163,8.287);
\gpcolor{rgb color={0.400,0.400,0.400}}
\gpsetlinetype{gp lt border}
\gpsetdashtype{gp dt solid}
\gpsetlinewidth{2.00}
\draw[gp path] (9.163,1.613)--(9.163,1.478);
\node[gp node center,font={\fontsize{4.0pt}{4.8pt}\selectfont}] at (9.163,1.016) {1536};
\gpcolor{rgb color={0.702,0.702,0.702}}
\gpsetlinetype{gp lt axes}
\gpsetdashtype{gp dt axes}
\gpsetlinewidth{1.00}
\draw[gp path] (11.103,1.613)--(11.103,8.287);
\gpcolor{rgb color={0.400,0.400,0.400}}
\gpsetlinetype{gp lt border}
\gpsetdashtype{gp dt solid}
\gpsetlinewidth{2.00}
\draw[gp path] (11.103,1.613)--(11.103,1.478);
\node[gp node center,font={\fontsize{4.0pt}{4.8pt}\selectfont}] at (11.103,1.016) {6144};
\draw[gp path] (2.115,8.287)--(2.115,1.613)--(11.671,1.613)--(11.671,8.287)--cycle;
\gpcolor{color=gp lt color border}
\node[gp node center,rotate=-270] at (1.105,4.950) {Speedup (strong)};
\node[gp node center] at (6.893,0.400) {Number of processes};
\gpcolor{rgb color={0.400,0.400,0.400}}
\gpsetlinewidth{1.00}
\draw[gp path] (2.391,6.028)--(2.391,8.107)--(5.331,8.107)--(5.331,6.028)--cycle;
\gpcolor{color=gp lt color border}
\node[gp node left] at (3.399,7.760) {$S_{\text{ass}}$};
\gpcolor{rgb color={0.800,0.000,0.063}}
\gpsetlinewidth{3.00}
\draw[gp path] (2.667,7.760)--(3.123,7.760);
\draw[gp path] (3.341,2.280)--(5.281,3.548)--(7.222,4.749)--(9.163,6.206)--(11.103,7.447);
\gpsetpointsize{5.20}
\gppoint{gp mark 1}{(3.341,2.280)}
\gppoint{gp mark 1}{(5.281,3.548)}
\gppoint{gp mark 1}{(7.222,4.749)}
\gppoint{gp mark 1}{(9.163,6.206)}
\gppoint{gp mark 1}{(11.103,7.447)}
\gppoint{gp mark 1}{(2.895,7.760)}
\gpcolor{color=gp lt color border}
\node[gp node left] at (3.399,7.298) {$S_{\text{init}}$};
\gpcolor{rgb color={0.800,0.000,0.663}}
\draw[gp path] (2.667,7.298)--(3.123,7.298);
\draw[gp path] (3.341,2.280)--(5.281,3.382)--(7.222,4.054)--(9.163,4.427)--(11.103,4.532);
\gppoint{gp mark 2}{(3.341,2.280)}
\gppoint{gp mark 2}{(5.281,3.382)}
\gppoint{gp mark 2}{(7.222,4.054)}
\gppoint{gp mark 2}{(9.163,4.427)}
\gppoint{gp mark 2}{(11.103,4.532)}
\gppoint{gp mark 2}{(2.895,7.298)}
\gpcolor{color=gp lt color border}
\node[gp node left] at (3.399,6.836) {$S_{\text{solve}}$};
\gpcolor{rgb color={0.329,0.000,0.800}}
\draw[gp path] (2.667,6.836)--(3.123,6.836);
\draw[gp path] (3.341,2.280)--(5.281,3.599)--(7.222,4.472)--(9.163,5.329)--(11.103,5.486);
\gppoint{gp mark 3}{(3.341,2.280)}
\gppoint{gp mark 3}{(5.281,3.599)}
\gppoint{gp mark 3}{(7.222,4.472)}
\gppoint{gp mark 3}{(9.163,5.329)}
\gppoint{gp mark 3}{(11.103,5.486)}
\gppoint{gp mark 3}{(2.895,6.836)}
\gpcolor{color=gp lt color border}
\node[gp node left] at (3.399,6.374) {$S_{\text{ideal}}$};
\gpcolor{rgb color={0.000,0.263,0.800}}
\draw[gp path] (2.667,6.374)--(3.123,6.374);
\draw[gp path] (3.341,2.280)--(5.281,3.615)--(7.222,4.950)--(9.163,6.285)--(11.103,7.620);
\gpdefrectangularnode{gp plot 1}{\pgfpoint{2.115cm}{1.613cm}}{\pgfpoint{11.671cm}{8.287cm}}
\end{tikzpicture}
 }
		\caption{Linear Elastic Problem Solver}
		\label{tab:strongElast}
	\end{subfigure}

	\begin{subfigure}[t]{\textwidth}
		\centering
		\scalebox{0.5}{\footnotesize\begin{tikzpicture}[gnuplot]
\tikzset{every node/.append style={scale=1.50}}
\path (0.000,0.000) rectangle (12.500,8.750);
\gpcolor{rgb color={0.702,0.702,0.702}}
\gpsetlinetype{gp lt axes}
\gpsetdashtype{gp dt axes}
\gpsetlinewidth{1.00}
\draw[gp path] (2.943,1.613)--(11.671,1.613);
\gpcolor{rgb color={0.400,0.400,0.400}}
\gpsetlinetype{gp lt border}
\gpsetdashtype{gp dt solid}
\gpsetlinewidth{2.00}
\draw[gp path] (2.943,1.613)--(2.808,1.613);
\node[gp node right,font={\fontsize{8.0pt}{9.6pt}\selectfont}] at (2.808,1.613) {0.01};
\gpcolor{rgb color={0.702,0.702,0.702}}
\gpsetlinetype{gp lt axes}
\gpsetdashtype{gp dt axes}
\gpsetlinewidth{1.00}
\draw[gp path] (2.943,2.039)--(11.671,2.039);
\gpcolor{rgb color={0.400,0.400,0.400}}
\gpsetlinetype{gp lt border}
\gpsetdashtype{gp dt solid}
\gpsetlinewidth{2.00}
\draw[gp path] (2.943,2.039)--(2.876,2.039);
\gpcolor{rgb color={0.702,0.702,0.702}}
\gpsetlinetype{gp lt axes}
\gpsetdashtype{gp dt axes}
\gpsetlinewidth{1.00}
\draw[gp path] (2.943,2.245)--(11.671,2.245);
\gpcolor{rgb color={0.400,0.400,0.400}}
\gpsetlinetype{gp lt border}
\gpsetdashtype{gp dt solid}
\gpsetlinewidth{2.00}
\draw[gp path] (2.943,2.245)--(2.876,2.245);
\gpcolor{rgb color={0.702,0.702,0.702}}
\gpsetlinetype{gp lt axes}
\gpsetdashtype{gp dt axes}
\gpsetlinewidth{1.00}
\draw[gp path] (2.943,2.382)--(11.671,2.382);
\gpcolor{rgb color={0.400,0.400,0.400}}
\gpsetlinetype{gp lt border}
\gpsetdashtype{gp dt solid}
\gpsetlinewidth{2.00}
\draw[gp path] (2.943,2.382)--(2.876,2.382);
\gpcolor{rgb color={0.702,0.702,0.702}}
\gpsetlinetype{gp lt axes}
\gpsetdashtype{gp dt axes}
\gpsetlinewidth{1.00}
\draw[gp path] (2.943,2.484)--(11.671,2.484);
\gpcolor{rgb color={0.400,0.400,0.400}}
\gpsetlinetype{gp lt border}
\gpsetdashtype{gp dt solid}
\gpsetlinewidth{2.00}
\draw[gp path] (2.943,2.484)--(2.876,2.484);
\gpcolor{rgb color={0.702,0.702,0.702}}
\gpsetlinetype{gp lt axes}
\gpsetdashtype{gp dt axes}
\gpsetlinewidth{1.00}
\draw[gp path] (2.943,2.566)--(11.671,2.566);
\gpcolor{rgb color={0.400,0.400,0.400}}
\gpsetlinetype{gp lt border}
\gpsetdashtype{gp dt solid}
\gpsetlinewidth{2.00}
\draw[gp path] (2.943,2.566)--(2.808,2.566);
\node[gp node right,font={\fontsize{8.0pt}{9.6pt}\selectfont}] at (2.716,2.566) {0.1};
\gpcolor{rgb color={0.702,0.702,0.702}}
\gpsetlinetype{gp lt axes}
\gpsetdashtype{gp dt axes}
\gpsetlinewidth{1.00}
\draw[gp path] (2.943,2.993)--(11.671,2.993);
\gpcolor{rgb color={0.400,0.400,0.400}}
\gpsetlinetype{gp lt border}
\gpsetdashtype{gp dt solid}
\gpsetlinewidth{2.00}
\draw[gp path] (2.943,2.993)--(2.876,2.993);
\gpcolor{rgb color={0.702,0.702,0.702}}
\gpsetlinetype{gp lt axes}
\gpsetdashtype{gp dt axes}
\gpsetlinewidth{1.00}
\draw[gp path] (2.943,3.198)--(11.671,3.198);
\gpcolor{rgb color={0.400,0.400,0.400}}
\gpsetlinetype{gp lt border}
\gpsetdashtype{gp dt solid}
\gpsetlinewidth{2.00}
\draw[gp path] (2.943,3.198)--(2.876,3.198);
\gpcolor{rgb color={0.702,0.702,0.702}}
\gpsetlinetype{gp lt axes}
\gpsetdashtype{gp dt axes}
\gpsetlinewidth{1.00}
\draw[gp path] (2.943,3.335)--(11.671,3.335);
\gpcolor{rgb color={0.400,0.400,0.400}}
\gpsetlinetype{gp lt border}
\gpsetdashtype{gp dt solid}
\gpsetlinewidth{2.00}
\draw[gp path] (2.943,3.335)--(2.876,3.335);
\gpcolor{rgb color={0.702,0.702,0.702}}
\gpsetlinetype{gp lt axes}
\gpsetdashtype{gp dt axes}
\gpsetlinewidth{1.00}
\draw[gp path] (2.943,3.438)--(11.671,3.438);
\gpcolor{rgb color={0.400,0.400,0.400}}
\gpsetlinetype{gp lt border}
\gpsetdashtype{gp dt solid}
\gpsetlinewidth{2.00}
\draw[gp path] (2.943,3.438)--(2.876,3.438);
\gpcolor{rgb color={0.702,0.702,0.702}}
\gpsetlinetype{gp lt axes}
\gpsetdashtype{gp dt axes}
\gpsetlinewidth{1.00}
\draw[gp path] (2.943,3.520)--(11.671,3.520);
\gpcolor{rgb color={0.400,0.400,0.400}}
\gpsetlinetype{gp lt border}
\gpsetdashtype{gp dt solid}
\gpsetlinewidth{2.00}
\draw[gp path] (2.943,3.520)--(2.808,3.520);
\node[gp node right,font={\fontsize{8.0pt}{9.6pt}\selectfont}] at (2.716,3.520) {1};
\gpcolor{rgb color={0.702,0.702,0.702}}
\gpsetlinetype{gp lt axes}
\gpsetdashtype{gp dt axes}
\gpsetlinewidth{1.00}
\draw[gp path] (2.943,3.946)--(11.671,3.946);
\gpcolor{rgb color={0.400,0.400,0.400}}
\gpsetlinetype{gp lt border}
\gpsetdashtype{gp dt solid}
\gpsetlinewidth{2.00}
\draw[gp path] (2.943,3.946)--(2.876,3.946);
\gpcolor{rgb color={0.702,0.702,0.702}}
\gpsetlinetype{gp lt axes}
\gpsetdashtype{gp dt axes}
\gpsetlinewidth{1.00}
\draw[gp path] (2.943,4.152)--(11.671,4.152);
\gpcolor{rgb color={0.400,0.400,0.400}}
\gpsetlinetype{gp lt border}
\gpsetdashtype{gp dt solid}
\gpsetlinewidth{2.00}
\draw[gp path] (2.943,4.152)--(2.876,4.152);
\gpcolor{rgb color={0.702,0.702,0.702}}
\gpsetlinetype{gp lt axes}
\gpsetdashtype{gp dt axes}
\gpsetlinewidth{1.00}
\draw[gp path] (2.943,4.288)--(11.671,4.288);
\gpcolor{rgb color={0.400,0.400,0.400}}
\gpsetlinetype{gp lt border}
\gpsetdashtype{gp dt solid}
\gpsetlinewidth{2.00}
\draw[gp path] (2.943,4.288)--(2.876,4.288);
\gpcolor{rgb color={0.702,0.702,0.702}}
\gpsetlinetype{gp lt axes}
\gpsetdashtype{gp dt axes}
\gpsetlinewidth{1.00}
\draw[gp path] (2.943,4.391)--(11.671,4.391);
\gpcolor{rgb color={0.400,0.400,0.400}}
\gpsetlinetype{gp lt border}
\gpsetdashtype{gp dt solid}
\gpsetlinewidth{2.00}
\draw[gp path] (2.943,4.391)--(2.876,4.391);
\gpcolor{rgb color={0.702,0.702,0.702}}
\gpsetlinetype{gp lt axes}
\gpsetdashtype{gp dt axes}
\gpsetlinewidth{1.00}
\draw[gp path] (2.943,4.473)--(11.671,4.473);
\gpcolor{rgb color={0.400,0.400,0.400}}
\gpsetlinetype{gp lt border}
\gpsetdashtype{gp dt solid}
\gpsetlinewidth{2.00}
\draw[gp path] (2.943,4.473)--(2.808,4.473);
\node[gp node right,font={\fontsize{8.0pt}{9.6pt}\selectfont}] at (2.716,4.473) {10};
\gpcolor{rgb color={0.702,0.702,0.702}}
\gpsetlinetype{gp lt axes}
\gpsetdashtype{gp dt axes}
\gpsetlinewidth{1.00}
\draw[gp path] (2.943,4.900)--(11.671,4.900);
\gpcolor{rgb color={0.400,0.400,0.400}}
\gpsetlinetype{gp lt border}
\gpsetdashtype{gp dt solid}
\gpsetlinewidth{2.00}
\draw[gp path] (2.943,4.900)--(2.876,4.900);
\gpcolor{rgb color={0.702,0.702,0.702}}
\gpsetlinetype{gp lt axes}
\gpsetdashtype{gp dt axes}
\gpsetlinewidth{1.00}
\draw[gp path] (2.943,5.105)--(11.671,5.105);
\gpcolor{rgb color={0.400,0.400,0.400}}
\gpsetlinetype{gp lt border}
\gpsetdashtype{gp dt solid}
\gpsetlinewidth{2.00}
\draw[gp path] (2.943,5.105)--(2.876,5.105);
\gpcolor{rgb color={0.702,0.702,0.702}}
\gpsetlinetype{gp lt axes}
\gpsetdashtype{gp dt axes}
\gpsetlinewidth{1.00}
\draw[gp path] (2.943,5.242)--(11.671,5.242);
\gpcolor{rgb color={0.400,0.400,0.400}}
\gpsetlinetype{gp lt border}
\gpsetdashtype{gp dt solid}
\gpsetlinewidth{2.00}
\draw[gp path] (2.943,5.242)--(2.876,5.242);
\gpcolor{rgb color={0.702,0.702,0.702}}
\gpsetlinetype{gp lt axes}
\gpsetdashtype{gp dt axes}
\gpsetlinewidth{1.00}
\draw[gp path] (2.943,5.345)--(11.671,5.345);
\gpcolor{rgb color={0.400,0.400,0.400}}
\gpsetlinetype{gp lt border}
\gpsetdashtype{gp dt solid}
\gpsetlinewidth{2.00}
\draw[gp path] (2.943,5.345)--(2.876,5.345);
\gpcolor{rgb color={0.702,0.702,0.702}}
\gpsetlinetype{gp lt axes}
\gpsetdashtype{gp dt axes}
\gpsetlinewidth{1.00}
\draw[gp path] (2.943,5.427)--(11.671,5.427);
\gpcolor{rgb color={0.400,0.400,0.400}}
\gpsetlinetype{gp lt border}
\gpsetdashtype{gp dt solid}
\gpsetlinewidth{2.00}
\draw[gp path] (2.943,5.427)--(2.808,5.427);
\node[gp node right,font={\fontsize{8.0pt}{9.6pt}\selectfont}] at (2.716,5.427) {100};
\gpcolor{rgb color={0.702,0.702,0.702}}
\gpsetlinetype{gp lt axes}
\gpsetdashtype{gp dt axes}
\gpsetlinewidth{1.00}
\draw[gp path] (2.943,5.853)--(11.671,5.853);
\gpcolor{rgb color={0.400,0.400,0.400}}
\gpsetlinetype{gp lt border}
\gpsetdashtype{gp dt solid}
\gpsetlinewidth{2.00}
\draw[gp path] (2.943,5.853)--(2.876,5.853);
\gpcolor{rgb color={0.702,0.702,0.702}}
\gpsetlinetype{gp lt axes}
\gpsetdashtype{gp dt axes}
\gpsetlinewidth{1.00}
\draw[gp path] (2.943,6.059)--(11.671,6.059);
\gpcolor{rgb color={0.400,0.400,0.400}}
\gpsetlinetype{gp lt border}
\gpsetdashtype{gp dt solid}
\gpsetlinewidth{2.00}
\draw[gp path] (2.943,6.059)--(2.876,6.059);
\gpcolor{rgb color={0.702,0.702,0.702}}
\gpsetlinetype{gp lt axes}
\gpsetdashtype{gp dt axes}
\gpsetlinewidth{1.00}
\draw[gp path] (2.943,6.195)--(11.671,6.195);
\gpcolor{rgb color={0.400,0.400,0.400}}
\gpsetlinetype{gp lt border}
\gpsetdashtype{gp dt solid}
\gpsetlinewidth{2.00}
\draw[gp path] (2.943,6.195)--(2.876,6.195);
\gpcolor{rgb color={0.702,0.702,0.702}}
\gpsetlinetype{gp lt axes}
\gpsetdashtype{gp dt axes}
\gpsetlinewidth{1.00}
\draw[gp path] (2.943,6.298)--(11.671,6.298);
\gpcolor{rgb color={0.400,0.400,0.400}}
\gpsetlinetype{gp lt border}
\gpsetdashtype{gp dt solid}
\gpsetlinewidth{2.00}
\draw[gp path] (2.943,6.298)--(2.876,6.298);
\gpcolor{rgb color={0.702,0.702,0.702}}
\gpsetlinetype{gp lt axes}
\gpsetdashtype{gp dt axes}
\gpsetlinewidth{1.00}
\draw[gp path] (2.943,6.380)--(11.671,6.380);
\gpcolor{rgb color={0.400,0.400,0.400}}
\gpsetlinetype{gp lt border}
\gpsetdashtype{gp dt solid}
\gpsetlinewidth{2.00}
\draw[gp path] (2.943,6.380)--(2.808,6.380);
\node[gp node right,font={\fontsize{8.0pt}{9.6pt}\selectfont}] at (2.716,6.380) {1000};
\gpcolor{rgb color={0.702,0.702,0.702}}
\gpsetlinetype{gp lt axes}
\gpsetdashtype{gp dt axes}
\gpsetlinewidth{1.00}
\draw[gp path] (2.943,6.806)--(7.075,6.806);
\draw[gp path] (11.395,6.806)--(11.671,6.806);
\gpcolor{rgb color={0.400,0.400,0.400}}
\gpsetlinetype{gp lt border}
\gpsetdashtype{gp dt solid}
\gpsetlinewidth{2.00}
\draw[gp path] (2.943,6.806)--(2.876,6.806);
\gpcolor{rgb color={0.702,0.702,0.702}}
\gpsetlinetype{gp lt axes}
\gpsetdashtype{gp dt axes}
\gpsetlinewidth{1.00}
\draw[gp path] (2.943,7.012)--(7.075,7.012);
\draw[gp path] (11.395,7.012)--(11.671,7.012);
\gpcolor{rgb color={0.400,0.400,0.400}}
\gpsetlinetype{gp lt border}
\gpsetdashtype{gp dt solid}
\gpsetlinewidth{2.00}
\draw[gp path] (2.943,7.012)--(2.876,7.012);
\gpcolor{rgb color={0.702,0.702,0.702}}
\gpsetlinetype{gp lt axes}
\gpsetdashtype{gp dt axes}
\gpsetlinewidth{1.00}
\draw[gp path] (2.943,7.149)--(7.075,7.149);
\draw[gp path] (11.395,7.149)--(11.671,7.149);
\gpcolor{rgb color={0.400,0.400,0.400}}
\gpsetlinetype{gp lt border}
\gpsetdashtype{gp dt solid}
\gpsetlinewidth{2.00}
\draw[gp path] (2.943,7.149)--(2.876,7.149);
\gpcolor{rgb color={0.702,0.702,0.702}}
\gpsetlinetype{gp lt axes}
\gpsetdashtype{gp dt axes}
\gpsetlinewidth{1.00}
\draw[gp path] (2.943,7.251)--(7.075,7.251);
\draw[gp path] (11.395,7.251)--(11.671,7.251);
\gpcolor{rgb color={0.400,0.400,0.400}}
\gpsetlinetype{gp lt border}
\gpsetdashtype{gp dt solid}
\gpsetlinewidth{2.00}
\draw[gp path] (2.943,7.251)--(2.876,7.251);
\gpcolor{rgb color={0.702,0.702,0.702}}
\gpsetlinetype{gp lt axes}
\gpsetdashtype{gp dt axes}
\gpsetlinewidth{1.00}
\draw[gp path] (2.943,7.334)--(7.075,7.334);
\draw[gp path] (11.395,7.334)--(11.671,7.334);
\gpcolor{rgb color={0.400,0.400,0.400}}
\gpsetlinetype{gp lt border}
\gpsetdashtype{gp dt solid}
\gpsetlinewidth{2.00}
\draw[gp path] (2.943,7.334)--(2.808,7.334);
\node[gp node right,font={\fontsize{8.0pt}{9.6pt}\selectfont}] at (2.716,7.334) {10000};
\gpcolor{rgb color={0.702,0.702,0.702}}
\gpsetlinetype{gp lt axes}
\gpsetdashtype{gp dt axes}
\gpsetlinewidth{1.00}
\draw[gp path] (2.943,7.760)--(7.075,7.760);
\draw[gp path] (11.395,7.760)--(11.671,7.760);
\gpcolor{rgb color={0.400,0.400,0.400}}
\gpsetlinetype{gp lt border}
\gpsetdashtype{gp dt solid}
\gpsetlinewidth{2.00}
\draw[gp path] (2.943,7.760)--(2.876,7.760);
\gpcolor{rgb color={0.702,0.702,0.702}}
\gpsetlinetype{gp lt axes}
\gpsetdashtype{gp dt axes}
\gpsetlinewidth{1.00}
\draw[gp path] (2.943,7.965)--(7.075,7.965);
\draw[gp path] (11.395,7.965)--(11.671,7.965);
\gpcolor{rgb color={0.400,0.400,0.400}}
\gpsetlinetype{gp lt border}
\gpsetdashtype{gp dt solid}
\gpsetlinewidth{2.00}
\draw[gp path] (2.943,7.965)--(2.876,7.965);
\gpcolor{rgb color={0.702,0.702,0.702}}
\gpsetlinetype{gp lt axes}
\gpsetdashtype{gp dt axes}
\gpsetlinewidth{1.00}
\draw[gp path] (2.943,8.102)--(7.075,8.102);
\draw[gp path] (11.395,8.102)--(11.671,8.102);
\gpcolor{rgb color={0.400,0.400,0.400}}
\gpsetlinetype{gp lt border}
\gpsetdashtype{gp dt solid}
\gpsetlinewidth{2.00}
\draw[gp path] (2.943,8.102)--(2.876,8.102);
\gpcolor{rgb color={0.702,0.702,0.702}}
\gpsetlinetype{gp lt axes}
\gpsetdashtype{gp dt axes}
\gpsetlinewidth{1.00}
\draw[gp path] (2.943,8.205)--(11.671,8.205);
\gpcolor{rgb color={0.400,0.400,0.400}}
\gpsetlinetype{gp lt border}
\gpsetdashtype{gp dt solid}
\gpsetlinewidth{2.00}
\draw[gp path] (2.943,8.205)--(2.876,8.205);
\gpcolor{rgb color={0.702,0.702,0.702}}
\gpsetlinetype{gp lt axes}
\gpsetdashtype{gp dt axes}
\gpsetlinewidth{1.00}
\draw[gp path] (2.943,8.287)--(11.671,8.287);
\gpcolor{rgb color={0.400,0.400,0.400}}
\gpsetlinetype{gp lt border}
\gpsetdashtype{gp dt solid}
\gpsetlinewidth{2.00}
\draw[gp path] (2.943,8.287)--(2.808,8.287);
\node[gp node right,font={\fontsize{8.0pt}{9.6pt}\selectfont}] at (2.716,8.287) {100000};
\gpcolor{rgb color={0.702,0.702,0.702}}
\gpsetlinetype{gp lt axes}
\gpsetdashtype{gp dt axes}
\gpsetlinewidth{1.00}
\draw[gp path] (4.062,1.613)--(4.062,8.287);
\gpcolor{rgb color={0.400,0.400,0.400}}
\gpsetlinetype{gp lt border}
\gpsetdashtype{gp dt solid}
\gpsetlinewidth{2.00}
\draw[gp path] (4.062,1.613)--(4.062,1.478);
\node[gp node center,font={\fontsize{8.0pt}{9.6pt}\selectfont}] at (4.062,1.016) {24};
\gpcolor{rgb color={0.702,0.702,0.702}}
\gpsetlinetype{gp lt axes}
\gpsetdashtype{gp dt axes}
\gpsetlinewidth{1.00}
\draw[gp path] (5.835,1.613)--(5.835,8.287);
\gpcolor{rgb color={0.400,0.400,0.400}}
\gpsetlinetype{gp lt border}
\gpsetdashtype{gp dt solid}
\gpsetlinewidth{2.00}
\draw[gp path] (5.835,1.613)--(5.835,1.478);
\node[gp node center,font={\fontsize{8.0pt}{9.6pt}\selectfont}] at (5.835,1.016) {96};
\gpcolor{rgb color={0.702,0.702,0.702}}
\gpsetlinetype{gp lt axes}
\gpsetdashtype{gp dt axes}
\gpsetlinewidth{1.00}
\draw[gp path] (7.607,1.613)--(7.607,6.490);
\draw[gp path] (7.607,8.107)--(7.607,8.287);
\gpcolor{rgb color={0.400,0.400,0.400}}
\gpsetlinetype{gp lt border}
\gpsetdashtype{gp dt solid}
\gpsetlinewidth{2.00}
\draw[gp path] (7.607,1.613)--(7.607,1.478);
\node[gp node center,font={\fontsize{8.0pt}{9.6pt}\selectfont}] at (7.607,1.016) {384};
\gpcolor{rgb color={0.702,0.702,0.702}}
\gpsetlinetype{gp lt axes}
\gpsetdashtype{gp dt axes}
\gpsetlinewidth{1.00}
\draw[gp path] (9.380,1.613)--(9.380,6.490);
\draw[gp path] (9.380,8.107)--(9.380,8.287);
\gpcolor{rgb color={0.400,0.400,0.400}}
\gpsetlinetype{gp lt border}
\gpsetdashtype{gp dt solid}
\gpsetlinewidth{2.00}
\draw[gp path] (9.380,1.613)--(9.380,1.478);
\node[gp node center,font={\fontsize{8.0pt}{9.6pt}\selectfont}] at (9.380,1.016) {1536};
\gpcolor{rgb color={0.702,0.702,0.702}}
\gpsetlinetype{gp lt axes}
\gpsetdashtype{gp dt axes}
\gpsetlinewidth{1.00}
\draw[gp path] (11.153,1.613)--(11.153,6.490);
\draw[gp path] (11.153,8.107)--(11.153,8.287);
\gpcolor{rgb color={0.400,0.400,0.400}}
\gpsetlinetype{gp lt border}
\gpsetdashtype{gp dt solid}
\gpsetlinewidth{2.00}
\draw[gp path] (11.153,1.613)--(11.153,1.478);
\node[gp node center,font={\fontsize{8.0pt}{9.6pt}\selectfont}] at (11.153,1.016) {6144};
\draw[gp path] (2.943,8.287)--(2.943,1.613)--(11.671,1.613)--(11.671,8.287)--cycle;
\gpcolor{color=gp lt color border}
\node[gp node center,rotate=-270] at (1.105,4.950) {Time [s]};
\node[gp node center] at (7.307,0.400) {Number of processes};
\gpcolor{rgb color={0.400,0.400,0.400}}
\gpsetlinewidth{1.00}
\draw[gp path] (7.075,6.490)--(7.075,8.107)--(11.395,8.107)--(11.395,6.490)--cycle;
\gpcolor{color=gp lt color border}
\node[gp node left] at (8.083,7.760) {$T_{\text{lin,init}}$};
\gpcolor{rgb color={0.800,0.000,0.063}}
\gpsetlinewidth{3.00}
\draw[gp path] (7.351,7.760)--(7.807,7.760);
\draw[gp path] (4.062,5.041)--(5.835,4.401)--(7.607,4.142)--(9.380,3.962)--(11.153,3.917);
\gpsetpointsize{5.20}
\gppoint{gp mark 1}{(4.062,5.041)}
\gppoint{gp mark 1}{(5.835,4.401)}
\gppoint{gp mark 1}{(7.607,4.142)}
\gppoint{gp mark 1}{(9.380,3.962)}
\gppoint{gp mark 1}{(11.153,3.917)}
\gppoint{gp mark 1}{(7.579,7.760)}
\gpcolor{color=gp lt color border}
\node[gp node left] at (8.083,7.298) {$T_{\text{lin,solve}}$};
\gpcolor{rgb color={0.800,0.000,0.663}}
\draw[gp path] (7.351,7.298)--(7.807,7.298);
\draw[gp path] (4.062,5.330)--(5.835,4.769)--(7.607,4.375)--(9.380,4.017)--(11.153,3.953);
\gppoint{gp mark 2}{(4.062,5.330)}
\gppoint{gp mark 2}{(5.835,4.769)}
\gppoint{gp mark 2}{(7.607,4.375)}
\gppoint{gp mark 2}{(9.380,4.017)}
\gppoint{gp mark 2}{(11.153,3.953)}
\gppoint{gp mark 2}{(7.579,7.298)}
\gpcolor{color=gp lt color border}
\node[gp node left] at (8.083,6.836) {$T_{\text{newton}}$};
\gpcolor{rgb color={0.329,0.000,0.800}}
\draw[gp path] (7.351,6.836)--(7.807,6.836);
\draw[gp path] (4.062,6.329)--(5.835,5.757)--(7.607,5.277)--(9.380,4.750)--(11.153,4.418);
\gppoint{gp mark 3}{(4.062,6.329)}
\gppoint{gp mark 3}{(5.835,5.757)}
\gppoint{gp mark 3}{(7.607,5.277)}
\gppoint{gp mark 3}{(9.380,4.750)}
\gppoint{gp mark 3}{(11.153,4.418)}
\gppoint{gp mark 3}{(7.579,6.836)}
\gpdefrectangularnode{gp plot 1}{\pgfpoint{2.943cm}{1.613cm}}{\pgfpoint{11.671cm}{8.287cm}}
\end{tikzpicture}
 }
		\scalebox{0.5}{\footnotesize\begin{tikzpicture}[gnuplot]
\tikzset{every node/.append style={scale=1.50}}
\path (0.000,0.000) rectangle (12.500,8.750);
\gpcolor{rgb color={0.702,0.702,0.702}}
\gpsetlinetype{gp lt axes}
\gpsetdashtype{gp dt axes}
\gpsetlinewidth{1.00}
\draw[gp path] (2.115,2.280)--(11.671,2.280);
\gpcolor{rgb color={0.400,0.400,0.400}}
\gpsetlinetype{gp lt border}
\gpsetdashtype{gp dt solid}
\gpsetlinewidth{2.00}
\draw[gp path] (2.115,2.280)--(1.980,2.280);
\node[gp node right,font={\fontsize{8.0pt}{9.6pt}\selectfont}] at (1.980,2.280) {1};
\gpcolor{rgb color={0.702,0.702,0.702}}
\gpsetlinetype{gp lt axes}
\gpsetdashtype{gp dt axes}
\gpsetlinewidth{1.00}
\draw[gp path] (2.115,3.123)--(11.671,3.123);
\gpcolor{rgb color={0.400,0.400,0.400}}
\gpsetlinetype{gp lt border}
\gpsetdashtype{gp dt solid}
\gpsetlinewidth{2.00}
\draw[gp path] (2.115,3.123)--(2.048,3.123);
\gpcolor{rgb color={0.702,0.702,0.702}}
\gpsetlinetype{gp lt axes}
\gpsetdashtype{gp dt axes}
\gpsetlinewidth{1.00}
\draw[gp path] (2.115,3.566)--(11.671,3.566);
\gpcolor{rgb color={0.400,0.400,0.400}}
\gpsetlinetype{gp lt border}
\gpsetdashtype{gp dt solid}
\gpsetlinewidth{2.00}
\draw[gp path] (2.115,3.566)--(2.048,3.566);
\gpcolor{rgb color={0.702,0.702,0.702}}
\gpsetlinetype{gp lt axes}
\gpsetdashtype{gp dt axes}
\gpsetlinewidth{1.00}
\draw[gp path] (2.115,3.868)--(11.671,3.868);
\gpcolor{rgb color={0.400,0.400,0.400}}
\gpsetlinetype{gp lt border}
\gpsetdashtype{gp dt solid}
\gpsetlinewidth{2.00}
\draw[gp path] (2.115,3.868)--(2.048,3.868);
\gpcolor{rgb color={0.702,0.702,0.702}}
\gpsetlinetype{gp lt axes}
\gpsetdashtype{gp dt axes}
\gpsetlinewidth{1.00}
\draw[gp path] (2.115,4.097)--(11.671,4.097);
\gpcolor{rgb color={0.400,0.400,0.400}}
\gpsetlinetype{gp lt border}
\gpsetdashtype{gp dt solid}
\gpsetlinewidth{2.00}
\draw[gp path] (2.115,4.097)--(2.048,4.097);
\gpcolor{rgb color={0.702,0.702,0.702}}
\gpsetlinetype{gp lt axes}
\gpsetdashtype{gp dt axes}
\gpsetlinewidth{1.00}
\draw[gp path] (2.115,4.283)--(11.671,4.283);
\gpcolor{rgb color={0.400,0.400,0.400}}
\gpsetlinetype{gp lt border}
\gpsetdashtype{gp dt solid}
\gpsetlinewidth{2.00}
\draw[gp path] (2.115,4.283)--(1.980,4.283);
\node[gp node right,font={\fontsize{8.0pt}{9.6pt}\selectfont}] at (1.888,4.283) {8};
\gpcolor{rgb color={0.702,0.702,0.702}}
\gpsetlinetype{gp lt axes}
\gpsetdashtype{gp dt axes}
\gpsetlinewidth{1.00}
\draw[gp path] (2.115,5.126)--(11.671,5.126);
\gpcolor{rgb color={0.400,0.400,0.400}}
\gpsetlinetype{gp lt border}
\gpsetdashtype{gp dt solid}
\gpsetlinewidth{2.00}
\draw[gp path] (2.115,5.126)--(2.048,5.126);
\gpcolor{rgb color={0.702,0.702,0.702}}
\gpsetlinetype{gp lt axes}
\gpsetdashtype{gp dt axes}
\gpsetlinewidth{1.00}
\draw[gp path] (2.115,5.568)--(11.671,5.568);
\gpcolor{rgb color={0.400,0.400,0.400}}
\gpsetlinetype{gp lt border}
\gpsetdashtype{gp dt solid}
\gpsetlinewidth{2.00}
\draw[gp path] (2.115,5.568)--(2.048,5.568);
\gpcolor{rgb color={0.702,0.702,0.702}}
\gpsetlinetype{gp lt axes}
\gpsetdashtype{gp dt axes}
\gpsetlinewidth{1.00}
\draw[gp path] (2.115,5.870)--(11.671,5.870);
\gpcolor{rgb color={0.400,0.400,0.400}}
\gpsetlinetype{gp lt border}
\gpsetdashtype{gp dt solid}
\gpsetlinewidth{2.00}
\draw[gp path] (2.115,5.870)--(2.048,5.870);
\gpcolor{rgb color={0.702,0.702,0.702}}
\gpsetlinetype{gp lt axes}
\gpsetdashtype{gp dt axes}
\gpsetlinewidth{1.00}
\draw[gp path] (2.115,6.100)--(2.391,6.100);
\draw[gp path] (6.435,6.100)--(11.671,6.100);
\gpcolor{rgb color={0.400,0.400,0.400}}
\gpsetlinetype{gp lt border}
\gpsetdashtype{gp dt solid}
\gpsetlinewidth{2.00}
\draw[gp path] (2.115,6.100)--(2.048,6.100);
\gpcolor{rgb color={0.702,0.702,0.702}}
\gpsetlinetype{gp lt axes}
\gpsetdashtype{gp dt axes}
\gpsetlinewidth{1.00}
\draw[gp path] (2.115,6.285)--(2.391,6.285);
\draw[gp path] (6.435,6.285)--(11.671,6.285);
\gpcolor{rgb color={0.400,0.400,0.400}}
\gpsetlinetype{gp lt border}
\gpsetdashtype{gp dt solid}
\gpsetlinewidth{2.00}
\draw[gp path] (2.115,6.285)--(1.980,6.285);
\node[gp node right,font={\fontsize{8.0pt}{9.6pt}\selectfont}] at (1.888,6.285) {64};
\gpcolor{rgb color={0.702,0.702,0.702}}
\gpsetlinetype{gp lt axes}
\gpsetdashtype{gp dt axes}
\gpsetlinewidth{1.00}
\draw[gp path] (2.115,7.128)--(2.391,7.128);
\draw[gp path] (6.435,7.128)--(11.671,7.128);
\gpcolor{rgb color={0.400,0.400,0.400}}
\gpsetlinetype{gp lt border}
\gpsetdashtype{gp dt solid}
\gpsetlinewidth{2.00}
\draw[gp path] (2.115,7.128)--(2.048,7.128);
\gpcolor{rgb color={0.702,0.702,0.702}}
\gpsetlinetype{gp lt axes}
\gpsetdashtype{gp dt axes}
\gpsetlinewidth{1.00}
\draw[gp path] (2.115,7.570)--(2.391,7.570);
\draw[gp path] (6.435,7.570)--(11.671,7.570);
\gpcolor{rgb color={0.400,0.400,0.400}}
\gpsetlinetype{gp lt border}
\gpsetdashtype{gp dt solid}
\gpsetlinewidth{2.00}
\draw[gp path] (2.115,7.570)--(2.048,7.570);
\gpcolor{rgb color={0.702,0.702,0.702}}
\gpsetlinetype{gp lt axes}
\gpsetdashtype{gp dt axes}
\gpsetlinewidth{1.00}
\draw[gp path] (2.115,7.872)--(2.391,7.872);
\draw[gp path] (6.435,7.872)--(11.671,7.872);
\gpcolor{rgb color={0.400,0.400,0.400}}
\gpsetlinetype{gp lt border}
\gpsetdashtype{gp dt solid}
\gpsetlinewidth{2.00}
\draw[gp path] (2.115,7.872)--(2.048,7.872);
\gpcolor{rgb color={0.702,0.702,0.702}}
\gpsetlinetype{gp lt axes}
\gpsetdashtype{gp dt axes}
\gpsetlinewidth{1.00}
\draw[gp path] (2.115,8.102)--(2.391,8.102);
\draw[gp path] (6.435,8.102)--(11.671,8.102);
\gpcolor{rgb color={0.400,0.400,0.400}}
\gpsetlinetype{gp lt border}
\gpsetdashtype{gp dt solid}
\gpsetlinewidth{2.00}
\draw[gp path] (2.115,8.102)--(2.048,8.102);
\gpcolor{rgb color={0.702,0.702,0.702}}
\gpsetlinetype{gp lt axes}
\gpsetdashtype{gp dt axes}
\gpsetlinewidth{1.00}
\draw[gp path] (2.115,8.287)--(11.671,8.287);
\gpcolor{rgb color={0.400,0.400,0.400}}
\gpsetlinetype{gp lt border}
\gpsetdashtype{gp dt solid}
\gpsetlinewidth{2.00}
\draw[gp path] (2.115,8.287)--(1.980,8.287);
\node[gp node right,font={\fontsize{8.0pt}{9.6pt}\selectfont}] at (1.888,8.287) {512};
\gpcolor{rgb color={0.702,0.702,0.702}}
\gpsetlinetype{gp lt axes}
\gpsetdashtype{gp dt axes}
\gpsetlinewidth{1.00}
\draw[gp path] (3.341,1.613)--(3.341,6.028);
\draw[gp path] (3.341,8.107)--(3.341,8.287);
\gpcolor{rgb color={0.400,0.400,0.400}}
\gpsetlinetype{gp lt border}
\gpsetdashtype{gp dt solid}
\gpsetlinewidth{2.00}
\draw[gp path] (3.341,1.613)--(3.341,1.478);
\node[gp node center,font={\fontsize{8.0pt}{9.6pt}\selectfont}] at (3.341,1.016) {24};
\gpcolor{rgb color={0.702,0.702,0.702}}
\gpsetlinetype{gp lt axes}
\gpsetdashtype{gp dt axes}
\gpsetlinewidth{1.00}
\draw[gp path] (5.281,1.613)--(5.281,6.028);
\draw[gp path] (5.281,8.107)--(5.281,8.287);
\gpcolor{rgb color={0.400,0.400,0.400}}
\gpsetlinetype{gp lt border}
\gpsetdashtype{gp dt solid}
\gpsetlinewidth{2.00}
\draw[gp path] (5.281,1.613)--(5.281,1.478);
\node[gp node center,font={\fontsize{8.0pt}{9.6pt}\selectfont}] at (5.281,1.016) {96};
\gpcolor{rgb color={0.702,0.702,0.702}}
\gpsetlinetype{gp lt axes}
\gpsetdashtype{gp dt axes}
\gpsetlinewidth{1.00}
\draw[gp path] (7.222,1.613)--(7.222,8.287);
\gpcolor{rgb color={0.400,0.400,0.400}}
\gpsetlinetype{gp lt border}
\gpsetdashtype{gp dt solid}
\gpsetlinewidth{2.00}
\draw[gp path] (7.222,1.613)--(7.222,1.478);
\node[gp node center,font={\fontsize{8.0pt}{9.6pt}\selectfont}] at (7.222,1.016) {384};
\gpcolor{rgb color={0.702,0.702,0.702}}
\gpsetlinetype{gp lt axes}
\gpsetdashtype{gp dt axes}
\gpsetlinewidth{1.00}
\draw[gp path] (9.163,1.613)--(9.163,8.287);
\gpcolor{rgb color={0.400,0.400,0.400}}
\gpsetlinetype{gp lt border}
\gpsetdashtype{gp dt solid}
\gpsetlinewidth{2.00}
\draw[gp path] (9.163,1.613)--(9.163,1.478);
\node[gp node center,font={\fontsize{8.0pt}{9.6pt}\selectfont}] at (9.163,1.016) {1536};
\gpcolor{rgb color={0.702,0.702,0.702}}
\gpsetlinetype{gp lt axes}
\gpsetdashtype{gp dt axes}
\gpsetlinewidth{1.00}
\draw[gp path] (11.103,1.613)--(11.103,8.287);
\gpcolor{rgb color={0.400,0.400,0.400}}
\gpsetlinetype{gp lt border}
\gpsetdashtype{gp dt solid}
\gpsetlinewidth{2.00}
\draw[gp path] (11.103,1.613)--(11.103,1.478);
\node[gp node center,font={\fontsize{8.0pt}{9.6pt}\selectfont}] at (11.103,1.016) {6144};
\draw[gp path] (2.115,8.287)--(2.115,1.613)--(11.671,1.613)--(11.671,8.287)--cycle;
\gpcolor{color=gp lt color border}
\node[gp node center,rotate=-270] at (1.105,4.950) {Speedup (strong)};
\node[gp node center] at (6.893,0.400) {Number of processes};
\gpcolor{rgb color={0.400,0.400,0.400}}
\gpsetlinewidth{1.00}
\draw[gp path] (2.391,6.028)--(2.391,8.107)--(6.435,8.107)--(6.435,6.028)--cycle;
\gpcolor{color=gp lt color border}
\node[gp node left] at (3.399,7.760) {$S_{\text{lin,init}}$};
\gpcolor{rgb color={0.800,0.000,0.063}}
\gpsetlinewidth{3.00}
\draw[gp path] (2.667,7.760)--(3.123,7.760);
\draw[gp path] (3.341,2.280)--(5.281,3.768)--(7.222,4.369)--(9.163,4.787)--(11.103,4.893);
\gpsetpointsize{5.20}
\gppoint{gp mark 1}{(3.341,2.280)}
\gppoint{gp mark 1}{(5.281,3.768)}
\gppoint{gp mark 1}{(7.222,4.369)}
\gppoint{gp mark 1}{(9.163,4.787)}
\gppoint{gp mark 1}{(11.103,4.893)}
\gppoint{gp mark 1}{(2.895,7.760)}
\gpcolor{color=gp lt color border}
\node[gp node left] at (3.399,7.298) {$S_{\text{lin,solve}}$};
\gpcolor{rgb color={0.800,0.000,0.663}}
\draw[gp path] (2.667,7.298)--(3.123,7.298);
\draw[gp path] (3.341,2.280)--(5.281,3.585)--(7.222,4.502)--(9.163,5.334)--(11.103,5.483);
\gppoint{gp mark 2}{(3.341,2.280)}
\gppoint{gp mark 2}{(5.281,3.585)}
\gppoint{gp mark 2}{(7.222,4.502)}
\gppoint{gp mark 2}{(9.163,5.334)}
\gppoint{gp mark 2}{(11.103,5.483)}
\gppoint{gp mark 2}{(2.895,7.298)}
\gpcolor{color=gp lt color border}
\node[gp node left] at (3.399,6.836) {$S_{\text{newton}}$};
\gpcolor{rgb color={0.329,0.000,0.800}}
\draw[gp path] (2.667,6.836)--(3.123,6.836);
\draw[gp path] (3.341,2.280)--(5.281,3.609)--(7.222,4.725)--(9.163,5.952)--(11.103,6.722);
\gppoint{gp mark 3}{(3.341,2.280)}
\gppoint{gp mark 3}{(5.281,3.609)}
\gppoint{gp mark 3}{(7.222,4.725)}
\gppoint{gp mark 3}{(9.163,5.952)}
\gppoint{gp mark 3}{(11.103,6.722)}
\gppoint{gp mark 3}{(2.895,6.836)}
\gpcolor{color=gp lt color border}
\node[gp node left] at (3.399,6.374) {$S_{\text{ideal}}$};
\gpcolor{rgb color={0.000,0.263,0.800}}
\draw[gp path] (2.667,6.374)--(3.123,6.374);
\draw[gp path] (3.341,2.280)--(5.281,3.615)--(7.222,4.950)--(9.163,6.285)--(11.103,7.620);
\gpdefrectangularnode{gp plot 1}{\pgfpoint{2.115cm}{1.613cm}}{\pgfpoint{11.671cm}{8.287cm}}
\end{tikzpicture}
 }
\caption{Shape Derivative Solver}
		\label{tab:strongShapeDer}
	\end{subfigure}
\vskip\baselineskip
	\begin{subfigure}[t]{\textwidth}
		\centering
		\begin{tabular}{rrrccc}
			\\
			\toprule
			Procs  & Refs  & DoFs  & Linear solver  & Newton solver   & Linear solver     \\
			&       &        &  (elasticity)      & (shape derivative)    & (shape derivative)\\ 
			\midrule
			24 & 4 & 6,085,035 & 159 & 16 & 293  \\
			96 & 4 & 6,085,035 & 158 & 16 & 295  \\
			384 & 4 & 6,085,035 & 161 & 16 & 289  \\
			1,536 & 4 & 6,085,035 & 163 & 16 & 293  \\
			6,144 & 4 & 6,085,035 & 162 & 16 & 296 \\
			\bottomrule
		\end{tabular}
		\caption{Accumulated iteration counts for the strong scaling study}
		\label{tab:StrongScaling}
	\end{subfigure}
	\caption{\textbf{Strong scaling:} For the first two optimization steps, the accumulated wallclock time is shown for: (a) the elastic solver, (b) the shape derivative solver. In (c), the accumulated iteration counts are presented for solving the elasticity PDE-constraint with geometric multigrid, the required Newton steps to compute the shape derivative, and the accumulated geometric multigrid steps to solve the linearization within the Newton algorithm}
	\label{fig:strong_scaling}
\end{figure}

\section{Conclusion and outlook}
\label{sec:conclusion}

In this article, we have extended a gradient-penalized shape optimization algorithm \cite{VogelSiebenborn2019} to a 3d test setting. The methodology is evaluated for a purely geometric benchmarking test searching for a space-filling design with minimal surface area. The numerical results reflect Kelvin's conjecture and show tetrakaidecahedral shapes. A second study combines the geometric constraints with a linear elastic model for a composite of materials with different properties. 
The optimization problem searches, starting from a reference configuration, for a stiffer material by relocating and reshaping the composite.
The numerical results show a cell arrangement similar to the structure of the upper layer of the human skin. Preventing degeneration of the mesh elements between interfaces and thus maintaining a certain thickness of the inter-cellular channels is shown to be achievable by descent directions with a gradient penalization.
In addition, we focused on the strong and weak scalability of the proposed algorithm. The strong scaling showed reasonable results taking into consideration the use of a serial LU base solver for the multigrid preconditioner. The weak scaling results were close to ideal for up to 12,288 cores and 3 billion DoFs. Overall, multigrid scalability was retained, which is crucial for future work on larger, more realistic models.  Although we used a linear elastic model, the proposed shape optimization methodology can be expanded to other material models employing the same penalization technique. The presented 3d results are a significant step towards applying the optimization algorithm to models with biologically realistic material properties. 

\subsection*{Acknowledgment}
The simulations were performed on the national supercomputer Cray XC40 {\it Hazel Hen} at the High Performance Computing Center Stuttgart (HLRS) under the grant {\it ShapeOptCompMat} (ACID 44171, Shape Optimization for 3d Composite Material Models). The authors gratefully acknowledge the HLRS for supporting this project by providing computing time on the GCS Supercomputer {\it Hazel Hen} and their kind support.

\bibliographystyle{abbrvnat}      \bibliography{update_citations}   

\end{document}